\documentclass[11pt]{article}
\usepackage{hyperref}
\usepackage{graphicx}
\usepackage{graphics}
\usepackage{dsfont}
\usepackage{epsfig}
\usepackage{bbm}
\usepackage{amsmath,amssymb,amsthm,amscd}
\usepackage[sort&compress,square,numbers]{natbib}
\usepackage{float}
\usepackage{braket}
\restylefloat{table}
\usepackage{mathdots}
\usepackage{blkarray}
\usepackage{multirow}
\usepackage{xspace}

\usepackage{tikz,textcomp}
\usetikzlibrary{decorations.pathreplacing,calc,fadings,fit}
\usetikzlibrary{shapes,arrows,positioning,chains,matrix}

\usepackage{listings}

\usepackage[OT2,T1]{fontenc}
\DeclareSymbolFont{cyrletters}{OT2}{wncyr}{m}{n}
\DeclareMathSymbol{\Sha}{\mathalpha}{cyrletters}{"58}

\newcommand{\one}[0]{\ensuremath{\mathbf{1} }\xspace}
\newcommand{\two}[0]{\ensuremath{\mathbf{2} }\xspace}

\newcommand{\three}[0]{\ensuremath{\mathbf{3} }\xspace}

\def\Mgn[#1]#2{{\overline{\cal M}_{#1,#2}}}
\def\pqs[#1,#2]{{\footnotesize{$\left[\begin{array}{c} #1\\#2  \end{array}\right]$}}} 
\def\pqsu[#1,#2]{\left[\begin{array}{c} #1\\#2  \end{array}\right]} 
\def\pqssu[#1,#2]{{\footnotesize{\left[\begin{array}{c} #1\\#2  \end{array}\right]}}} 
\def\pqh[#1,#2]{{\footnotesize{$\left[\begin{array}{c} #1\\#2  \end{array}\right]$}}} 
\def\pqhu[#1,#2]{\left[\begin{array}{c} #1\\#2  \end{array}\right]}

\newcommand{\ba}{\begin{eqnarray*}}
\newcommand{\ea}{\end{eqnarray*}}
\newcommand{\ban}{\begin{eqnarray}}
\newcommand{\ean}{\end{eqnarray}}
\newcommand{\be}{\begin{equation}}
\newcommand{\ee}{\end{equation}}
\newcommand{\ben}{\begin{equation}}
\newcommand{\een}{\end{equation}}
\numberwithin{equation}{section}

\numberwithin{equation}{section}

\begin{document}
\begin{titlepage}

\begin{center}
\hfill UWThPh-2019-39\\
\vspace*{ 2.0cm}
{\Large {\bf GV-Spectroscopy for F-theory on genus-one fibrations}}\\[12pt]
\vspace{-0.1cm}
\bigskip
\bigskip 
 {
{{Paul-Konstantin~Oehlmann}$^{\,\text{a}}$} and {{Thorsten~Schimannek}$^{\,\text{b}} $}
\bigskip }\\[3pt]
\vspace{0.cm}
{
	{\it${}^{\text{a}}$ Department of Physics and Astronomy,~Uppsala University,\\~Regementsv\"agen 1, 77120 Uppsala,~Sweden}\\paul-konstantin.oehlmann@physics.uu.se\\[.5em]
	{\it ${}^{\text{b}}$ Faculty of Physics,~University of Vienna,\\~Boltzmanngasse 5,~1090 Vienna,~Austria}\\thorsten.schimannek@univie.ac.at
}
\\[2.0cm]
\end{center}

\begin{abstract}
	We present a novel technique to obtain base independent expressions for the matter loci of fibrations of complete intersection Calabi-Yau onefolds in toric ambient spaces.
	These can be used to systematically construct elliptically and genus one fibered Calabi-Yau $d$-folds that lead to desired gauge groups and spectra in F-theory. 
	The technique, which we refer to as GV-spectroscopy, is based on the calculation of fiber Gopakumar-Vafa invariants using the Batyrev-Borisov construction of mirror pairs
	and application of the so-called Frobenius method to the data of a parametrized auxiliary polytope.
	In particular for fibers that generically lead to multiple sections, only multi-sections or that are complete intersections in higher codimension, our technique is vastly more efficient than
	classical approaches.
	As an application we study two Higgs chains of six-dimensional supergravities that are engineered by fibrations of codimension two complete intersection fibers.
	Both chains end on a vacuum with $G=\mathbb{Z}_4$ that is engineered by fibrations of bi-quadrics in $\mathbb{P}^3$.
	We use the detailed knowledge of the structure of the reducible fibers that we obtain from GV-spectroscopy to comment on the corresponding Tate-Shafarevich group.
	We also show that for all fibers the six-dimensional supergravity anomalies including the discrete anomalies generically cancel.
\end{abstract}

\end{titlepage}
\clearpage
\setcounter{footnote}{0}
\setcounter{tocdepth}{2}
\tableofcontents
\clearpage

\section{Introduction}
F-theory provides a powerful dictionary between non-perturbative type IIB string compactifications and the geometry of genus one fibered Calabi-Yau manifolds~\cite{Vafa:1996xn,Morrison:1996na,Morrison:1996pp}.
In particular, for any given Calabi-Yau manifold the physical consistency conditions that arise, for example due to potential anomalies in the corresponding effective theories, are satisfied automatically.
Various limits and dualities connect F-theory to all of the perturbative string theories as well as M-theory.
This ubiquity of F-theoretic descriptions within the string theory landscape together with the powerful geometrization make F-theory an ideal study ground for questions about string universality and the so-called swampland.
It is therefore of particular interest to develop tools to construct and analyze genus one fibered Calabi-Yau manifolds.

Toric geometry can be used to construct an abundant amount of genus one fibered Calabi-Yau manifolds.
In particular, the so-called \textit{fiber based approach} allows to systematically engineer geometries such that the corresponding F-theory vacua exhibit particular physical features~\cite{Cvetic:2013nia,Cvetic:2013uta,Klevers:2014bqa,Braun:2014qka}.
To this end one starts with a family of Calabi-Yau onefolds, i.e. tori, that are realized as complete intersections in toric ambient spaces.
The coefficients of the defining equations can then be promoted to sections of line bundles over a base $B$.
If the fiber is a complete intersection in codimension $r$ then the family of fibrations is determined by the choice of $B$ together with $2r$ line bundles.

Assuming that the sections are sufficiently generic, the gauge group as well as the spectrum can be expressed entirely in terms of inequalities among the Chern classes of the line bundles.
Moreover, the multiplicities of hypermultiplets in any given representation are integrals over polynomials in the Chern classes over $B$~\footnote{For Calabi-Yau fourfolds one also has to choose
the $G_4$ flux that enters the calculation.}.
In the following we will refer to the set of inequalities and polynomials as the \textit{fibration data}.

The fibration data can, at least in principle, be obtained in a completely base independent fashion.
In this way the fiber based approach allows to systematically engineer genus one fibered Calabi-Yau manifolds from building blocks that can be studied and classified independently.
However, it can be extremely difficult to determine the fibration data for a given family of fibers.
This is true in particular for families of fibers that generically lead to multiple sections, multi-sections or fibers that are complete intersections in higher codimension.
In this paper we present a novel technique to calculate the base independent fibration data for complete intersection fibers in toric ambient spaces.

We will argue in Section~\ref{sec:count} that certain Gopakumar-Vafa (GV) invariants, which we refer to as \textit{fiber GV invariants}, encode the structure of the singular fibers of a genus one fibered Calabi-Yau threefold.
Although the direct calculation of GV invariants is a hard problem in itself, it is often relatively straightforward to obtain the invariants via mirror symmetry.
For complete intersections in toric ambient spaces the Batyrev-Borisov construction~\cite{Batyrev:1994pg} associates a mirror pair to every nef partition of a reflexive polytope, see Section~\ref{sec:batbori}.
Moreover, techniques to calculate the GV invariants for such mirror pairs have been developed in~\cite{Hosono:1993qy,Hosono:1994ax}.
In Section~\ref{sec:gvcalculation_baseindependent} we argue, that the methods from~\cite{Hosono:1994ax} can be applied to a certain auxiliary polytope that is parametrized by the choices of line bundles on $B=\mathbb{P}^2$.

The resulting fiber GV invariants are polynomials in the Chern classes of the line bundles and the base and provide the \textit{base independent} fibration data.
This technique, which we refer to as \textit{GV-spectroscopy}, provides a very efficient way to obtain the fibration data for complete intersection fibers in toric ambient spaces.
In particular, using a complete intersection fiber that realizes the gauge group $G=SU(3)\times SU(2)\times\mathbb{Z}_2$ we illustrate that the Dynkin labels of matter curves can be immediately read off from the degrees of the fiber invariants.
It is then almost trivial to assemble them into representations.
The corresponding family of fibers also leads to non-flat fibers and the corresponding supergravities will generically be coupled to superconformal matter.
We demonstrate the viability of the technique with many other examples and in each case we check that all of the anomalies of the associated six-dimensional F-theory vacua cancel.
However, note that the resulting expressions can also be used to engineer Calabi-Yau $n$-folds with $n>3$.

The application of GV-spectroscopy is particularly easy when the fiber leads to a small gauge group $G$.
This applies in particular to the bi-quadrics in $\mathbb{P}^3$ which lead to genus one fibrations with four-sections and $G=\mathbb{Z}_4$~\footnote{This fiber has
also been discussed in~\cite{Braun:2014qka,Kimura:2019bzv,Kimura:2019qxf,Kimura:2019syr} but the matter loci have not been calculated.}.
We explicitly calculate the corresponding fiber GV invariants and determine the base independent fibration data in Section~\ref{sec:gvcalculation_baseindependent}.
Let us stress that these results would be extremely hard, perhaps even impossible to obtain using classical approaches.
The main reason is that the necessary ideal decompositions require too much computing time and memory.

We then study two sets of complete intersection fibers that are related via extremal transitions to the generic bi-quadrics.
Analogous to the situation for hypersurfaces discussed in~\cite{Klevers:2014bqa}, the corresponding effective theories are connected via Higgs transitions.
In particular, we find both a Higgs transition from a model with $G=U(1)$ and a transition from $G=U(1)\times\mathbb{Z}_2$ to the vacua realized from bi-quadrics with $G=\mathbb{Z}_4$.
The family with $G=U(1)$ can in turn be related via (un)Higgsings to the model with $G=U(1)^3$ presented in~\cite{Cvetic:2013qsa}.
Again, GV-spectroscopy turns out to be vastly more efficient than earlier approaches and we provide the complete base independent spectra for all fibers.
We thereby correct some of the previous results for $G=U(1)^3$.
Another family of fibers that seems to engineer $G=SU(2)\times\mathbb{Z}_4$ and also Higgses to $G=\mathbb{Z}_4$ will, due to subtleties associated with multiple fibers, be discussed in Appendix~\ref{sec:multi}.
Note that the maps into the Weierstrass form and the generic gauge groups have been determined in~\cite{Braun:2014qka}~\footnote{F-theory compactifications on complete intersections in toric ambient spaces
have also been studied from a non-fiber based perspective in~\cite{Anderson:2016ler}. GV-spectroscopy can also be applied to those geometries.}.
We follow their convention and label nef partitions by (\#poly, \#nef) where the indices of polytopes are taken with respect to the ordering of the Kreuzer-Skarke list of three-dimensional reflexive polytopes~\cite{Kreuzer:1998vb}.
The indices of nef partitions follow the order generated by PALP~\cite{Braun:2012vh}.

On a more formal note, let us point out the similarity to the enumerative geometry of K3 surfaces.
Ordinary Gromov-Witten invariants vanish for K3 surfaces because the latter can always be deformed into surfaces that do not have any complex submanifolds.
However, one can define a reduced Gromov-Witten theory that allows only deformations in certain subfamilies.
The corresponding invariants are non-trivial and directly related to the fiber Gopakumar-Vafa invariants of K3-fibered Calabi-Yau threefolds~\cite{maulik2007gromovwitten,klemm2008noetherlefschetz}.

Another interesting property of the fiber invariants that arise from genus one fibered Calabi-Yau threefolds is that they combine into a Jacobi form~\cite{Cota:2019cjx}.
This also holds for the base independent invariants which then lead to families of Jacobi forms that are parametrized by the choice of a base and the classes of line bundles.
Let us also recall that Batyrev-Borisov mirror symmetry for families of fibers is conjectured to exchange fibers that generically lead to genus one fibrations with multi-sections
and fibers that lead to torsional sections in the Mordell-Weil group~\cite{Klevers:2014bqa,Braun:2014qka,Oehlmann:2016wsb}.
All of this suggests that enumerative geometry and mirror symmetry for families of Calabi-Yau onefolds is surprisingly non-trivial and warrants further exploration.

The structure of the paper is as follows:
In Section~\ref{sec:ftpre} we review some generalities about F-theory compactifications.
Of particular importance will be the geometric origin of BPS states in the circle compactified F-theory and, for the analysis of the Higgs chains, the relation between genus one fibrations and discrete symmetries.
Then, in Section~\ref{sec:fibsfromfams}, we describe the fiber based approach to the construction of elliptic and genus one fibrations.
There we will introduce the problem that GV-spectroscopy is going to solve.

Section~\ref{sec:GVspec} is the technical heart of the paper.
First, in Section~\ref{sec:count}, we review the definition of Gopakumar-Vafa invariants and how they relate to the F-theory spectrum.
This is followed in Section~\ref{sec:batbori} by a brief summary of the Batyrev-Borisov construction of mirror pairs of Calabi-Yau complete intersections in toric ambient spaces.
In Section~\ref{sec:gvcalculation_general} we review how the Gopakumar-Vafa invariants for Batyrev-Borisov Calabi-Yaus can be calculated using mirror symmetry.
Some time is spent to explain the individual steps of the calculation.
The procedure is then illustrated at the hand of a fibration of bi-quadrics over $\mathbb{P}^2$ in Section~\ref{sec:ex1bq}.
We are then fully equipped to generalize this technique to enable the base independent calculation of fiber GV invariants.
This is explained in Section~\ref{sec:gvcalculation_baseindependent} at the hand of an example.
The crucial step is to apply the machinery from Section~\ref{sec:gvcalculation_general} to a parametrized auxilliary polytope.
In Section~\ref{sec:su3su2z2model} we provide another example with gauge group $G=SU(3)\times SU(2)\times\mathbb{Z}_2$ and non-flat fibers.
The example also illustrates that the individual weights of states from reducible fibers can be read off directly from the GV invariants.

In Section~\ref{sec:CICYFI} we are then studying families of codimension two complete intersection fibers that arrange into two Higgs chains.
The field theoretic analysis and in particular a discussion of the discrete anomalies as well as the implications for the Tate-Shafarevich group will be performed in Section~\ref{sec:FieldTheory}.

During our work on this project we learned of~\cite{Kashani-Poor:2019jyo} which has some overlap.

 \section*{Acknowledgments}
We would like to thank Albrecht Klemm, Dami$\acute{\text{a}}$n Kaloni Mayorga Pe$\tilde{\text{n}}$a, Cesar Fierro Cota, Nikhil Raghuram and Fabian Ruehle for interesting discussions.
 P. K. O. would like to acknowledge the supported by the individual DFG grant OE 657/1-1 during the main part of this work. P. K. O. also acknowledges support by a stipend of the Carl Tryggers Foundation no. CTS 18:213 during the completion of this work. 
The work of T.S. is supported by the Austrian Science Fund (FWF): P30904-N27.
P. K. O. would like to gratefully acknowledge the hospitality of the Simons Center for Geometry and Physics (and the semester long program, \emph{The Geometry and Physics of Hitchin Systems}) during the completion of this work.

\section{F-theory preliminaries}
\label{sec:ftpre}
Before we turn towards Gopakumar-Vafa invariants, let us briefly review the relevant entries of the F-theory dictionary that relate physical properties of six-dimensional $\mathcal{N}=1$ supergravities to the geometry of genus one fibered Calabi-Yau threefolds.
The connection is most transparent when the theory is compactified further to five dimensions.
It is then dual to M-theory compactified on the Calabi-Yau.
As will become clear when we review their definition, this is also the physical reason why Gopakumar-Vafa invariants encode the F-theory spectrum.
But first, let us discuss the geometric origin of gauge symmetries, matter representations and discrete symmetries.
Note that everything in this section is standard and for more details we refer the reader to~\cite{Weigand:2018rez,Cvetic:2018bni}.

\subsection{Gauge symmetry and matter from reducible fibers}
From a six-dimensional perspective the gauge symmetry arises from stacks of seven-branes and their position is encoded in the complex structure of the fiber, the so-called $\tau$-profile, of an elliptically or genus one fibered Calabi-Yau threefold $M$.
More precisely, the branes are wrapping curves over which the fiber becomes singular and the type of singularity determines the type of brane stack.
The union of those curves is called the \textit{discriminant locus}.
Matter is located on intersections of brane stacks or at loci where the curve wrapped by the branes itself becomes singular.
It is possible to derive the gauge group and the matter representations from this perspective but requires significant effort.

The relation between geometry and physics is clearer when F-theory is further compactified on a circle.
Then a Coulomb branch opens up and at a generic point the theory is dual to M-theory on the Calabi-Yau.
The Wilson line parameters are identified with K\"ahler moduli of curves that resolve the singularities in the fibers.
This is the situation that we will now consider.

It has been shown by Kodaira that singularities of the genus one fiber, over components of the discriminant locus that are in codimension one of the base, admit an ADE-classification.
We can decompose the discriminant into irreducible curves $\mathcal{S}_{\mathfrak{g}_I}^b\subset B$ such that the singularity type over $\mathcal{S}_{\mathfrak{g}_I}^b$ corresponds to the ADE type Lie-algebra $\mathfrak{g}_I$.
Singularities from curves of singular fibers can be crepantly resolved and on a generic point of the K\"ahler moduli space the generic fibers over $\mathcal{S}_{\mathfrak{g}_I}^b$ become reducible with $\text{rk}(\mathfrak{g}_I)+1$ rational components.
Moreover, the latter intersect like the affine Dynkin diagram of $\mathfrak{g}_I$.
We will denote components of reducible fibers collectively as \textit{fibral curves}.

The fibral curves can experience monodromies along closed paths around special points in $\mathcal{S}_{\mathfrak{g}_I}^b$ and this leads to a corresponding action on the Dynkin diagram.
Let us denote the Lie-algebra that corresponds to the folded Dynkin diagram by $\tilde{\mathfrak{g}}_I$.
The total spaces of monodromy invariant combinations of fibral curves fibered over $\mathcal{S}_{\mathfrak{g}_I}$ lead to \textit{fibral divisors} in the Calabi-Yau.

The M-theory three-form can be reduced along the harmonic forms that correspond to the fibral divisors resolving $\pi^{-1}(\mathcal{S}_{\mathfrak{g}_I})$ and this leads to vector bosons that generate the Cartan algebra of $\tilde{\mathfrak{g}}_I$.
Additional vector bosons arise from M2-branes that wrap the fibral curves and, in the limit where the volumes of the fibral curves that do not correspond to the affine node of the Dynkin diagram go to zero, those become massless.
This leads to an enhanced gauge symmetry with the algebra of the gauge group given by $\tilde{\mathfrak{g}}_I$.
Note that if in addition the volume of the curve representing the affine node is sent to zero this amounts to the decompactification limit of F-theory.
The M2-branes that wrap the fibral curve also yield $2g_I$ half-hypermultiplets that are BPS saturated, where $g_I$ is the genus of $\mathcal{S}_{\mathfrak{g}_I}$~\cite{Witten:1996qb,Katz:1996ht}.
In the F-theory limit they combine with $g_I\cdot\text{rk}(\tilde{\mathfrak{g}})$ neutral half-hypermultiplets, see below, into massless hypermultiplets that form $g_I$ adjoint representations of $\tilde{\mathfrak{g}}_I$.

The Shioda-Tate-Wazir theorem~\cite{TateWazir} states that for flat elliptic fibrations the group of divisors is generated by fibral divisors, sections and vertical divisors.
Recall that \textit{vertical divisors} are preimages of divisors in the base under the projection that determines the fibration.
The theorem can be extended to genus one fibrations with multi-sections~\cite{Braun:2014oya} and the corresponding divisors replace those of sections as generators.
In M-theory all of the corresponding harmonic forms lead to vector bosons via reduction of the three-form.
However, only those that arise from fibral divisors and particular linear combinations of (multi-)sections are compatible with the F-theory limit.
The other vector fields result from Kaluza-Klein reductions of six-dimensional two-form fields.

Those non-fibral divisors for which the corresponding vector bosons lift to six dimensions can be obtained by an orthogonalization procedure, the so-called \textit{Shioda map} $\sigma$, that acts on the set of multi-sections~\footnote{Note
that this terminology is somewhat non-standard and the Shioda map is originally defined as a homomorphism from the group of sections of an elliptic fibration into the group of divisors.
The interpretation as an orthogonalization procedure in F-theory make it natural to generalize it to genus one fibrations that only exhibit multi-sections.
However, note that the multi-sections do not form a group and the generalized Shioda map is therefore not a homomorphism (see however~\cite{Grimm:2015wda}).}.
To define the Shioda map one first has to pick a multi-section $E_0$ that essentially corresponds to the Kaluza-Klein vector from the graviton.
In a slight abuse of notation we refer to such an $N$-section as the \textit{zero $N$-section}.
The image of another multi-section $E_1$ under the Shioda map is defined as
\begin{align}
	\sigma(E_1)=E_1-E_0+\tilde{D}\,,
\end{align}
where $\tilde{D}$ is the unique linear combination of vertical and fibral divisors which ensures that $\sigma(E_1)$ does not intersect non-affine fibral curves and that it is compatible with the F-theory limit.
The precise form of $\tilde{D}$ will not be relevant to our discussion and for details we refer the reader to~\cite{Klevers:2014bqa}.
If there are, up to linear equivalence and multiples of vertical or fibral divisors, $n$ independent (multi-)sections, this leads to an additional $U(1)^{n-1}$ factor of the gauge group.
Genus one fibrations that do not have a section also lead to discrete symmetries~\cite{Braun:2014oya}.
We will discuss this in more detail in the next section.

Charged matter arises when the singularities of the fiber are enhanced over isolated points of the base.
We will assume that the Calabi-Yau still admits a crepant resolution but note that this is not guaranteed to be the case.
The geometry of threefolds where this assumption is violated has been discussed in~\cite{Grassi:2018rva}.
If the Calabi-Yau is fully resolved then the enhanced singularities are resolved by isolated fibral curves and M2-branes that wrap those curves lead to BPS states in the effective theory~\cite{Witten:1996qb,Katz:1996ht}.
The weights and $U(1)$-charges of the state that corresponds to an isolated fibral curve $C$ are respectively determined by the intersections of this curve with fibral divisors and images of multi-sections under the Shioda map.
The spectrum also contains $2h^{2,1}(M)+2$ massless and uncharged half-hypermultiplets from the reduction of the M-theory three-form along harmonic three-forms on $M$.

Finally, another source of BPS states are M2-branes that wrap curves in the base of the Calabi-Yau.
Under the duality with F-theory those can be identified with excitations of D3-branes that wrap the curve in the base as well as the compactification circle.
From a six-dimensional perspective the D3-branes appear as non-critical strings~\cite{Witten:1996qb,Klemm:1996hh} which are charges of tensor fields.

In order to streamline the exposition we have omitted many details e.g. about the correct choice and normalization of the fibral divisors and for those we refer to the review~\cite{Weigand:2018rez}.
None of the omissions will be essential to our discussion.

\subsection{Discrete symmetries and genus one fibrations}
We are now giving a brief overview of the origin of discrete symmetries in F-theory and their connection to M-theory on genus-one fibered Calabi-Yau 3-folds~\cite{deBoer:2001wca,Braun:2014oya,Morrison:2014era,Mayrhofer:2014haa,Anderson:2014yva}~\footnote{Similar arguments hold for higher dimensional Calabi-Yau geometries.}.
This is not needed for the discussion of GV-spectroscopy but will be relevant for our analysis in Section~\ref{sec:FieldTheory}.

A genus one fibered Calabi-Yau threefold can have a section which then singles out a point on the generic fiber.
Such a fibration is also called elliptic.
More generally, a genus one fibration $M$ might only have multi-sections which intersect the generic fiber multiple times and the intersection points experience monodromies.
In that case one can still find an elliptic fibration that exhibits the same $\tau$-profile over the same base $B$ by considering the fibration of the Jacobians of the fibers.
This is correpsondingly called the Jacobian fibration or the Jacobian $\text{Jac}(M)$ associated to $M$.
Since in F-theory only the base $B$, the $\tau$-profile and possibly flux choices are relevant, one  expects that both fibrations lead to the same physics.

It turns out that one can define an action of the Mordell-Weil group of the Jacobian fibration on the genus one fibration.
Moreover, the genus-one fibrations with isomorphic Jacobian fibrations together with a choice of action form a group that is called the Weil-Ch$\hat{\text{a}}$talet group~\cite{dolgachev1992elliptic} or the group of {\it elliptic Calabi-Yau torsors}~\cite{Bhardwaj:2015oru}.
In the absence of multiple fibers~\footnote{Constructions including those phenomena in a physics context can be found in~\cite{deBoer:2001wca,Bhardwaj:2015oru,Anderson:2018heq,Anderson:2019kmx}.} this group coincides with the Tate-Shafarevich group (TS) $\Sha(M)$.
While F-theory on those geometries should be the same, the M-theory compactification, which is related to F-theory by a circle reduction, differs dramatically for each TS element as these are topologically distinct manifolds.
Field theoretically, this distinction is realized by a discrete Wilson line $\xi$ that is switched on along the compactification circle~\cite{Cvetic:2015moa}.   

Let us ellaborate on this point.
One can relate both $M$ and $\text{Jac}(M)$ to a third fibration $M^\prime$ that is elliptic and therefore exhibits a zero section but also a second section $s_1$.
This geometry is related to $M$ and $\text{Jac}(M)$ via conifold transitions, and possibly flops that do not change the base $B$~\cite{Morrison:2014era}.
The second section is a free generator of the Mordell-Weil group $MW(M^\prime)$ that leads to a $U(1)$ gauge symmetry both in F- and M-theory.
In physics terms we have unHiggsed the $\mathbb{Z}_N$ into a $U(1)$ gauge symmetry that was broken by the non-zero vev of  a non-minimally charged hyper multiplet.
In the following we will denote this as $U(1)_F$ to distinguish it from the Kaluza-Klein $U(1)$ that appears in five dimensions.

Indeed, the mere presence of the second section $s_1$ implies $I_2$ fibers at codimension two~\cite{Morrison:2012ei} in the base.
These fibers naturally lead to (half-)hypermultiplets in the six- and five-dimensional theory.
The corresponding charges can be calculated by intersecting the irreducible components of the $I_2$ fiber with the image under the Shioda map
\begin{align}
\sigma(s_1)=[s_1]-[s_0]+\pi(s_0 \cdot s_1) - c_1(B) \, ,
\end{align}
and they have to take the following form for some $a,b\in\mathbb{Z}$:
\begin{align}
\begin{array}{|c|c|c|}\cline{2-3}
\multicolumn{1}{c|}{}& C_{1}&C_{2} \\ \hline
s_0 & a & -a+1 \\ \hline
s_{1}& -b & b+1 \\ \hline
\sigma(s_1)& a+b & -a-b \\ \hline 
\end{array} 
\end{align} 
Note that the intersections can be negative when sections degenerate and wrap components of the fiber~\footnote{See e.g.~\cite{Lawrie:2015hia} for a general treatment of $U(1)$ charges in $SU(5)$ models.}.
With respect to the six-dimensional $U(1)_F$ the corresponding hypermultiplets have a charge $q=|a+b|=N$ and this is non-minimal relative to the charge of other hypermultiplets.

Recall that in M-theory compactified on the same geometry the corresponding half-hypermultiplets are massive and originate from M2 branes that wrap the reducible fibral curves.
Moreover there is an additional $U(1)$ symmetry present, that can be obtained from the expansion of the M-theory three-form along the divisor class of the zero-section $s_{0}$.
This is naturally identified with the Kaluza-Klein $U(1)_{KK}$ that arises from the circle compactification in F-theory.
Hence intersections of fibral curves with $s_0$ are to be interpreted as Kaluza-Klein charges of the massive hypermultiplet states.

Higgsing the $U(1)_{F}$ to a discrete symmetry is achieved by giving a vev to the charged $N$ hypermultiplets and this geometrically corresponds to the aforementioned conifold transition.
First one shrinks down a set of isolated fibral curves $C$ to zero size such that the half-hypermultiplets become massless in M-theory.
Then switching on a vev amounts to a complex structure deformation that deforms the singularity.

The key point is now, that the respective shrunken fibral curve $C_1$ might have a non-trivial intersection $a$ with $s_0$.
In fact, when the zero-section $s_0$ is not holomorphic, neither of the fibral components will have a trivial intersection.
This, however, is at odds with the perspective of the circle reduced theory, as those states are massive KK states.
However field theoretically, the additional freedom to switch on a Wilson line $\xi$ along the KK circle, that is embedded into $U(1)_F$ as
\begin{align}
\xi = \int_{S^1} A_F \, ,
\end{align} 
helps to remedy the situation.
The presence of $\xi$ shifts the mass of the n-th KK-tower state with charges $q$ as 
 \begin{align}
 m_{(q,n)} =  |q \xi +n| \, .
 \end{align}
For certain fractional flux choices, new massless modes can appear that are able to attain a vev and provide the Goldstone mode needed to break the gauge symmetry.  

If no flux is necessary, the transition will lead to a $\mathbb{Z}_N\times U(1)_{KK}$ theory in five-dimensions which is engineered by the Jacobian fibration.
However, if $\xi$ is not zero (modulo $1$), the unbroken gauge symmetry will be a $U(1)$ that we denote by $U(1)_E$.
This situation, together with the correct linear combination of $U(1)_{F}$ and $U(1)_{KK}$ that remains unbroken, is summarized in the following table:
\begin{align}
\begin{array}{|l|l|}\hline
$flux choice $\xi  & $5d Symmetry$ \\ \hline
\xi = -\frac{n}{q}\text{ mod 1}&  {U(1)_E} =n \cdot U(1)_{6d}-q\cdot U(1)_{KK}   \\
\xi = 0 & \mathbb{Z}_N  \times U(1)_{KK}    \\ \hline
\end{array}
\end{align}
In total we find $q$ inequivalent flux choices that lead to a corresponding number of different modes in the KK tower that become massless.

Let us stress again that only the fluxless case preserves the discrete symmetry in five-dimensions.
Hence, interpreting this reduction from the geometric point of view, we only expect a discrete symmetry when shrinking a curve that leads to a half-hypermultiplet with non-minimal $U(1)_F$ charge but does not intersect $s_0$.
Shrinking that curve and deforming the corresponding singularity therefore leaves the zero-section untouched and results in the Jacobian $\text{Jac}(M)$.
The mixing of  $U(1)_F$ with of $U(1)_{KK}$ motivates the notion of the {\it discrete Shioda map}
\begin{align}
\sigma(s^{(N)}_0) = [s^{(N)}_{0}]  + {D} \, ,
\end{align}
with ${D}$ being again a fractional linear combination of vertical and fibral divisors~\footnote{If this term contains additional fibral divisors, it can lead to a mixing of the $\mathbb{Z}_N$ symmetry with the discrete center of other gauge groups.}.
The intersection of a fibral curve with $\sigma(s^{(N)}_0)$ determines the discrete charge in the F-theory uplift.

A careful analysis~\cite{Mayrhofer:2014laa} shows that the volume of ${C_1}$ also controls that of other $I_2$ fiber components that correspond to hypermultiplets of different $U(1)$ charges $\widetilde{q}$.
However, while the conifold deforms the singular fiber that leads to the Higgs field into a smooth three sphere, this does not happen for the other $I_2$ fibers.
Those fibers develop terminal singularities that do not admit a crepant resolution~\cite{Arras:2016evy,Grassi:2018rva}.
Hence, while the Jacobian fibration admits a section, it is utterly singular.

In the genus one geometries, however, those loci are generically smooth which makes them the preferred geometry for our GV computations.
The Jacobian $\text{Jac}(M)$ is also remarkable in the sense that it admits torsional two-and three-cycles \cite{Mayrhofer:2014laa}
\begin{align}
	\text{Tors}(H_3(\text{Jac}(M), \mathbb{Z}))= \mathbb{Z}_N \, .
\end{align}
In the M-theory picture it is the corresponding non-harmonic two-form that yields a massive $U(1)$ and is responsible for the discrete symmetry.
In the cases with non-trivial discrete Wilson line $\xi$ however, the same shrinking process will involve both sections  $s_0$ and $s_1$ and merge them into a multi-section, without leading to the aforementioned torsion cycles.

\section{Fibrations from families of fibers}
\label{sec:fibsfromfams}
We will now describe the fiber based approach to construct genus one fibered Calabi-Yau manifolds~\cite{Klevers:2014bqa,Cota:2019cjx}
and in that context define the problem that GV-spectroscopy is going to solve.

It is well known that every elliptically fibered manifold is birationally equivalent to a fibration of Weierstrass curves where the generic fiber is realized as a hypersurface $\{p=0\}$, with
\begin{align}
	p=y^2-(x^3+fxz^4+gz^6)\,,
\end{align}
in an appropriate fibration of the weighted projective space $\mathbb{P}_{123}$ over the base $B$.
Recall that birational equivalence means that the varieties are isomorphic on a dense subset.
In this case they are isomorphic at least away from the discriminant locus $\{\Delta=0\}\subset B$ in the base, where
\begin{align}
	\Delta=4f^3+27g^3\,.
\end{align}
On the ambient space of the Weierstrass fibration we can write the divisor classes that are associated to the homogeneous coordinates as
\begin{align}
	[x]=2E_0+\mathcal{S}_1\,,\quad [y]=3E_0+\mathcal{S}_2\,,\quad [z]=E_0+\mathcal{S}_3\,,
\end{align}
where $\mathcal{S}_i,\,i\in\{1,2,3\}$ are vertical divisors.
Moreover, we can use the equivalence relation among the coordinates to set $\mathcal{S}_3\sim0$.
The Calabi-Yau condition for the elliptically fibered hypersurface then implies that $\mathcal{S}_1=2c_1$ and $\mathcal{S}_2=3c_1$, where $c_1$ is the pre-image of the anticanonical divisor on $B$.
It also follows that $f\in\Gamma(B,-4K_B)$ and $g\in\Gamma(B,-6K_B)$.

One drawback of the Weierstass form is, that it requires intricate specializations in order to engineer $\tau$-profiles that lead to desired gauge groups and matter representations.
Even then it is difficult to explicitly resolve the geometry, i.e. to go to a generic point on a Coulomb branch, and to study the structure of the reducible fibers.
Let us instead consider the Tate form in the resolved space $\widehat{\mathbb{P}_{123}}$ and allow for more general coefficients.
The fiber is then a generic hypersurface in a smooth toric variety with toric data
\begin{align}
	\begin{blockarray}{rrrrrrrl}
		\begin{block}{r(rr|rrrr)l}
			x& 1& 0& 0& 0& 0& 3&\\
			y& 0& 1& 0& 0& 1&-2&\\
			z&-2&-3& 0& 1&-2& 1&\\
			e_1&-1& 0& 1&-2& 3& 0&\\
			e_2&-1&-2&-2& 1& 0& 0&\\
			e_3& 0&-1& 1& 0& 0& 0&\\
			& 0& 0& 0& 0&-2&-2&\\
		\end{block}
	\end{blockarray}\,.
	\label{eqn:p123resdata}
\end{align}
Here the first two columns contain the points that generate the one-dimensional cones and the last four columns contain the exponents of the corresponding homogeneous coordinates in the scaling relations.
The generic anticanonical hypersurface takes the form
\begin{align}
	\begin{split}
	p=&s_1\cdot e_1^3e_2^4e_3^2 z^6 +s_2\cdot e_1^2e_2^3e_3^2 x z^4 +s_3\cdot  e_1 e_2^2 e_3^2 x^2 z^2  +s_4\cdot e_2 e_3^2  x^3 \\
	&+s_5\cdot e_1^2 e_2^2e_3  yz^3   +s_6\cdot  e_1  e_2 e_3  x y z + s_7\cdot e_1 y^2\,.
	\end{split}
\end{align}

Using the scaling relations we can introduce divisor classes
\begin{align}
	[x]=2E_0+\mathcal{S}_1\,,\quad [y]=3E_0+\mathcal{S}_2\,,\quad [z]=E_0\,,\quad [e_i]=E_i\,,\, i\in\{1,2,3\}\,,
\end{align}
such that $\mathcal{S}_i,\,i\in\{1,2\}$ are again vertical divisors on the base.
Now the Calabi-Yau condition does not restrict those divisors but only imposes that the coefficients are sections
\begin{align}
	\begin{split}
		s_1&\sim \mathcal{S}_1+\mathcal{S}_2+c_1\,,\quad s_2\sim \mathcal{S}_2+c_1\,,\quad s_3\sim \mathcal{S}_2-\mathcal{S}_1+c_1\,,\\
		s_4&\sim \mathcal{S}_2-2\mathcal{S}_1+c_1\,,\quad s_5\sim \mathcal{S}_1+c_1\,,\quad s_6\sim c_1\,,\quad s_7\sim \mathcal{S}_1-\mathcal{S}_2+c_1\,,
	\end{split}
\end{align}
where $s\sim D$ is a shorthand for $s\in\Gamma(B,\mathcal{O}(D))$.
It is easy to see that the fiber becomes reducible over $\{s_4=0\}$ and $\{s_7=0\}$.
Moreover, using the Stanley-Reisner ideal
\begin{align}
	\mathcal{SRI}=\langle x z,\,x e_1,\,xe_2,\, y e_3,\,yz,\,ye_2,\,e_1z,\,e_1e_3,\,e_2e_3\rangle\,,
\end{align}
one finds that the components respectively intersect like the affine Dynkin diagrams of $SU(2)$ and $SU(3)$.
The generic Weierstrass form can be recovered by imposing $s_4\sim 0$ and $s_7\sim 0$, i.e. those coefficients have to be constants, which again fixes $\mathcal{S}_1=2c_1$ and $\mathcal{S}_2=3c_1$.
However, for generic choices of bundles $\mathcal{S}_1,\mathcal{S}_2$ and sections $s_i,\,i=1,...,7$, the gauge group associated to the fibration will be $G=SU(3)\times SU(2)$.

One can still find a birational map into Weierstrass form and this expresses $f,g$ and $\Delta$ in terms of the coefficients $s_i,\,i=1,...,7$.
The map can be used to find the cycles in the base over which the singularity of the fiber enhances and to determine the corresponding matter spectrum~\cite{Klevers:2014bqa}.
The simplest matter locus is $\{s_4=s_7=0\}$ with correspoding class $(\mathcal{S}_2-2\mathcal{S}_1+c_1)(\mathcal{S}_1-\mathcal{S}_2+c_1)$ where the section $\{z=0\}$ degenerates into a rational curve and this leads to a hypermultiplet in the bi-fundamental representation.
Other matter loci can also be determined with relative ease because they are charged under the non-Abelian gauge group.
This implies that they are also complete intersections in the base.

Other families of Calabi-Yau onefold hypersurfaces and complete intersections in toric ambient spaces can be studied in an analogous fashion.
In particular, the generic gauge groups have been determined for all hypersurfaces and codimension two complete intersections in toric ambient spaces~\cite{Klevers:2014bqa,Braun:2014qka}.
This provides a rich set of building blocks to construct genus one fibered Calabi-Yau manifolds of arbitrary dimensions.
However, base independent expressions for the matter loci have, with the exception of the complete intersection studied in~\cite{Cvetic:2013qsa,Oehlmann:2016wsb}, only been determined for hypersurfaces.

In general there will be matter that is charged only under the Abelian or discrete part of the gauge group.
The corresponding loci have been determined for hypersurfaces~\cite{Klevers:2014bqa} by exploiting special properties of the family of fibers and ideal decomposition techniques from computational algebraic geometry.
However, the latter require too much computing time and memory for all but the simplest examples und they can not be parallelized.
Determining the fibration data for fibrations of cubics in $\mathbb{P}^2$ that lead to $G=\mathbb{Z}_3$ was an arduous task.

In the next section we develop a technique to efficiently calculate base independent expressions for the classes of all matter loci.
It can be applied to all generic families of complete intersection fibers in toric ambient spaces.
The cancellation of six-dimensional anomalies then provides a strong independent check of consistency.
An elementary example will be the complete intersection of two quadrics in $\mathbb{P}^3$ which generically leads to genus one fibrations with four-sections and corresponding gauge group $G=\mathbb{Z}_4$.
\section{GV-spectroscopy for complete intersection fibers}
\label{sec:GVspec}
In this section we will first review the definition of Gopakumar-Vafa invariants and describe how they encode the F-theory spectrum.
We will then describe the Batyrev-Borisov construction of mirror pairs of Calabi-Yau complete intersections in toric ambient spaces and explain how the corresponding GV invariants can be calculated by exploiting mirror symmetry.
This will be illustrated at the hand of a fibration of bi-quadrics over $\mathbb{P}^2$.
We then describe how base indepdendent expressions for the fiber GV invariants can be calculated by using a parametrized auxiliary polytope and nef partition.

\subsection{Counting fibral curves with Gopakumar-Vafa invariants}
\label{sec:count}
Gopakumar-Vafa invariants are weighted sums of multiplicities of BPS states in the five dimensional effective theory from an M-theory compactification on the Calabi-Yau $M$.
The BPS states arise from wrapped M2-branes and after circle compactification to four dimensions they can be interpreted as bound states of D2 and D0 branes.
At genus zero the Gopakumar-Vafa invariants are actually identical to the so-called instanton numbers which have already been introduced in~\cite{CANDELAS199121}, see also~\cite{Cox:2000vi}.
However, the geometric interpretation that will be relevant for us is based on the analysis via D2-D0 bound states in~\cite{Gopakumar:1998jq}.

The little group of massive states in the five dimensional theory is
\begin{align}
	SO(4)=SU(2)_1\times SU(2)_2\,,
\end{align}
and the charges of the BPS states can be labelled by classes $\beta\in H_2(M)$.
It is possible to write the BPS spectrum of the five dimensional theory as
\begin{align}
	\sum\limits_{j_1,j_2}N^\beta_{j_1,j_2}\left[\left(\frac12,0\right)\oplus 2\left(0,0\right)\right]\otimes(j_1,j_2)\,,
\end{align}
and then the Gopakumar-Vafa invariants $n^r_\beta$ are defined by tracing over the $j_2$ quantum numbers
\begin{align}
	\sum_r n_\beta^r I_r=\sum\limits_{j_1,j_2}(-1)^{2j_2}(2j_2+1)N^\beta_{j_1,j_2}\cdot [j_1]\,,
\end{align}
where $I_r=(\left[\frac12\right]+2\left[0\right])^r$.

The crucial result by Gopakumar and Vafa was~\cite{Gopakumar:1998ii,Gopakumar:1998jq} that the topological string partition function $Z_{\text{top.}}$ encodes these manifestly integral invariants via
the relation
\begin{align}
	\log(Z_{\text{top.}})=\sum\limits_{\beta\in H_2(M,\mathbb{Z})}\sum\limits_{r=0}^\infty\sum\limits_{m=1}^\infty \frac{n^r_\beta}{m}\left(2\sin\left(\frac{m\lambda}{2}\right)\right)^{2r-2}q^{m\beta}\,,
\end{align}
where $\lambda$ is the topological string coupling constant and $q=\exp(2\pi i\omega\cdot \beta)$ with $\omega$ being the complexified K\"ahler form on $M$.
The topological string partition function can also be expressed in terms of the free energies $F_g$
\begin{align}
	\log(Z_{\text{top.}})=\sum\limits_{g=0}^\infty\lambda^{2g-2}F_g\,,
\end{align}
where each $F_g$ encodes the topological string amplitudes at genus $g$.
To calculate $n^r_\beta$ one only needs to know the free energies $F_g$ for $g\le r$.
In particular, the invariants $n^0_\beta$ are determined by the genus zero topological string amplitudes and the corresponding free energy $F_0$.
The genus zero free energy is also called the \textit{prepotential} due to its role in the effective four dimensional $N=2$ supergravity.

Let us assume that $D_i,\,i=1,...,h^{1,1}(B)$ form a basis of vertical divisors on a genus one fibered Calabi-Yau threefold $M$ and introduce $\tilde{n}^0_\beta$ to denote the restriction of $n^0_\beta$ to classes $\beta\in H_2(M)$ that satisfy
\begin{align}
	\beta\cdot D_i=0\,,\quad i=1,...,h^{1,1}(B)\,.
\end{align}
In other words, the corresponding curves are linear combinations of rational fibral curves and multiples of the generic fiber.
We will refer to $\tilde{n}^0_\beta$ as the \textit{fiber Gopakumar-Vafa invariants}.
Moreover, we will denote the class of the generic fiber by $F$ and assume that $\beta-F$ is not contained in the Mori cone of $M$, i.e. it does not correspond to the class of an actual curve inside $M$.
This implies that curves in the class $\beta$ will be a union of rational fibral curves.

By bound states of D2 and D0 branes it was argued in~\cite{Gopakumar:1998jq} that the invariants $n^0_\beta$ are given by
\begin{align}
	n^0_\beta=(-1)^d\chi(\widehat{\mathcal{M}}_\beta)\,,\quad d=\text{dim}\left(\widehat{\mathcal{M}}_\beta\right)\,,
\end{align}
where $\widehat{\mathcal{M}}_\beta$ is the moduli space of curves in the class $\beta$ together with a choice of flat $U(1)$ connection.
We have assumed that $\beta$ is a union of rational curves and therefore the flat $U(1)$ connections are trivial.
Then $\widehat{\mathcal{M}}_\beta=\mathcal{M}_\beta$ where $\mathcal{M}_\beta$ is just the moduli space of curves.
We can then distinguish three possible cases for the curves in the class $\beta$:
\begin{itemize}
	\item They are isolated fibral curves. In this case the interpretation of the enumerative invariants is straightforward and $\tilde{n}^0_\beta$ counts the number of such curves.
		$M2$ branes that wrap these isolated curves lead to half-hyper multiplets in the effective action.
	\item They form a family of fibral curves over a curve $C$ of genus $g$ in the base.
		The interpretation is then also clear and $-\tilde{n}^0_\beta$ corresponds to the Euler characteristic of $C$.
		Reduction of the M-theory three form along the fibral divisors and the families of $M2$ branes that wrap the fibral curves lead to vector and half-hyper multiplets in the
		effective theory.
	\item They correspond to the $I_1$ degenerations of the generic fiber, i.e. $\beta=F$. Then $-\tilde{n}^0_\beta$ is the Euler characteristic of $M$.
		See~\cite{Candelas:1994hw} for a discussion of this phenomenon.
\end{itemize}
All other invariants $\tilde{n}^0_\beta$ are determined by the periodicity
\begin{align}
	\tilde{n}^0_\beta=\tilde{n}^0_{\beta+F}\,.
\end{align}
The periodicity property is highly non-trivial from a geometrical perspective.
However, from the dual F-theory perspective it accounts for the Kaluza-Klein tower that arises after compactifying from six to five dimensions.
A geometric discussion in the context of elliptic fibrations with at most $I_1$ singularities in the fiber can be found in~\cite{Candelas:1994hw}
and for more general genus one fibrations it was observed in~\cite{Cota:2019cjx}.

The information in $\tilde{n}^0_\beta$ is sufficient to completely determine the structure of the degenerate fibers and thus the spectrum of the corresponding F-theory vacuum.
Moreover, for genus one fibers that are complete intersections in toric ambient spaces, we develop a formalism to calculate those invariants in a base independent manner.
This leads to expressions for the multiplicities of representations in the corresponding F-theory vacua that are polynomials in the line bundles which parametrize the choice of fibration over any given base.
In particular for fibers that engineer Abelian or discrete gauge groups, those expressions are extremely hard to obtain with traditional methods from computational algebraic geometry~\cite{Klevers:2014bqa}.

Let us note that all of the fiber GV invariants $\tilde{n}^g_\beta$ with $g=1$ are identical and equal to minus the Euler characteristic of the base of the fibration while for $g>1$ they vanish.
It has been shown in~\cite{Huang:2015sta,Cota:2019cjx} that the genus one fiber invariants combine with the invariants at genus zero into a Jacobi form.
Furthermore, the GV invariants $n^g_\beta$ for classes $\beta$ that satisfy $\beta\cdot D\ne0$ for some vertical divisor $D$ encode the BPS states that arise from non-critical strings in six-dimensions wrapping the compactification circle.
The corresponding contributions to the topological string partition function encode the elliptic genera of those strings~\cite{Klemm:1996hh,Haghighat:2013gba,Haghighat:2014vxa,Haghighat:2015ega}.

\subsection{The Batyrev-Borisov construction of CICYs}
\label{sec:batbori}
In the following we will review the Batyrev-Borisov construction of mirror pairs of Calabi-Yau manifolds that are complete intersections in toric ambient spaces~\cite{Batyrev:1994pg}.
This is a generalization of the construction of mirror pairs of hypersurfaces in terms of dual pairs of reflexive polyhedra~\cite{batyrev1993dual}.
It will allow us to construct families of genus one curves as was done in~\cite{Braun:2014qka} and to calculate Gopakumar-Vafa invariants via mirror symmetry following~\cite{Hosono:1994ax}.

Consider a $d+r$-dimensional toric variety $\mathbb{P}_\Delta$ that corresponds to a fine regular star triangulation (FRST) of a reflexive polytope $\Delta^\circ$.
Normally it is sufficient to consider only those points of $\Delta^\circ$ that are not in the interior of facets, i.e. faces of codimension one.
The corresponding toric variety will exhibit orbifold singularities but those do not intersect a generic Calabi-Yau complete intersection.
However, since we want to use complete intersection curves to construct fibrations, we have to use all points of the polytope.
Denote the fan associated to $\mathbb{P}_\Delta$ by $\Sigma_\Delta$ and the set of generators of $1$-dimensional cones by $\Sigma_\Delta(1)$.
Let us denote the generators of the Mori cone by $C_s,\,s=1,...,k$ and introduce the Mori vectors
\begin{align}
	l^{(s)}_\rho=D_\rho\cdot C_s\,,\quad s=1,...,k\,,\quad \rho\in\Sigma_\Delta(1)\,,
\end{align}
where $D_\rho=[x_\rho]$ is the divisor associated to the homogeneous coordinate $x_\rho$ which in turn corresponds to a generator $\rho\in\Sigma_\Delta(1)$.
Furthermore, let the polytopes $\Delta_i,\nabla_i,\,i=1,...,r$ define a nef-partition such that $\Delta=\Delta_1+\dots+\Delta_r$, where polytopes are added via the Minkowski sum, and $\Delta^\circ=\langle\nabla_1,\dots,\nabla_r\rangle_{\text{conv}}$ as well as
\begin{align}
	\langle m,n\rangle\ge -\delta_{ij}\,,\quad\forall\,m\in\Delta_i,\,n\in\nabla_j\,.
\end{align}
We denote the divisor that corresponds to $\Delta_i$ by $D_{\Delta_i}$ and introduce
\begin{align}
	l^{(s)}_{0,i}=D_{\Delta_i}\cdot C_s\,.
\end{align}
An FRST of the reflexive polytope $\nabla^\circ=\langle\Delta_1,\dots,\Delta_r\rangle$ corresponds to a toric variety $\mathbb{P}_{\nabla}$.

The intersection inside $\mathbb{P}_{\nabla}$ of the vanishing loci of the polynomials
\begin{align}
	P_i=\sum\limits_{m\in\nabla_i}a_{i,m}\prod\limits_{\rho\in\Sigma_W(1)}y_{\rho}^{\langle m,\rho\rangle+1}\,,
\end{align}
where $y_\rho$ is the homogeneous coordinate associated to $\rho\in\Sigma_{\nabla}(1)$, is a Calabi-Yau $d$-fold that we call $W$.
An analogous construction leads to a mirror family of Calabi-Yau $d$-folds $M$ inside $\mathbb{P}_{\Delta}$.
Note that any $\rho\in\Sigma_\Delta(1)$ is contained in $\nabla_j$ for exactly one $j\in\{1,...,r\}$ and let us define $a_\rho\equiv a_{j,\rho}$.
Then one can introduce Batyrev coordinates
\begin{align}
	z_s=\prod\limits_{\rho\in\Sigma_\Delta(1)} a_\rho^{l^{(s)}_\rho}\cdot\prod\limits_{i=1}^r a_{i,0}^{l^{(s)}_{0,i}}\,,
	\label{eqn:batyrev}
\end{align}
that parametrize the complex structure of $W$. 
Let us stress that every Batyrev coordinate corresponds to a Mori vector and the corresponding curve is dual to a generator of the K\"ahler cone. 
For technical reasons we will always assume that the Mori cone is simplicial.

\subsection{Calculating Gopakumar-Vafa invariants with mirror symmetry}
\label{sec:gvcalculation_general}
Let us now explain how the Gopakumar-Vafa invariants $n^0_\beta$ of a Calabi-Yau threefold $M$ can be calculated using the topological B-model on the mirror Calabi-Yau $W$.
To this end we will assume that $M$ has been obtained from the Batyrev or, more generally, the Batyrev-Borisov construction.
For a detailed review of mirror symmetry and the calculation of enumerative invariants we refer the reader to~\cite{Hosono:1994av,Alim:2012gq} while the application to complete intersections was developed in~\cite{Hosono:1994ax}.

Topological string theories can be obtained from an $N=(2,2)$ non-linear sigma model into a Calabi-Yau threefold $M$ by coupling the spin connection on the worldsheet
to one of the two $U(1)$ R-symmetries.
The two possible choices lead to the so-called topological A- and B-model.
The observables of the A-model, at least in the vicinity of the large volume limit, only depend on the complexified K\"ahler moduli $\vec{t}$ of $M$ and the path integral localizes on holomorphic maps into $M$~\footnote{Deep in the moduli space
the geometric interpretation via the target space $M$ breaks down and a description in terms of a Landau-Ginzberg theory, another non-linear sigma model or a hybrid theory becomes valid, see e.g.~\cite{Witten:1993yc}.}.
On the other hand, the B-model is only sensitive to the complex structure parameters $\vec{z}$ of $M$ and the path integral localizes on constant maps.
This makes the observables in the B-model particularly easy to calculate.

However, the two theories are related by mirror symmetries and the A-model on a Calabi-Yau threefold $M$ is dual to the B-model on the mirror Calabi-Yau $W$.
To obtain the genus zero invariants we therefore need to calculate the genus zero free energy $F_0(\vec{z})$ of the B-model and identify the mirror map $z(t)$ that relates the moduli spaces of the two theories,
i.e. the complex structure moduli space of $W$ with the stringy K\"ahler moduli space of $M$.

Both, the genus zero free energy of the B-model and the mirror map, are encoded in the periods over the holomorphic $3$-form $\Omega$ on $W$.
In particular, the genus zero free energy can be interpreted as the prepotential that underlies the special geometry of the moduli spaces.
In the B-model this manifests as follows.
One can choose a symplectic basis of $3$-cycles $A^I,B_J\in H_3(W),\,I,J=0,...,h^{2,1}(W)$ and a dual basis of differentials $\alpha^I,\beta_J\in H^3(W)$ such that
\begin{align}
	\int_{A^I}\alpha_J=\int_{B^I}\beta_J=A^J\cap B_I=\delta^J_I\,,\quad \int_{A^I}\beta_J=\int_{B^I}\alpha_J=A^J\cap A_I=B^J\cap A_I=0\,.
\end{align}
Then the $(3,0)$-form $\Omega$ can be expanded as
\begin{align}
	\Omega(z)=X^I(z)\alpha_I+\mathcal{F}_J(z)\beta^J\,.
\end{align}
The coefficients $X^I(z)$ can be used as projective coordinates on the complex structure moduli space and, around a point $z_0$ where $X^0(z_0)\ne 0$, affine coordinates are given by
\begin{align}
	t^i=\frac{X^i}{X^0}\,,\quad i=1,...,k\,,
\end{align}
where again $k=h^{1,1}(M)=h^{2,1}(W)$.
Special geometry implies, that in affine coordinates the expansion of $\Omega$ takes the form
\begin{align}
	\left(X^0\right)^{-1}\Omega=\alpha_0+t^i\alpha_i+\partial_{t^i}F_0(t)\beta^i+\left(2F_0(t)-t^j\partial_{t^j}F_0(t)\right)\beta^0\,,
	\label{eqn:periodsaffine}
\end{align}
and $F_0(t)$ is the prepotential, i.e. the topological string free energy at genus zero.

Based on the monodromies that act on a basis of brane charges one can argue, that the large volume limiting point of the A-model, where a classical geometric interpretation is valid, can be identified
with a point of maximally unipotent monodromy (MUM) in the complex structure moduli space of the B-model.
In an expansion around the MUM-point, the monodromy implies that the a basis of periods takes the form
\begin{align}
	\begin{split}
		\omega_0=&1+\mathcal{O}(z)\,,\quad \omega_{1,i}=\frac{1}{2\pi i}\omega_0\log(z_i)+\mathcal{O}(z)\,,\\
		\omega_{2,i}=&\omega_0\cdot p_{2,i}\left(\log(z_1),...,\log(z_{k})\right)+\mathcal{O}(z)\,,\\
		\omega_3=&\omega_0\cdot p_3\left(\log(z_1),...,\log(z_{k})\right)+\mathcal{O}(z)\,,
	\end{split}
	\label{eqn:omegaperiods}
\end{align}
where $i=1,...,k$ and $p_d(x_1,\dots,x_k)$ are polynomials of degree $d$.

Let us assume that $\vec{z}$ are Batyrev coordinates and therefore correspond to generators of the Mori cone and in turn also to the dual generators $\{J_1,...,J_k\}$ of the K\"ahler cone on $M$.
Then one can show, that $\vec{z}=0$ is a point of maximally unipotent monodromy.
In particular, the normalized periods $t^i=(\omega_0)^{-1}\omega_{1,i}$ provide the mirror map and can be identified with the coefficients in the expansion of the complexified K\"ahler form on $M$,
\begin{align}
	\omega=t^1\cdot J_1+\dots+t^{k}\cdot J_k\,.
\end{align}
Note that the Mori cone on $M$ and therefore also the limiting point $\vec{z}=0$ depend on the triangulation of the reflexive polytope that defines the toric ambient space of $M$.
In general the induced K\"ahler cones from different triangulations of $\Delta^\circ$ allow distinct large volume limits of $M$ which are dual to different limiting points in the complex structure moduli space of $W$.

In terms of the affine coordinates the prepotential can be choosen such that
\begin{align}
	\begin{split}
	\partial_{t^j}F_0(t)=&\frac12 c_{ijk}t^jt^k+\frac12 (c_{iij}+\tilde{c}_j)t^j+\frac16 c_{iii}+b_i+\mathcal{O}(e^{2\pi i t^i})\,,
	\label{eqn:triplelogleading}
	\end{split}
\end{align}
where we introduced
\begin{align}
	c_{ijk}=\int_M J_iJ_jJ_k\,,\quad b_i=\frac{1}{24}\int_M c_2(M)\cdot J_i\,.
	\label{eqn:cijkbi}
\end{align}
There is an ambguity in the choice of $\tilde{c}_j\in\mathbb{Z}$ which will be irrelevant for the calculation of the Gopakumar-Vafa invariants.
The right hand side of~\eqref{eqn:triplelogleading} corresponds to the central charges of $4$-branes on $M$.
The leading behaviour, that is the terms polynomial in $\vec{t}$, can be calculated using the so-called $\Gamma$-class formula, see e.g.~\cite{Cota:2019cjx} for a review.
However, we still need to obtain the basis of periods~\eqref{eqn:omegaperiods}.

This can be done by using a closed expression for $\omega_0(z)$ and then applying the Frobenius method~\cite{Hosono:1994ax}.
In summary, one introduces
\begin{align}
	\omega(z,\kappa)=\sum\limits_{\lambda\in\mathbb{N}^k}\frac{\prod\limits_{i=1}^r\Gamma\left[1+\sum\limits_{s=1}^k l^{(s)}_{0,i}(\lambda_s+\kappa_s)\right]}{\prod\limits_{\rho\in\Sigma_V(1)}\Gamma\left[1+\sum\limits_{s=1}^k l^{(s)}_\rho(\lambda_s+\kappa_s)\right]}\prod\limits_{s=1}^kz_s^{\lambda_s+\kappa_s}\,,
	\label{eqn:periodgenerating}
\end{align}
where $\kappa\in\mathbb{C}^k$, such that $\omega_0=\omega(z,0)$ is the regular period.
The other periods can be obtained by taking  linear combinations of derivatives
\begin{align}
	\frac{\partial}{\partial \kappa_{i_1}}\cdot\dots\cdot\frac{\partial}{\partial \kappa_{i_k}}\omega(z,\kappa)\bigg|_{\kappa=0}\,,
	\label{eqn:omegaderivatives}
\end{align}
that reproduce the correct leading behaviour.
In particular, the mirror map is determined by the periods with leading behaviour $\omega_{1,s}\sim\log(z_s)$,
\begin{align}
	\omega_{1,s}=\frac{1}{2\pi i}\frac{\partial}{\partial\kappa_s}\omega(z,\kappa)\bigg|_{\kappa=0}\,.
	\label{eqn:rholinearlog}
\end{align}
Note that when setting $\kappa=0$ one has to take care about cancellation of poles in the gamma and polygamma functions that arise from the derivatives with respect to $\kappa_i$.

To calculate the Gopakumar-Vafa invariants we only need the derivatives of the prepotential up to contributions that are polynomial in $\vec{t}$.
We can therefore use~\eqref{eqn:triplelogleading} and identify $(\partial_{t^i}F_0)(z)$ with
\begin{align}
	(\partial_{t^i}F_0)(z)\sim-\frac12\frac{1}{\omega_0}\frac{1}{(2\pi i)^2}\sum\limits_{i,j,k=1}^sc_{ijk}\frac{\partial^2}{\partial\kappa_j\partial\kappa_k}\omega(z,\kappa)\bigg|_{\kappa=0}\,.
	\label{eqn:prepotentialz}
\end{align}
The result will contain powers of $\pi$ and the Euler-Mascheroni constant but all of them are multiplying linear combinations of $\omega_{1,s}$ and $\omega_0$, i.e. classical contributions, and they can be safely dropped for the purpose
of calculating enumerative invariants.

Combining all of the pieces, one can calculate the Gopakumar-Vafa invariants at genus zero by performing the following steps:
\begin{enumerate}
	\item Fix a reflexive polytope $\Delta^\circ$, a nef-partition $\nabla_1,...,\nabla_r$ and a fine regular star triangulation of the points of $\Delta^\circ$.
		This determines a fan $\Sigma_\Delta$ and a corresponding toric ambient space $\mathbb{P}_\Delta$ as well as divisors $D_{\nabla_1},\dots,D_{\nabla_r}$ such that
		\begin{align}
			-K_{\mathbb{P}_{\Delta}}=D_{\nabla_1}+\dots+D_{\nabla_r}\,,
		\end{align}
		and $D_{\nabla_1}\cdot\dots\cdot D_{\nabla_r}$ is the class
		of a codimension $r$ complete intersection Calabi-Yau $M$ inside $\mathbb{P}_\Delta$.
		Denote the basis of divisors that is dual to the generators of the Mori cone of $\mathbb{P}_\Delta$ by $D_1,...,D_k$.
		The corresponding divisors on $M$ will be denoted by $J_1,...,J_k$.
	\item Calculate the triple intersection numbers on $M$
		\begin{align}
			c_{ijk}=\int\limits_MJ_iJ_jJ_k=\int\limits_{\mathbb{P}_\Delta}D_iD_jD_k\cdot\prod\limits_{i=1}^r D_{\nabla_r}\,.
			\label{cijkcalculate}
		\end{align}
	\item The Mori cone on the ambient space determines linear relations $l^{(s)}_\rho$ among the generators of rays in $\Sigma_\Delta(1)$ and Batyrev coordinates~\eqref{eqn:batyrev} $z_s,\,s=1,...,k$ that parametrize
		the complex structure on the mirror $W$.
		 One can then write down the generating function of the periods over the holomorphic $3$-form $\Omega$ on $W$~\eqref{eqn:periodgenerating}.
		 This determines the mirror map and via~\eqref{eqn:prepotentialz} also the derivatives of the prepotential $(\partial_{t^i}F_0)(z)$ up to contributions that are, after plugging in the mirror map $\vec{z}(\vec{t})$, linear in $\vec{t}$.
	\item One can then immediately calculate the instanton contributions to the prepotential $F_0(t)$ which admits an expansion
		\begin{align}
			F_0(t)=\text{class.}+\sum\limits_{\beta\in H_2(M)}n^0_\beta\cdot\text{Li}_3(q^\beta)\,,\quad \text{Li}_3(q)=\sum\limits_{k=1}^\infty\frac{q^k}{k^3}\,,
		\end{align}
		where $q^\beta=\exp(2\pi i\omega\cdot\beta)$ and the K\"ahler class $\omega$ on $M$ can be expanded as
		\begin{align}
			\omega=t^1\cdot J_1+\dots+t^{k}\cdot J_k\,.
		\end{align}
		The classical terms are polynomials in $t_i,\,i=1,...,k$ and do not carry enumerative information while the coefficients $n^0_\beta$ are genus zero Gopakumar-Vafa invariants of degree $\beta$.
\end{enumerate}
As a first step we will illustrate this procedure at the example of a concrete genus one fibered Calabi-Yau threefold that is a codimension two complete intersection
in a toric ambient space.
We will then present a technique to perform the calculation of the fiber Gopakumar-Vafa invariants $\tilde{n}^0_\beta$ in a base independent manner.

\subsection{Example: Bi-quadric in $\mathbb{P}^3$ over $B=\mathbb{P}^2$}
\label{sec:ex1bq}
Let us now illustrate the procedure to calculate Gopakumar-Vafa invariants for complete intersection Calabi-Yau threefolds in toric ambient spaces at the hand of an example.
We will then see that the invariants encode the multiplicities of degenerations of the fiber and directly encode the spectrum of the corresponding F-theory vacuum.

To this end we consider a bi-quadric inside $\mathbb{P}^3$ that is fibered over the base $B=\mathbb{P}^2$.
We choose the toric data as follows:
\begin{align}
	\begin{blockarray}{rrrrrrrrl}
		\begin{block}{r(rrrrr|rr)l}
			x& 1& 0& 0& 0& 0& 1& 0&\rho_1\\
			y& 0& 1& 0& 0& 0& 1& 0&\rho_2\\
			z& 0& 0& 1& 0& 0& 1&-1&\rho_3\\
			w&-1&-1&-1& 0& 0& 1&-1&\rho_4\\
			b_1& 0& 0& 0& 1& 0& 0& 1&\rho_5\\
			b_2& 0& 0& 0& 0& 1& 0& 1&\rho_6\\
			b_3&-1&-1& 0&-1&-1& 0& 1&\rho_7\\
			& 0& 0& 0& 0& 0&-4&-1&\\
		\end{block}
	\end{blockarray}
	\begin{array}{rrrrr|rrr}
	\end{array}
	\label{eqn:p3exdata}
\end{align}
The first five columns in~\eqref{eqn:p3exdata} contain the coordinates of the ray generators $\rho_i,\,i=1,...,7$ of the toric fan $\Sigma_{\Delta}$ that is associated to a reflexive polytope $\Delta^\circ$.
The last two columns contain the linear relations
\begin{align}
	l^{(1)}=(1,1,1,1,0,0,0)\,,\quad l^{(2)}=(0,0,-1,-1,1,1,1)\,,
	\label{eqn:ex1lin}
\end{align}
among those generators that correspond to curves $C_1,C_2$ that form a basis of the Mori cone on the toric ambient space $\mathbb{P}_{\Delta}$.
The corresponding fine regular star triangulation also determines the Stanley-Reisner ideal
\begin{align}
	\mathcal{SRI}=\langle wxyz,\,b_1b_2b_3\rangle\,.
\end{align}
In particular, it is compatible with the fibration structure over the base $\mathbb{P}^2$ that corresponds to $\rho_5,\rho_6,\rho_7$.

The decomposition $\Delta^\circ=\langle \nabla_1,\nabla_2\rangle$ with
\begin{align}
	\nabla_1=\langle \rho_1,\rho_3,\rho_5,\rho_6,0\rangle\,,\quad\nabla_2=\langle \rho_2,\rho_4,\rho_7,0\rangle\,,
\end{align}
defines a nef-partition $-K_{\mathbb{P}_\Delta}=D_{\nabla_1}+D_{\nabla_2}$ such that
\begin{align}
	l^{(1)}_{0,i}=D_{\nabla_i}\cdot C_1=(-2,\,-2)\,,\quad l^{(2)}_{0,i}=D_{\nabla_i}\cdot C_2=(0,-1)\,.
	\label{eqn:ex1lnef}
\end{align}
We will denote the corresponding complete intersection Calabi-Yau threefold in $\mathbb{P}_{\Delta}$ by $M$.
By construction it exhibits a genus one fibration with generic fiber a bi-quadric in $\mathbb{P}^3$.
Moreover, from~\eqref{eqn:ex1lin} and~\eqref{eqn:ex1lnef} one can read off that in the conventions that we introduce in Section~\ref{sec:nef(0,0)} this fibration corresponds to the choice of bundles
\begin{align}
	\mathcal{S}_2=2\cdot \pi^{-1}(H)\,,\quad\mathcal{S}_6=0\,,\quad\mathcal{S}_7=\pi^{-1}(H)\,,\quad\mathcal{S}_9=2\cdot \pi^{-1}(H)\,,
\end{align}
where $H$ is the hyperplane class in $\mathbb{P}^2$ and $\pi^{-1}(H)=[b_i]$.
The reader is encouraged to check this identification using~\eqref{eqn:biquadbundles}.

The divisors $J_1=[\rho_1]$ and $J_2=[\rho_5]$ form a basis of the K\"ahler cone of the toric ambient space that is dual to $C_1,C_2$ and the restrictions to $M$ form a
basis of the K\"ahler cone on $M$.
By abuse of notation we will denote the restriction of the divisors also by $J_1,J_2$.
The intersection numbers~\eqref{cijkcalculate} on $M$ are encoded in the coefficients of the intersection polynomial
\begin{align}
	\mathcal{J}=16\cdot J_1^3+10\cdot J_1^2J_2+4J_1J_2^2\,.
	\label{eqn:ex1intpoly}
\end{align}
The linear relations~\eqref{eqn:ex1lin} and~\eqref{eqn:ex1lnef} determine the generating function~\eqref{eqn:periodgenerating} of periods
\begin{align}
	\begin{split}
		&\omega(z_1,z_2,\kappa_1,\kappa_2)=\\
		&\sum\limits_{\lambda_1,\lambda_2=0}^\infty\frac{\Gamma\left[1+2(\lambda_1+\kappa_1)\right]\Gamma\left[1+2(\lambda_1+\kappa_1)+\lambda_2+\kappa_2\right]}{\Gamma(1+\lambda_1+\kappa_1)^2\Gamma(1+\lambda_1+\kappa_1-\lambda_2-\kappa_2)^2\Gamma(1+\lambda_2+\kappa_2)^3}z_1^{\lambda_1+\kappa_1}z_2^{\lambda_2+\kappa_2}\,.
	\end{split}
\end{align}
One can now easily determine the regular and the logarithmic period~\eqref{eqn:rholinearlog}
\begin{align}
	\begin{split}
		\omega_0=&1 + 4 z_1 + 36 z_1^2 + 400 z_1^3 + 12 z_1 z_2 + 720 z_1^2 z_2+\mathcal{O}(z^4)\,,\\
		(2\pi i)\cdot\omega_{1,1}=&\omega_0\cdot\log(z_1)+8 z_1 + 84 z_1^2 + \frac{2960}{3}z_1^3 + 56 z_1 z_2 + 2688 z_1^2 z_2+\mathcal{O}(z^4)\,,\\
		(2\pi i)\cdot\omega_{1,2}=&\omega_0\cdot\log(z_2)+14 z_1 + 183 z_1^2 + \frac{7340}{3}z_1^3 - 14 z_1 z_2 + 924 z_1^2 z_2+\mathcal{O}(z^4)\,.\\
	\end{split}
\end{align}
Again let us point out that setting $\kappa_i,\,i=1,2$ to zero requires some care to cancel poles and zeroes in the gamma and polygamma functions.
The mirror map is given by
\begin{align}
	q_i(z_1,z_2)=\exp\left(2\pi i\omega_0^{-1}\cdot\omega_{1,i}\right)\,,
\end{align}
and can be inverted to obtain
\begin{align}
	\begin{split}
		z_1(q_1,q_2)=&q_1 - 8 q_1^2 + 44 q_1^3 - 192 q_1^4 - 56 q_1^2 q_2 - 240 q_1^3 q_2+\mathcal{O}(q^5)\,,\\
		z_2(q_1,q_2)=&q_2 - 14 q_1 q_2 + 83 q_1^2 q_2 - 266 q_1^3 q_2 + 14 q_1 q_2^2 - 532 q_1^2 q_2^2+\mathcal{O}(q^5)\,.
	\end{split}
	\label{eqn:ex1invmirror}
\end{align}
Using~\eqref{eqn:prepotentialz} and~\eqref{eqn:ex1intpoly} as well as the inverted mirror map~\eqref{eqn:ex1invmirror} we can calculate the derivative of the prepotential 
\begin{align}
	\begin{split}
	\partial_{t^1}F_0=& 8 \log(q_1)^2 + 10 \log(q_1) \log(q_2) + 2 \log(q_2)^2\\
	&+140 q_1 + 307 q_1^2 + \frac{3920}{9} q_1^3 + \frac{2291}{4} q_1^4 + 372 q_1 q_2 + 13868 q_1^2 q_2\\
	&+ 192372 q_1^3 q_2 + 5409 q_1^2 q_2^2+\mathcal{O}(q^5)\,.
	\end{split}
\end{align}
This encodes the genus zero Gopakumar-Vafa invariants $n^0_\beta$ as
\begin{align}
	\begin{split}
		\partial_{t^1}F_0=q_1\partial_{q_1}\sum\limits_{\beta\in H_2(X)}n^0_\beta\cdot\text{Li}_3(q^\beta)\,,
	\end{split}
\end{align}
and one can extract
\begin{align}
	\tilde{n}^0_{1}=140\,,\quad \tilde{n}^0_{2}=136\,,\quad \tilde{n}^0_{3}=140\,,\quad \tilde{n}^0_{4}=124\,,
	\label{eqn:ex1inv}
\end{align}
where we write $\tilde{n}^0_{d_F}$ for $n^0_\beta$ with
\begin{align}
	d_F=J_1\cdot\beta\,,\quad J_2\cdot\beta=0\,.
\end{align}
It is easy to calculate the invariants also for higher degrees and indeed one observes that they are periodic under $\beta\rightarrow\beta+F$ where $F$ is the class of the generic genus one fiber.

Since the intermediate expressions for the following base independent calculation can become quite large, it is important to note that at every step we could have
dropped all monomials that contain $z_2$.
The same is true for the corresponding mirror coordinate $q_2$, except that in $z_2(q_1,q_2)$~\eqref{eqn:ex1invmirror} we have to keep terms linear in $q_2$.

Now how are the numbers~\eqref{eqn:ex1inv} to be interpreted?
The divisor classes on $M$ are generated by the class of the 4-section $J_1$ and the vertical divisor $J_2$.
In particular there are no fibral divisors and reducible fibers only appear over isolated points in the base.
Therefore $\tilde{n}^0_{d_F}$ with $d_F<4$ counts the number of rational fibral curves that intersect the 4-section $d_F$ times.
We can then immediately conclude that there are $\frac12\tilde{n}^0_{1}=\frac12\tilde{n}^0_{3}=70$ fibers of $I_2$ type that lead to hyper multiplets with discrete charge $1$ and $3$ while
another $\frac12\tilde{n}^0_{2}=68$ fibers lead to $68$ hyper multiplets with charge $2$.
Moreover, $d_F=4$ corresponds to the class of the generic fiber and $-\tilde{n}^0_{4}=-124$ is the Euler characteristic of $M$.

\subsection{Base independent spectra from GV-invariants}
\label{sec:gvcalculation_baseindependent}
We are now going to demonstrate that the calculation of the Gopakumar-Vafa invariants at base degree zero in a genus one fibered complete intersection Calabi-Yau threefold inside a toric ambient space can be performed
base independently.
From the result we can then read off the cohomology classes of the positions of rational fibral curves in terms of the first Chern classes of the line bundles that parametrize the choice of fibration.
To be concrete we consider geometries where the fiber is a generic bi-quadric inside $\mathbb{P}^3$ but the technique directly generalizes to arbitrary genus one fibers that are complete intersections in toric ambient spaces.

To quote~\cite{Katz:1999xq}, our viewpoint is that since we expect to derive formulas which are generally valid, we are free to make extra assumptions in order to derive them.
The trick is that we can pretend that $B=\mathbb{P}^2$ but still extract base independent expressions.
A general toric variety that is a fibration of $\mathbb{P}^3$ over $\mathbb{P}^2$ will have the toric data
\begin{align}
	\begin{blockarray}{rrrrrrrrl}
		& & & & & & T^2& C_B\\
		\begin{block}{r(rrrrr|rr)l}
			x& 1& 0& 0& 0& 0& 1& s_1&\rho_1\\
			y& 0& 1& 0& 0& 0& 1& s_2&\rho_2\\
			z& 0& 0& 1& 0& 0& 1& s_3&\rho_3\\
			w&-1&-1&-1& 0& 0& 1& 0&\rho_4\\
			& 0& 0& 0& 1& 0& 0& 1&\rho_5\\
			& 0& 0& 0& 0& 1& 0& 1&\rho_6\\
			&-s_1&-s_2&-s_3&-1&-1& 0& 1&\rho_7\\
		\end{block}
	\end{blockarray}
	\begin{array}{rrrrr|rrr}
	\end{array}\,,
	\label{eqn:p3exdata}
\end{align}
for some values $s_1,s_2,s_3$. The linear relations
\begin{align}
	l^{(1)}=(1,1,1,1,0,0,0)\,,\quad l^{(2)}=(s_1,s_2,s_3,0,1,1,1)\,,
	\label{eqn:gvex2lrel}
\end{align}
do not necessarily correspond to generators of the Mori cone but that will not be important for our calculation.
In particular we do not need to fix a triangulation.
It is crucial to note that the scaling relations~\eqref{eqn:gvex2lrel} imply
\begin{align}
	[x]=[w]+s_1\cdot \pi^{-1}(H)\,,\quad [y]=[w]+s_2\cdot \pi^{-1}(H)\,,\quad [z]=[w]+s_3\cdot \pi^{-1}(H)\,,
\end{align}
where $H$ is again the hyperplane class on $\mathbb{P}^2$.
We also need to choose a nef partition which we can parametrize as
\begin{align}
	D_{\nabla_1}=D_{\rho_1}+D_{\rho_2}+\tilde{s}_1\cdot \pi^{-1}(H)\,,\quad D_{\nabla_2}=D_{\rho_1}+D_{\rho_2}+\tilde{s}_2\cdot \pi^{-1}(H)\,,
\end{align}
and this determines
\begin{align}
	l^{(1)}_{0,i}=(-2,\,-2)\,,\quad l^{(2)}_{0,i}=(\tilde{s}_1,\tilde{s}_2)\,.
\end{align}
We know that $c_{222}=0$ and $c_{122}=4$~\footnote{This implicitly singles out the triangulation that is compatible with the fibration structure.} but can leave all other triple intersection numbers $c_{ijk}$ generic and fix them by imposing the periodicity of the fiber invariants.
Anticipating that we will drop subleading terms in $z_2$ we can write down the \textit{reduced generating function}
\begin{align}
	\omega(z_1,\kappa_1,\kappa_2)=\sum\limits_{n=0}^\infty\frac{\Gamma\left[1+2(n+\kappa_1)+\tilde{s}_1\kappa_2\right]\Gamma\left[1+2(n+\kappa_1)+\tilde{s}_2\kappa_2\right]}{\Gamma(1+n+\kappa_1)\prod_{i=1}^3\Gamma(1+n+\kappa_1+s_i\kappa_2)}z_1^{n+\kappa_1}z_2^{\kappa_2}\,.
\end{align}
This leads to the regular and single logarithmic periods
\begin{align}
	\begin{split}
		\omega_0=&1+4z_1+36z_2^2+400z_1^3+4900z_1^4+63504z_1^5+\mathcal{O}(z^6)\,,\\
		(2\pi i)\omega_{1,1}=&\log(z_1)+8z_1+52z_1^2+\frac{1472}{3}z_1^3+5402z_1^4+\frac{323648}{5}z_1^5+\mathcal{O}(z^6)\,,\\
		(2\pi i)\omega_{1,2}=&\log(z_2)-2 {z_1} (2 {s_1}+2 {s_2}+2 {s_3}-3 {\tilde{s}_1}-3 {\tilde{s}_2})\\
		&+{z_1}^2 (-38 {s_1}-38 {s_2}-38 {s_3}+51 {\tilde{s}_1}+51 {\tilde{s}_2})+\mathcal{O}(z^3)\,.\\
	\end{split}
\end{align}
Inverting the mirror map and leaving the triple intersection numbers generic we can calculate $\partial_{t^1}F_0$ and obtain the fiber invariants $\tilde{n}^0_\beta$.
Imposing the periodicity $\tilde{n}^0_1=\tilde{n}^0_5$ fixes
\begin{align}
	\begin{split}
	c_{111}=& \frac12(-8 s_1 s_2 - 8 s_1 s_3 - 8 s_2 s_3 + 4 s_1 \tilde{s}_1 + 4 s_2 \tilde{s}_1 + 4 s_3 \tilde{s}_1 - 2 \tilde{s}_1^2 + 4 s_1 \tilde{s}_2 \\
		&+ 4 s_2 \tilde{s}_2 + 4 s_3 \tilde{s}_2 - 2 \tilde{s}_1 \tilde{s}_2 - 2 \tilde{s}_2^2 +c_{112}\left[- 2 s_1  - 2 s_2  - 2 s_3  + \tilde{s}_1  + \tilde{s}_2 \right])\,,
	\end{split}
\end{align}
and we obtain
\begin{align}
	\begin{split}
		\tilde{n}^0_1=&4 (2 s_1 + 2 s_2 + 2 s_3 - \tilde{s}_1 - 3 \tilde{s}_2) (2 s_1 + 2 s_2 + 2 s_3 - 3 \tilde{s}_1 - \tilde{s}_2)\,,\\
		\tilde{n}^0_2=&2 (6 s_1^2 + 16 s_1 s_2 + 6 s_2^2 + 16 s_1 s_3 + 16 s_2 s_3 + 6 s_3^2 - 15 s_1 \tilde{s}_1 - 15 s_2 \tilde{s}_1\\
			&- 15 s_3 \tilde{s}_1 + 10 \tilde{s}_1^2 - 15 s_1 \tilde{s}_2 - 15 s_2 \tilde{s}_2 - 15 s_3 \tilde{s}_2 + 10 \tilde{s}_1 \tilde{s}_2 + 10 \tilde{s}_2^2)\,,\\
		\tilde{n}^0_3=&4 (2 s_1 + 2 s_2 + 2 s_3 - \tilde{s}_1 - 3 \tilde{s}_2) (2 s_1 + 2 s_2 + 2 s_3 - 3 \tilde{s}_1 - \tilde{s}_2)\,,\\
		\tilde{n}^0_4=&2 (8 s_1^2 + 12 s_1 s_2 + 8 s_2^2 + 12 s_1 s_3 + 12 s_2 s_3 + 8 s_3^2 - 13 s_1 \tilde{s}_1 - 13 s_2 \tilde{s}_1\\
		&- 13 s_3 \tilde{s}_1 + 8 \tilde{s}_1^2 - 13 s_1 \tilde{s}_2 - 13 s_2 \tilde{s}_2 - 13 s_3 \tilde{s}_2 + 10 \tilde{s}_1 \tilde{s}_2 + 8 \tilde{s}_2^2)\,.
	\end{split}
\end{align}
At his point we can forget that we fixed $B=\mathbb{P}^2$.
In fact, let us assume that we have a biquadric inside $\mathbb{P}^3$ fibered over any base $B$
and use the parametrization of the fibration in terms of four line bundles $\mathcal{S}_i,\,i\in\{2,6,7,9\}$ on the base that is introduced in Section~\ref{sec:nef(0,0)}.
This amounts to replacing
\begin{align}
	\begin{split}
	s_1\rightarrow& 2c_1-(\mathcal{S}_2+\mathcal{S}_6+\mathcal{S}_7+\mathcal{S}_9)\,,\\
	s_2\rightarrow& c_1-(\mathcal{S}_6+\mathcal{S}_9)\,,\\
	s_3\rightarrow& c_1-(\mathcal{S}_7+\mathcal{S}_9)\,,\\
	\tilde{s}_1\rightarrow& 2c_1-(\mathcal{S}_6+\mathcal{S}_7+\mathcal{S}_9)\,,\\
	\tilde{s}_2\rightarrow& 3c_1-(\mathcal{S}_2+\mathcal{S}_6+\mathcal{S}_7+2\mathcal{S}_9)\,,
	\end{split}
	\label{eqn:ex2sident}
\end{align}
and we obtain
\begin{align}
	\begin{split}
		\tilde{n}^0_1=&4 (3 c_1 - \mathcal{S}_2 - \mathcal{S}_9) (c_1 + \mathcal{S}_2 + \mathcal{S}_9)\,,\\
		\tilde{n}^0_2=&2(6 c_1^2 - c_1 \mathcal{S}_2 + \mathcal{S}_2^2 + 4 c_1 \mathcal{S}_6 - 2 \mathcal{S}_2 \mathcal{S}_6 - 2 \mathcal{S}_6^2 + 4 c_1 \mathcal{S}_7 - 2 \mathcal{S}_2 \mathcal{S}_7 \\
		&- 2 \mathcal{S}_7^2 - c_1 \mathcal{S}_9 + 4 \mathcal{S}_2 \mathcal{S}_9 - 2 \mathcal{S}_6 \mathcal{S}_9 - 2 \mathcal{S}_7 \mathcal{S}_9 + \mathcal{S}_9^2)\,,\\
		\tilde{n}^0_3=&4 (3 c_1 - \mathcal{S}_2 - \mathcal{S}_9) (c_1 + \mathcal{S}_2 + \mathcal{S}_9)\,,\\
		\tilde{n}^0_4=&2 (12 c_1^2 - 7 c_1 \mathcal{S}_2 + 3 \mathcal{S}_2^2 - 4 c_1 \mathcal{S}_6 + 2 \mathcal{S}_2 \mathcal{S}_6 + 2 \mathcal{S}_6^2 - 4 c_1 \mathcal{S}_7 + 2 \mathcal{S}_2 \mathcal{S}_7 \\&+ 2 \mathcal{S}_7^2 - 7 c_1 \mathcal{S}_9 + 4 \mathcal{S}_2 \mathcal{S}_9 + 2 \mathcal{S}_6 \mathcal{S}_9 + 2 \mathcal{S}_7 \mathcal{S}_9 + 3 \mathcal{S}_9^2)\,.
	\end{split}
	\label{eqn:ex2gv}
\end{align}
We claim that these expressions capture the multiplicity of reducible fibers for generic fibrations over any base $B$.
Again, $\tilde{n}^0_1$ is the number of hyper multiplets of charge $\pm 1$ while $\frac12\tilde{n}^0_2$ is the number of hyper multiplets of charge $\pm2$.
Note that the $\tilde{n}^0_3$ with charge $3$ are contained in the hypermultiplets with charge $\pm 1$.
Moreover, $\chi=-\tilde{n}^0_4$ is the base independent expression for the Euler characteristic.
The structure of the reducible fibers is shown in Table~\ref{tab:nef00Spectrum}.

A first, highly non-trivial check can be performed using the anomalies of the six-dimensional effective theory.
The condition for cancellation of the pure gravitational anomaly of the six-dimensional $\mathcal{N}=1$ supergravity reads
\begin{align}
	H-V+29T=273\,,
	\label{eqn:ex2gravan}
\end{align}
where for a gauge group $G=\mathbb{Z}_4$ the number of vector multiplets is $V=0$.
The number of neutral hyper multiplets $H_{\text{neut.}}$ is then
\begin{align}
	H_{\text{neut.}}=T+3-\frac12 \chi(M)\,,
\end{align}
and the total number of hyper multiplets is
\begin{align}
	\frac12\left(\tilde{n}^0_1+\tilde{n}^0_2+\tilde{n}^0_3\right)+H_{\text{neut.}}\,.	
\end{align}
The number of tensor multiplets can be expressed in terms of the first Chern class of the base as $T=h^{1,1}(B)-1=9-c_1(B)^2$.
Indeed we find that~\eqref{eqn:ex2gravan} is satisfied for any choice of bundles $\mathcal{S}_i,\,i\in\{2,6,7,9\}$. 

Note that we will explain in Appendix~\ref{app:baseindependentintersection} how the triple intersection numbers as well as the Euler characteristic can be calculated directly in a base independent manner.
This provides another strong consistency check of the results that we obtain from the fiber Gopakumar-Vafa invariants.

It is now straightforward to apply this procedure to other complete intersection fibers.
Let us stress that there is always a freedom in the choice of parametrization in terms of line bundles.
The first step is therefore to fix a parametrization and then one can write down the corresponding linear relation that enters the generating functions of periods.

\subsection{Example: $G=SU(3)\times SU(2)\times\mathbb{Z}_2$}
\label{sec:su3su2z2model}
To demonstrate that GV-spectroscopy can also be applied to more complex fibrations, let us now discuss a fiber that generically leads to a gauge group $G=SU(2)\times SU(3)\times\mathbb{Z}_2$.
This will also illustrate how Dynkin labels of states and the corresponding representations can be read off directly form the Gopakumar-Vafa invariants.

We consider the nef partition $(190,2)$ which is equivalent to $(190,5),\,(139,0)$ and $(139,0)$~\cite{Braun:2014qka,Oehlmann:2016wsb}.
The corresponding auxilliary data is given by,
\begin{align}
	\begin{blockarray}{rrrrrrrrrrrrl}
		\begin{block}{r(rrrrr|rrrrrr)l}
			X&-1& 1& 0& 0& 0&-1& 1& 1& 0& 0&s_x&\\
			Y&-1& 1& 0& 0& 0& 0& 0& 1& 0& 0&s_y&\\
			Z& 1& 0& 0& 0& 0& 0& 0& 0& 0& 1& 0&\\
			e_1&-1& 0& 0& 0& 0& 1& 1&-2&-1& 0& 0&\\
			e_2&-1& 0& 1& 0& 0&-1& 0& 0& 1& 1& 0&\\
			e_3&-1& 1& 1& 0& 0& 1&-1& 0& 0& 0& 0&\\
			e_4& 0& 0& 1& 0& 0& 0& 1& 0& 0&-1& 0&\\
			e_5& 0& 0&-1& 0& 0& 0& 0& 0& 1& 0&s_5&\\
			& 0& 0& 0& 1& 0& 0& 0& 0& 0& 0& 1&\\
			& 0& 0& 0& 0& 1& 0& 0& 0& 0& 0& 1&\\
			& *& *& *&-1&-1& 0& 0& 0& 0& 0& 1&\\
		\end{block}
	\end{blockarray}
	\begin{array}{rrrrr|rrrrrrr}
	\end{array}\,.
	\label{eqn:1902aux}
\end{align}
We have introduced the homogeneous coordinates $X,Y,Z,e_2,...,e_5$ to parametrize the ambient space of the fiber and in order
to calculate Gopakumar-Vafa invariants we consider the basis of divisors
\begin{align}
	\begin{split}
		J_1=&[e_3]+[e_4]+[Z]\,,\quad J_2=[e_4]+[Z]\,,\quad J_3=[Y]-s_y\cdot \pi^{-1}(H)\,,\\
		J_4=&[e_5]-s_5\cdot \pi^{-1}(H)\,,\quad J_5=[Z]\,,
	\end{split}
	\label{eqn:1902J}
\end{align}
to parametrize the K\"ahler cone.
The integers $s_x,s_y,s_5$ parametrize the fibration of the ambient space over the auxilliary $\mathbb{P}^2$ base and $H$ denotes the hyperplane class of the latter.
The nef partition determines the polynomials
\begin{align}
	\begin{split}
		p_1=& e_2 e_3^3 e_1^2 (d_{1, 1} e_2 e_3 e_4 + d_{1, 2} e_5) X^4 + e_2 e_3^2 e_1^2 (d_{2, 1} e_2 e_3 e_4 + d_{2, 2} e_5) X^3 Y \\
		&+ e_2 e_3 e_1^2 (d_{3, 1} e_2 e_3 e_4 + d_{3, 2} e_5) X^2 Y^2 + e_2 e_1^2 (d_{4, 1} e_2 e_3 e_4 + d_{4, 2} e_5) X Y^3\\
		&+ d_{5} e_2^2 e_1^2 e_4 Y^4 + e_3 e_1 (d_{6, 1} e_2 e_3 e_4 + d_{6, 2} e_5) X^2 Z + e_1 (d_{7, 1} e_2 e_3 e_4 + d_{7, 2} e_5) X Y Z \\
		&+ d_{8} e_2 e_1 e_4 Y^2 Z + d_{9} e_4 Z^2\,,\\
		p_2=&d_{10, 1} e_2 e_3 e_4 + d_{10, 2} e_5\,,
	\end{split}
\end{align}
and the classes of the coefficients are determined by
\begin{align}
	[p_1]=J_2+J_5+s_a\cdot H\,,
\end{align}
where we have introduced another auxilliary parameter $s_a$.
The Stanley-Reisner ideal of the ambient space of the fiber is given by
\begin{align}
	\mathcal{SRI}=\langle X Y,\, X e_4,\, Y e_3,\, Z e_2,\, Z e_1,\, e_2 e_5,\, e_3 e_1,\, e_3 e_5,\, e_1 e_4,\, e_4 e_5 \rangle\,.
\end{align}

Note that away from the locus $d_{10,2}=0$ the fibration can be described by a generic hypersurface in $\mathbb{P}_{112}$ which has been discussed in~\cite{Klevers:2014bqa}.
It is then easy to see that $[X]$ and $[Y]$ restrict to two-sections on the complete intersection Calabi-Yau threefold, while $[e_1],...,[e_4]$ descend to fibral divisors.
In particular, over a generic point of the curve $\{d_9=0\}\subset B$ the fiber is of $I_2$ type and resolved by the fibral divisor $[e_1]$ which leads to an $SU(2)\times \mathbb{Z}_2$ gauge symmetry.
The complete intersection generically enhances this gauge symmetry into $SU(2)\times SU(3)\times\mathbb{Z}_2$.
This happens because over a generic point of the curve $\{d_{10,2}=0\}\subset B$ the fiber splits into three rational components that correspond to the fibers of the fibral divisors $[e_2],\,[e_3]$ and $[e_4]$.

To understand the behaviour over the points $\{d_{10,1}=d_{10,2}=0\}$ let us first note that the ambient space of the fiber itself is fibered over a $\mathbb{P}^1$.
When the coefficients $d_{10,1}$ and $d_{10,2}$ both vanish, the second equation $p_2=0$ is trivially satisfied and the fiber over these points is itself a genus one fibration over
$\mathbb{P}^1$.
The fibration thus generically contains non-flat fibers.

In order to determine the spectrum we calculate the fiber Gopakumar-Vafa invariants $\tilde{n}^0_{n_1,\dots,n_5}=\tilde{n}^0_\beta$ with $d_i=J_i\cdot \beta,\,i=1,\dots,5$ up to degree $d_1+\dots +d_5=9$.
The degrees of the class of the generic fiber are $(d_1,d_2,d_3,d_4,d_5)=(4,4,2,0,4)$ and therefore we can use the periodicity and the reflection symmetry of the invariants to determine all fiber GV invariants from the result.
One problem, however, is that we can not use the periodicity of the invariants to fix the necessary triple intersection numbers and we can not obtain the Euler characteristic which first appears at degree $14$.
We therefore employ the base independent intersection calculus introduced in Appendix~\ref{app:baseindependentintersection}.

Recall that to apply the technique described in Section~\ref{sec:gvcalculation_general} it is necessary to parametrize the K\"ahler form by divisors~\eqref{eqn:1902J} that ideally form a basis of the K\"ahler cone on the ambient space.
To interpret the invariants it is better to use a different basis, namely $[X],[Y],[e_1],[e_2],[e_4]$.
The corresponding degrees
\begin{align}
	d_X=[X]\cdot \beta\,,\quad \dots\,,\quad d_{e_4}=[e_4]\cdot \beta\,,
\end{align}
are related to $d_1,\dots,d_5$ via
\begin{align}
	\begin{split}
		d_X=&-d_1 + d_2 + d_4\,,\quad d_Y= d_4\,,\quad d_{e_1}= -d_1 + d_3 + d_5\,,\\ d_{e_2}=& d_1 - d_3 - 2 d_4\,,\quad d_{e_4}= d_2 - d_5\,.
	\end{split}
\end{align}
At this point we can forget about the auxilliary base and interpret the parameters $s_x,s_y,s_5,s_a$ as Chern classes of line bundles that parametrize the choice of fibration over any base $B$.
To underline this shift in the interpretation we replace the parameters by classes $\mathcal{S}_*$ and introduce $c_1$ to denote the first Chern class of $B$.
The results are listed in Table~\ref{tab:1902gv}.

As expected, the degrees $d_{e_1},d_{e_2},d_{e_4}$ correspond to the Dynkin labels of the states that arise from M2 branes that wrap the curves in the corresponding class.
One can then group the invariants according to the locus and immediately read off the representations.
In particular, we obtain the class of the positions of isolated $I_2$ fibers that do not intersect any of the fibral divisors and lead to hyper multiplets that are not charged under the non-Abelian gauge group.

\begin{table}
\begin{align*}
	\begin{array}{|rrrrr|c|}
	\hline
		\multirow{2}{*}{$d_X$}&\multirow{2}{*}{$d_Y$}&\multirow{2}{*}{$d_{e_1}$}&\multirow{2}{*}{$d_{e_2}$}&\multirow{2}{*}{$d_{e_4}$}&\multirow{2}{*}{$\tilde{n}^0_{d_X,d_Y,d_{e_1},d_{e_2},d_{e_4}}$}\\
		&&&&&\\\hline
	\multicolumn{5}{|c}{SU(3) \text{ adjoint}}&\\\hline
		2&1&0&-1&2&\multirow{4}{*}{$(c_1+\mathcal{S}_x+\mathcal{S}_y-\mathcal{S}_a)(\mathcal{S}_x+\mathcal{S}_y-\mathcal{S}_a)$}\\
	0&1&0& 1&-2&\\
	2&0&0& 1&1&\\
	0&1&0&-2&1&\\\hline
	\multicolumn{5}{|c}{SU(2) \text{ adjoint}}&\\\hline
	1&1&-2&0&0&\mathcal{S}_a(\mathcal{S}_a-c_1)\\\hline
	\multicolumn{5}{|c}{(\two,\three)}&\\\hline
	-1& 0& 1&-1& 0&\multirow{7}{*}{$\mathcal{S}_a(c_1+\mathcal{S}_x+\mathcal{S}_y-\mathcal{S}_a)$}\\
	 0& 1&-1&-1& 0&\\
	 0& 0& 1&-1& 1&\\
	 1& 1&-1&-1& 1&\\
	 1& 0& 1& 0& 1&\\
	 2& 1&-1& 0& 1&\\
	 0& 1& 1& 0&-1&\\\hline
	\multicolumn{5}{|c}{(\two,\one)}&\\\hline
		0& 0& 1& 0& 0&\multirow{2}{*}{$\mathcal{S}_a (5 c_1 + \mathcal{S}_a - 3 \mathcal{S}_x - 3 \mathcal{S}_y)$}\\
	 1& 1&-1& 0& 0&\\\hline
	\multicolumn{5}{|c}{(\one,\three)}&\\\hline
	2& 1& 0& 1& 0&\multirow{5}{*}{$(c_1+\mathcal{S}_x+\mathcal{S}_y-\mathcal{S}_a)(4 \mathcal{S}_a - 7 \mathcal{S}_x - 3 \mathcal{S}_y-4 \mathcal{S}_5)$}\\
	 0& 1& 0&-1& 0&\\
	 0& 0& 0& 1&-1&\\
	 0& 1& 0& 0&-1&\\
	 2& 1& 0& 0& 1&\\\hline
	\multicolumn{5}{|c}{(\one,\three)}&\\\hline
		1& 0& 0& 1& 0&\multirow{4}{*}{$(c_1+\mathcal{S}_x+\mathcal{S}_y-\mathcal{S}_a)( 3 \mathcal{S}_a - 2 \mathcal{S}_x - 6 \mathcal{S}_y-2 \mathcal{S}_5)$}\\
	 1& 1& 0&-1& 1&\\
	 1& 1& 0& 1&-1&\\
	 1& 0& 0& 0& 1&\\\hline
	\multicolumn{5}{|c}{(\one,\one)}&\\\hline
		1& 1& 0& 0& 0&\begin{array}{c}2 (-6 c_1 \mathcal{S}_5 + 16 c_1 \mathcal{S}_a + 6 \mathcal{S}_5 \mathcal{S}_a - 6 \mathcal{S}_a^2 - 18 c_1 \mathcal{S}_x - 6 \mathcal{S}_5 \mathcal{S}_x \\+ 9 \mathcal{S}_a \mathcal{S}_x - 6 \mathcal{S}_x^2 - 22 c_1 \mathcal{S}_y - 6 \mathcal{S}_5 \mathcal{S}_y + 13 \mathcal{S}_a \mathcal{S}_y - 10 \mathcal{S}_y^2)\end{array}\\\hline
	\multicolumn{5}{|c}{\text{Non-flat fibers}}&\\\hline
	 0& 1& 0& 0& 0&20 (c_1+ \mathcal{S}_x + \mathcal{S}_y - \mathcal{S}_a ) (c_1 + \mathcal{S}_5 - \mathcal{S}_a + \mathcal{S}_x + \mathcal{S}_y)\\\hline
	0& 0& 0& 1& 0&\multirow{3}{*}{$6 (c_1 + \mathcal{S}_x + \mathcal{S}_y- \mathcal{S}_a ) (c_1 + \mathcal{S}_5 - \mathcal{S}_a + \mathcal{S}_x + \mathcal{S}_y)$}\\
	 0& 1& 0&-1& 1&\\
	 0& 0& 0& 0& 1&\\\hline
		-1& 0& 0& 0& 0&\multirow{3}{*}{$2 (c_1 + \mathcal{S}_x + \mathcal{S}_y- \mathcal{S}_a ) (c_1 + \mathcal{S}_5 - \mathcal{S}_a + \mathcal{S}_x + \mathcal{S}_y)$}\\
	 1& 0& 0& 1& 1&\\
	 1& 1& 0&-1& 2&\\\hline
	 0& 0& 0& 1& 1&-2 (c_1 + \mathcal{S}_x + \mathcal{S}_y- \mathcal{S}_a ) (c_1 + \mathcal{S}_5 - \mathcal{S}_a + \mathcal{S}_x + \mathcal{S}_y)\\\hline
	\end{array}
\end{align*}
\caption{Base independent fiber Gopakumar-Vafa invariants for fibrations with generic fiber realized as nef partition $(190,2)$.}
\label{tab:1902gv}
\end{table}

In order to obtain the final degrees of freedom, namely the neutral singlets, we need the number of complex structure moduli or, equivalently, the Euler characteristic of the Calabi-Yau. 
This can be calculated following Appendix~\ref{app:baseindependentintersection} and reads
 \begin{align}
	\begin{split}
	\chi=&2 \left(6 c_1 \mathcal{S}_5 - 11 c_1 \mathcal{S}_a - 6 \mathcal{S}_5 \mathcal{S}_a + 3 \mathcal{S}_a^2 + 18 c_1 \mathcal{S}_x + 6 \mathcal{S}_5 \mathcal{S}_x  \right.\\
	&\left.- 9 \mathcal{S}_a \mathcal{S}_x+ 14 c_1 \mathcal{S}_y + 6 \mathcal{S}_5 \mathcal{S}_y - 5 \mathcal{S}_a \mathcal{S}_y + 12 \mathcal{S}_x \mathcal{S}_y - 4 \mathcal{S}_y^2\right)\,.
	\end{split}
	\label{eqn:1902euler}
\end{align}
From this we can easily compute the neutral singlets but have to take care to count the non-flat fiber contributions as non-toric K{\"a}hler classes.
We note that each non-flat fiber contributes exactly one K{\"a}hler parameter, identifying them as simple rank one E-string theories with multiplicity
\begin{align}
n_{\text{E-string}}= (c_1 + \mathcal{S}_x + \mathcal{S}_y- \mathcal{S}_a ) (c_1 + \mathcal{S}_5 - \mathcal{S}_a + \mathcal{S}_x + \mathcal{S}_y) \, .
\end{align}
This allows to compute the complex structures and neutral singlets as
\begin{align}
	\begin{split}
n_{1}=& 15-c_1^2 +n_{\text{E-string}} -\frac12 \chi \nonumber \\
=&15 - 2 \mathcal{S}_a^2 + 7 \mathcal{S}_a \mathcal{S}_x + \mathcal{S}_x^2 + 5 \mathcal{S}_5 (\mathcal{S}_a - \mathcal{S}_x - \mathcal{S}_y) + 3 \mathcal{S}_a \mathcal{S}_y\\
	&- 10 \mathcal{S}_x \mathcal{S}_y + 5 \mathcal{S}_y^2 - c_1 (5 \mathcal{S}_5 - 9 \mathcal{S}_a + 16 \mathcal{S}_x + 12 \mathcal{S}_y)
	\end{split}
\end{align}
We then have all that we need in order to check anomaly cancellation.
By picking a suitable GV representative from Table~\ref{tab:1902gv} we associate the following six-dimensional multiplicities to the representations:
\begin{align}
	\begin{split}
n_{(\mathbf{1},\mathbf{8})} =& 1+\frac12 \tilde{n}^0_{2,1,0,-1,2} \, ,\quad  n_{(\mathbf{3},\mathbf{1})} = 1+\frac12 \tilde{n}^0_{1,1,-2,0,0} \, ,\\
n_{(\mathbf{2},\mathbf{ 1})}=&\tilde{n}^0_{0,0,1,0,0}  \, ,\quad n_{(\mathbf{2}, \overline{\mathbf{3}})}=\tilde{n}^0_{-1,0,1,-1,0}  \, , \\
n_{(\mathbf{1},\mathbf{3}),a}=&\tilde{n}^0_{-1,0,1,-1,0}  \, ,\quad n_{(\mathbf{1},\mathbf{3}),b}=\tilde{n}^0_{1,0,0,1,0}  \, ,  \\
n_{(\mathbf{1},\mathbf{1})}=&\frac12 \tilde{n}^0_{1,1,0,0,0}  \, ,\quad n_{\text{E-string}}=\frac16 \tilde{n}^0_{0,0,0,1,0} \,.
	\end{split}
\end{align}

To check consistency of the six-dimensional supergravity theory we start with the $SU(2)$  and $SU(3)$ gauge anomalies.
For those we simply have to check the conditions
\begin{align}
\label{eq:gaugeAnomaly}
\begin{split}
\frac16 \left(  \sum_\mathbf{R} n_\mathbf{R}  A_\mathbf{R}-A_{\text{adj}} \right)=  c_1 \cdot b \, ,  \qquad
-\frac13 \left(  \sum_\mathbf{R}  n_\mathbf{R} C_\mathbf{R}-C_{\text{adj}} \right)=   b \cdot b  \, ,
\end{split}
\end{align}
with group theory coefficients
\begin{align}
	A_F=1\, ,\quad  A_{\text{adj}}= 2n\, ,\quad C_{F}=\frac12\, ,\quad C_{ \text{adj}}=6+n\,,
\end{align}
for $SU(n)$ $n=2,3$.
The divisors $b$ specify the $SU(2)$ and $SU(3)$ divisor classes and respectively read
\begin{align}
b_{SU(2)} =  \mathcal{S}_a \, , \quad b_{SU(3)} = c_1 - \mathcal{S}_a + \mathcal{S}_x + \mathcal{S}_y \, . 
\end{align}
This is enough in order to show cancellation of the $SU(2)$ gauge anomalies. 
Care has to be taken though, when including the E-string points that intersect the $SU(3)$ locus.

One E-string point contributes to the anomaly the equivalent of one tensor or 29 hypermultiplets and we can count them as those.
Moreover, in~\cite{Dierigl:2018nlv} it has been shown that the 29 hypers from an E-string theory with an $SU(3)$ subgroup of the $E_8$ flavor group being gauged, the hypermultiplets arrange into the $SU(3)$ representations
\begin{align}
29 \rightarrow 6 \times \mathbf{3} \oplus 11 \times \mathbf{1} \, .
\end{align}
Thus we have to include six fundamental representations of $SU(3)$ for every E-string point to cancel the gauge anomaly~\eqref{eq:gaugeAnomaly}.
This is enough to show the cancellation of all gauge anomalies.
It is worth to point out that we find precisely three fibral curves inside the non-flat fibers with matching $SU(3)$ weights to form a fundamental representation.
Couriously those weights even come with the correct multiplicity of six, as predicted in \cite{Dierigl:2018nlv}.
Finally, there is only the gravitational anomaly left which can also be shown to cancel when accounting for the 29 hypermultiplets per E-string point.

\section{CICY fibers in the restricted bi-quadric}
\label{sec:CICYFI}
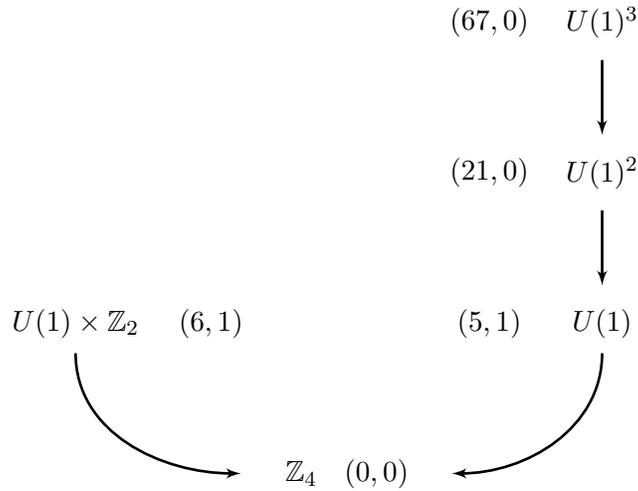
\begin{figure}[h!]
\centering
\begin{tikzpicture}[node distance=4mm, >=latex',block/.style = {draw, rectangle, minimum height=65mm, minimum width=83mm,align=center},]
\begin{scope}[shift={(0,0)},scale=1]
	\node at (1.5,4) {$U(1)\times\mathbb{Z}_2$};
	\node at (3.3,4) {${(6,1)}$};
	\node at (5.5,2) {${(0,0)}$};
	\node at (4.5,2) {$\mathbb{Z}_4$};
	\node at (7,8) {${(67,0)}$};
	\node at (7,6) {${(21,0)}$};
	\node at (7,4) {${(5,1)}$};
	\node at (8.5,8) {$U(1)^3$};
	\node at (8.5,6) {$U(1)^2$};
	\node at (8.5,4) {$U(1)$};
	\draw [line width=1,->] (8.5,7.5)  -- (8.5,6.5);
	\draw [line width=1,->] (8.5,5.5)  -- (8.5,4.5);

	\draw [line width=1,->] (8.5,3.6) to[out=270,in=0] (6.5,2);

	\draw [line width=1,->] (1.5,3.6) to[out=270,in=180] (3.7,2);
\end{scope}
\end{tikzpicture}
	\caption{The two chains of extremal transitions among complete intersection fibers that end on the generic family of bi-quadrics in $\mathbb{P}^3$.
	For each family we list the index of the polytope, the index of the nef partition and the generic gauge group.}
\label{fig:fiberchains}
\end{figure}

In this section we will study five families of fibers that are codimension two complete intersections in three-dimensional toric ambient spaces.
The corresponding fibrations are connected via extremal transitions which in the corresponding supergravities manifest as Higgs transitions.
As shown in Figure~\ref{fig:fiberchains}, the theories arrange into two Higgs chains that both end on a vacuum with $G=\mathbb{Z}_4$.
The latter is being engineered by fibrations of bi-quadrics in $\mathbb{P}^3$.
However, the field theoretic aspects will be studied in Section~\ref{sec:FieldTheory}.

Here we are going to summarize the geometric data and the associated matter loci that we obtain using GV-spectroscopy.
Note that the calculation of the matter loci follows the generic recipe that we described in the previous section and, since the intermediate results
are unwieldly and hardly illuminating, we will only state the results.
We will also determine some of the matter loci using the classical approach which provides yet another check of our technique.
\subsection{nef (0,0): $G=\mathbb{Z}_4$}
\label{sec:nef(0,0)}
\begin{figure}[ht!] 
\begin{picture}(0,200)
\put(0,0){
\includegraphics[scale=0.4]{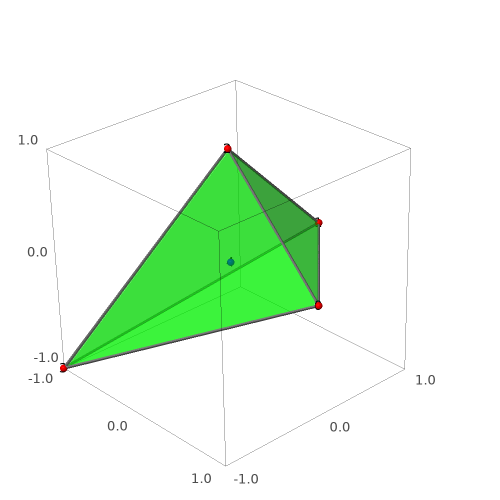}} 
\put(180,100){
{\footnotesize
\begin{tabular}{|c|c|c|}\hline
vertex & coord. & divisor \\ \hline
(-1,-1,-1)&w& $H$ \\ 
	(1,0,0)&x& $H +2 c_1-(\mathcal{S}_2+\mathcal{S}_6 + \mathcal{S}_7+\mathcal{S}_9)$ \\
	(0,1,0)&y& $H+c_1-(\mathcal{S}_6+\mathcal{S}_9)$ \\
	(0,0,1)&z& $H+c_1 -(\mathcal{S}_7+\mathcal{S}_9)  $ \\ \hline 
\multicolumn{2}{|c|}{Intersections} & $H^3 =1$ \\ \hline
\end{tabular}
}
}
\end{picture}  
 \caption{\label{fig:Nef00Data}{The $\mathbb{P}^3$ ambient space polytope of nef (0,0) is depicted on the left. The table shows the respective coordinates and the bundles they transform under.}}
\end{figure} 
We will consider the generic family of bi-quadrics in $\mathbb{P}^3$ and some restrictions that can be torically resolved.
Hence we will investigate the generic family first and this will set the conventions for our concrete analysis in the other cases\footnote{Geometries of this kind have also been considered in~\cite{Kimura:2019bzv,Kimura:2019qxf,Kimura:2019syr}. The Jacobian of nef (0,0) can be found in~\url{http://wwwth.mpp.mpg.de/members/jkeitel/weierstrass/data/0_0.txt}.}.

The tetrahedron $\Delta^\circ$ with vertices
\begin{align}
\rho_1=(-1,\,-1,\,-1)\,,\quad \rho_2=(1,\,0,\,0)\,,\quad \rho_3=(0,\,1,\,0)\,,\quad \rho_4=(0,\,0,\,1)\,,
\end{align}
admits a unique triangulation that corresponds to the three-dimensional projective space $\mathbb{P}_\Delta=\mathbb{P}^3$.
The vertices correspond to the homogeneous coordinates $[w:x:y:z]\in\mathbb{P}^3$ with Stanley-Reisner ideal
\begin{align}
\mathcal{SRI} = \langle w x y z \rangle \, . 
\end{align}
The anticanonical divisor of $\mathbb{P}^3$ admits a nef-partition with $\nabla_1=\langle 0,\rho_1,\rho_2\rangle,\,\nabla_2=\langle0,\rho_3,\rho_4\rangle$ that gives the
complete intersection $V=\{p_1=0\}\cap\{p_2=0\}\in\mathbb{P}^3$ where $p_1,p_2$ are the two quadrics
\begin{align}
\label{eq:bibiquadric}
\begin{split}
p_1=&s_{1,1}\cdot w^2+s_{1,2}\cdot wx+s_{1,3}\cdot wy+s_{1,4}\cdot wz+s_{1,5}\cdot x^2\\
+&s_{1,6}\cdot xy+s_{1,7}\cdot xz+s_{1,8}\cdot y^2+s_{1,9}\cdot yz+s_{1,10}\cdot z^2\,,\\
p_2=&s_{2,1}\cdot w^2+s_{2,2}\cdot wx+s_{2,3}\cdot wy+s_{2,4}\cdot wz+s_{2,5}\cdot x^2\\
+&s_{2,6}\cdot xy+s_{2,7}\cdot xz+s_{2,8}\cdot y^2+s_{2,9}\cdot yz+s_{2,10}\cdot z^2\,.
\end{split}
\end{align}
The coefficients $s_{i,j}$ parametrize the complex structure (redundantly) and, for a generic choice of values, $V$ is a smooth curve of genus one~\footnote{Note that for specific choices of the fibration, certain sections $s_{i,j}$ might be constants and can be absorbed in coordinate redefinitions. Such cases we refer to  as  {\it non-generic} fibrations as here also the gauge symmetry might be non-torically enhanced (see e.g. Section~4.1.3 in \cite{Klevers:2014bqa}).}.
It is easy to see that $[w]=[x]=[y]=[z]\equiv H$, with intersection property $H^3=1$ and $[w]\cdot[p_1]\cdot[p_2]=4$.
Therefore the fiber has no toric sections and one toric four-section .

We adopt the parametrization from~\cite{Schimannek:2018ttm} of fibrations over a base $B$ in terms of four divisors $\mathcal{S}_2,\mathcal{S}_6,\mathcal{S}_7,\mathcal{S}_9$
such that 
\begin{align}
\begin{split}
	[w]=& H\,,\quad [x]= H+2\cdot c_1(B)-(\mathcal{S}_2+\mathcal{S}_6+\mathcal{S}_7+\mathcal{S}_9)\,,\\
	[y]=& H+c_1(B)-(\mathcal{S}_6+\mathcal{S}_9)\,,\quad [z]= H+c_1(B)-(\mathcal{S}_7+\mathcal{S}_9)\,,
	\label{eqn:biquadbundles}
\end{split}
\end{align}
and the classes of the quadrics~\eqref{eq:bibiquadric} are
\begin{align}
[p_1] = 2H + 2 c_1 - (\mathcal{S}_6 + \mathcal{S}_7 + \mathcal{S}_9) \, , \qquad [p_2] = 2H + 3 c_1 - (\mathcal{S}_2+\mathcal{S}_6 + \mathcal{S}_7 +2 \mathcal{S}_9)\,.
\end{align} 
The coefficients are then sections of the line bundles associated to the divisors listed in Table~\ref{tab:s1p3_coefficients} and a summary of the line bundle data of the ambient $\mathbb{P}^3$ is given in Figure~\ref{fig:Nef00Data}. 
\begin{table}[ht!]
\begin{minipage}{.45\linewidth}
\begin{tabular}{c|c}
\text{Coefficient}&\text{Divisor}\\\hline
$s_{1,1}$&$2c_1-\mathcal{S}_6-\mathcal{S}_7-\mathcal{S}_9$\\
$s_{1,2}$&$\mathcal{S}_2$\\
$s_{1,3}$&$c_1-\mathcal{S}_7$\\
$s_{1,4}$&$c_1-\mathcal{S}_6$\\
$s_{1,5}$&$-2c_1+2\mathcal{S}_2+\mathcal{S}_6+\mathcal{S}_7+\mathcal{S}_9$\\
$s_{1,6}$&$-c_1+\mathcal{S}_2+\mathcal{S}_6+\mathcal{S}_9$\\
$s_{1,7}$&$-c_1+\mathcal{S}_2+\mathcal{S}_7+\mathcal{S}_9$\\
$s_{1,8}$&$\mathcal{S}_6-\mathcal{S}_7+\mathcal{S}_9$\\
$s_{1,9}$&$\mathcal{S}_9$\\
$s_{1,10}$&$-\mathcal{S}_6+\mathcal{S}_7+\mathcal{S}_9$\\
\end{tabular}
\end{minipage}
\begin{minipage}{.45\linewidth}
\begin{tabular}{c|c}
\text{Coefficient}&\text{Divisor}\\\hline
$s_{2,1}$&$3c_1-\mathcal{S}_2-\mathcal{S}_6-\mathcal{S}_7-2\mathcal{S}_9$\\
$s_{2,2}$&$c_1-\mathcal{S}_9$\\
$s_{2,3}$&$2c_1-\mathcal{S}_2-\mathcal{S}_7-\mathcal{S}_9$\\
$s_{2,4}$&$2c_1-\mathcal{S}_2-\mathcal{S}_6-\mathcal{S}_9$\\
$s_{2,5}$&$-c_1+\mathcal{S}_2+\mathcal{S}_6+\mathcal{S}_7$\\
$s_{2,6}$&$\mathcal{S}_6$\\
$s_{2,7}$&$\mathcal{S}_7$\\
$s_{2,8}$&$c_1-\mathcal{S}_2+\mathcal{S}_6-\mathcal{S}_7$\\
$s_{2,9}$&$c_1-\mathcal{S}_2$\\
$s_{2,10}$&$c_1-\mathcal{S}_2-\mathcal{S}_6+\mathcal{S}_7$\\
\end{tabular}
\end{minipage}
\caption{{Divisor classes of the coefficients for the biquadric.}}
\label{tab:s1p3_coefficients}
\end{table} 
The matter loci have been determined in Section~\ref{sec:gvcalculation_baseindependent}.
Let us point out that from the GV invariants~\eqref{eqn:ex2gv} even the intersections of the individual components of reducible
fibers with (multi-)sections can be deduced.
We summarize the spectrum and the intersections of the corresponding reducible fibers with the four-section $s_0^{(4)}$ in Table~\ref{tab:nef00Spectrum}.

We are now in the position to study certain specializations of this fiber that are realized as complete intersections in three-dimensional toric ambient spaces. These specializations will come with additional divisor classes that correspond to additional sections or multi-sections.

\begin{table}[h!]
\begin{center}
 \begin{tabular}{|c|c|c|}\hline
Representations &  Multiplicity  & Fiber \\ \hline 
$ \mathbf{1}_1 $&$  4 (3 c_1 - \mathcal{S}_2 - \mathcal{S}_9) (c_1 + \mathcal{S}_2 + \mathcal{S}_9) $ &      
   \rule{0pt}{1.2cm}\parbox[c]{1.8cm}{    \includegraphics[scale=0.25]{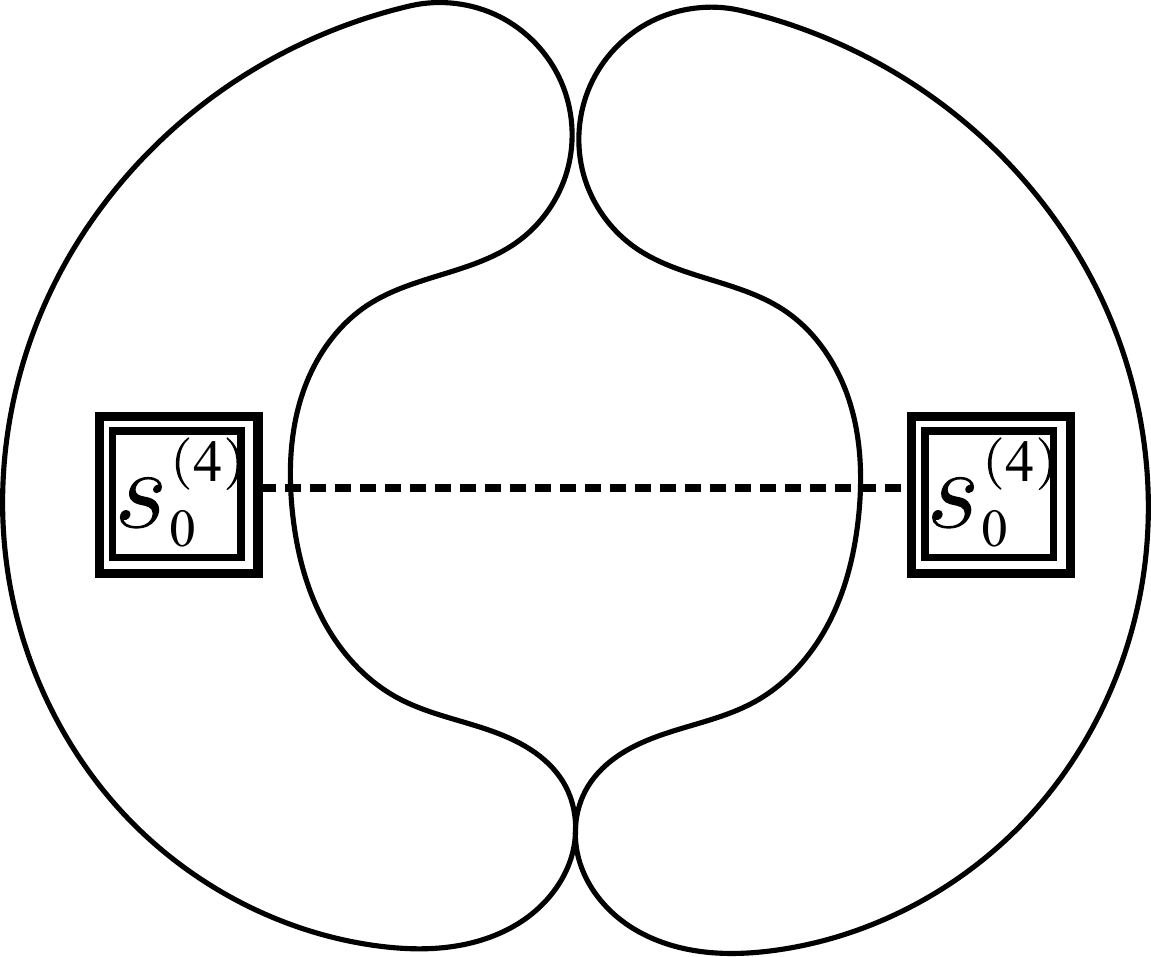}}\hspace{1cm}  
      \\ \hline
	$	\mathbf{1}_2 $& $ \begin{array}{l}
	6 c_1^2 + \mathcal{S}_2^2 - 2 (\mathcal{S}_6^2 + \mathcal{S}_7^2) - 2 \mathcal{S}_2 (\mathcal{S}_6 + \mathcal{S}_7 - 2 \mathcal{S}_9)\\   - 
 2 (\mathcal{S}_6 + \mathcal{S}_7) \mathcal{S}_9 + \mathcal{S}_9^2 - c_1 (\mathcal{S}_2 - 4 (\mathcal{S}_6 + \mathcal{S}_7) + \mathcal{S}_9)
	   \end{array}$       &  
	   \rule{0pt}{1.2cm}\parbox[c]{1.8cm}{    \includegraphics[scale=0.25]{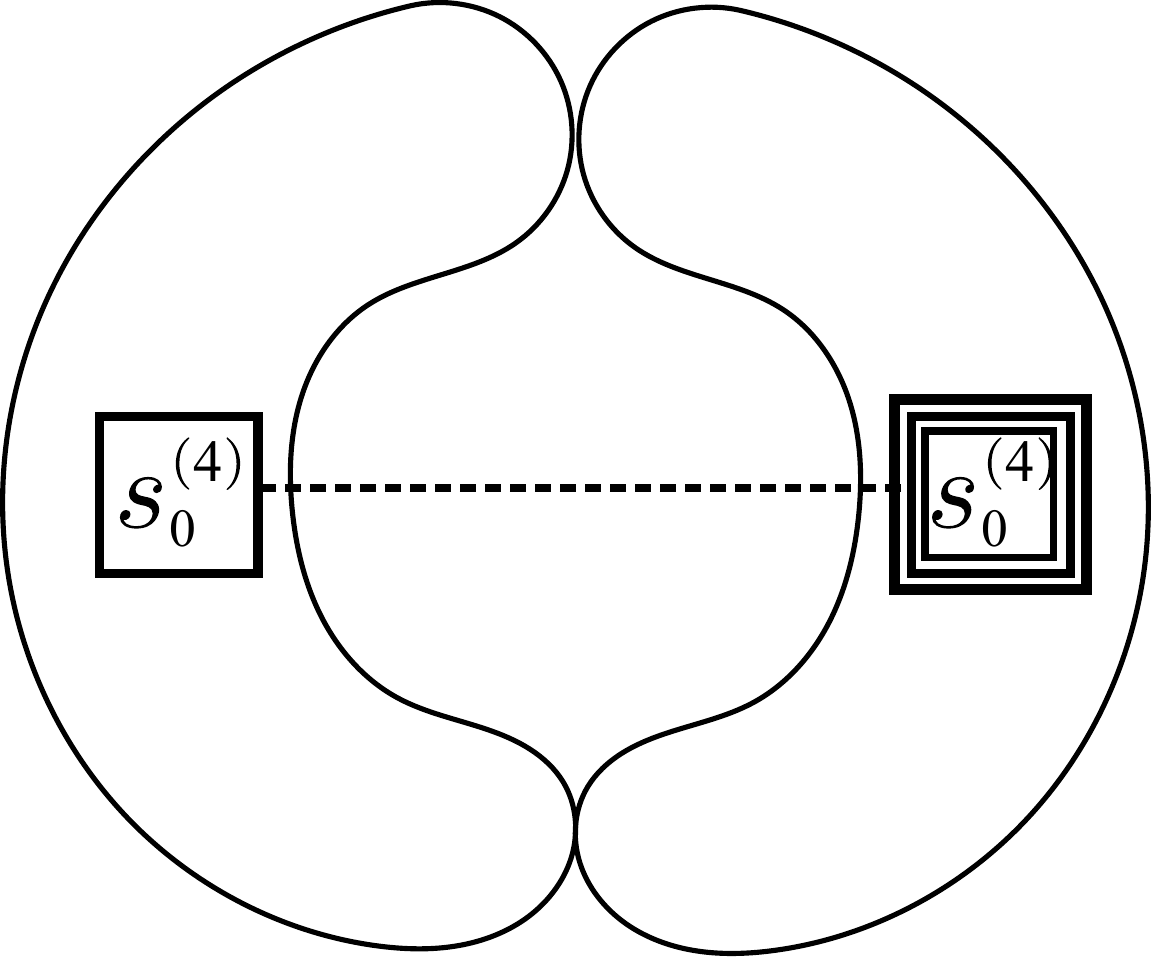}}\hspace{1cm}

	    \\              \hline  
 	$ \mathbf{1}_0$ & $\begin{array}{l} 
 	 11 c_1^2 + 3 \mathcal{S}_2^2 + 2 (6 + \mathcal{S}_6^2 + \mathcal{S}_7^2) +  2 (\mathcal{S}_6 + \mathcal{S}_7) \mathcal{S}_9 + 3 \mathcal{S}_9^2  \\+ 
 2 \mathcal{S}_2 (\mathcal{S}_6 + \mathcal{S}_7 + 2 \mathcal{S}_9) - c_1 (7 \mathcal{S}_2 + 4 (\mathcal{S}_6 + \mathcal{S}_7) + 7 \mathcal{S}_9)  \end{array}$ & \\ \hline
 \end{tabular}
 \caption{\label{tab:nef00Spectrum} {Summary of the charged matter representations and multiplicities under $  \mathbb{Z}_4$ and codimension two fibers of nef $(0,0)$.                  }}
 \end{center}
\end{table}

\subsection{nef (6,1): $G=\mathbb{Z}_2 \times U(1)$}
\begin{figure}[th!] 
\begin{picture}(0,200)
\put(0,0){
\includegraphics[scale=0.4]{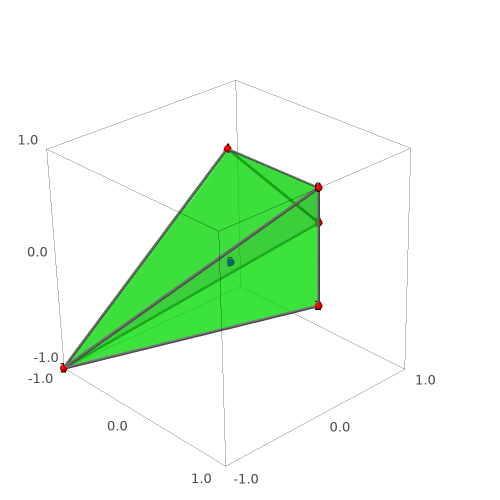}} 
\put(180,100){
{\footnotesize
\begin{tabular}{|c|c|c|}\hline
vertex & coord. & divisor \\ \hline
(-1,-1,0)&w& $H_1$ \\ 
 (0,0,1)&x& $H_1 -H_2+2 c_1-(\mathcal{S}_2+\mathcal{S}_6 + \mathcal{S}_7+\mathcal{S}_9)$ \\
(0,1,0)&y& $H_1+c_1-(\mathcal{S}_6+\mathcal{S}_9)$ \\
(1,0,-1)&z& $H_1-H_2+c_1 -(\mathcal{S}_7+\mathcal{S}_9)  $ \\  
(1,0,0)&e$_2$& $ H_2 $ \\ \hline
Intersections&  \multicolumn{2}{|c|}{ $H_1^3=1\, ,\, H_2^3=-2\, , \,  H_1 H_2^2=-1\, ,\, H_1^2 H_2=0$} \\ \hline
\end{tabular} 
}
} 
\end{picture} \caption{\label{fig:Nef61}{The ambient space polytope of nef (6,1). The table shows the respective coordinates and the bundles they transform under.}}
\end{figure} 

The second codimension two nef partition of the three dimensional reflexive polytope with PALP index 6 has been shown in~\cite{Braun:2014qka} to generically not exhibit a toric section and also no fibral divisors~\footnote{\url{http://wwwth.mpp.mpg.de/members/jkeitel/weierstrass/data/6_1.txt}}.
The defining equations are given by
\begin{align}
\label{eq:nef61}
\begin{split}
p_1=&e_2^2 (s_{1,5}\cdot x^2 + s_{1,7}\cdot x z + s_{1,10} \cdot z^2)+ e_2 (s_{1,2} \cdot w x + s_{1,6 } \cdot x y + s_{1,4}\cdot w z + s_{1,9}\cdot y z)\\ & + s_{1,1}\cdot w^2 + s_{1,3}\cdot w y + s_{1,8}\cdot y^2  \, , \\ 
p_2=& e_2 (s_{2,5}\cdot x^2 + s_{2,7}\cdot x z + s_{2,10}\cdot z^2)+  s_{2,6}\cdot xy + s_{2,2}\cdot w x + s_{2,4}\cdot w z + s_{2,9}\cdot y z \, .
\end{split}
\end{align}
and they can be viewed as a specialization of the bi-quadric of nef (0,0), by setting the coefficients $s_{2,1}, s_{2,3},s_{2,8}$ in Equation~\ref{eq:bibiquadric} to zero. A picture of the polytope, the homogeneous coordinates $w,x,y,z,e_2$ that correspond to the vertices and the corresponding divisor classes can be found in Figure~\ref{fig:Nef61}.
The tetrahedron of $\mathbb{P}^3$ can be obtained by dropping the point that corresponds to $e_2$ and applying the
lattice automorphism
\begin{align}
	M_{6}\rightarrow M_{0}:\quad\vec{p}\mapsto\left(\begin{array}{ccc}
		1&0&1\\
		0&1&0\\
		1&0&0
	\end{array}\right)\vec{p}\,,
\end{align}
to the remaining rays.
Using a triangulation of the ambient space, we obtain the Stanley-Reisner ideal
\begin{align}
\label{eq:SRInef61}
\mathcal{SRI} = \langle x z, e_2 w y \rangle \, .
\end{align}
The divisors $[e_2],[x],[z]$ intersect the generic fiber twice while $[w]$ and $[y]$ intersect it four times.  
A generic fibration constructed with this fiber has no sections but only two sections.

There are two linearly independent divisors on the ambient space and those lead to independent two-sections.
Hence we will have a $\mathbb{Z}_2 \times U(1)$ gauge symmetry in F-theory.
We choose two two-sections as follows:
\begin{align}
\label{eq:2sections}
	\begin{split}
		s_0^{(2)}:\,&   \{e_2 = 0\} \cap\left\{ \begin{array}{l} \hat{p}_1=s_{1,1}w^2 + s_{1,3}w y + s_{1,8}y^2=0\\\hat{p}_2= s_{2,6}xy + s_{2,2}w x + s_{2,4}w z + s_{2,9}y z=0  \end{array} \right.  \\
			s_1^{(2)}:\,&  \{z= 0\} \cap\left\{ \begin{array}{l} \tilde{p}_1=s_{1,5} e_2^2 +   s_{1,2 } e_2 w + s_{1,1} w^2 + e_2 s_{1,6} y + s_{1,3} w y + s_{1,8} y^2=0\\ \tilde{p}_2= s_{2,5} e_2 + s_{2,2} w + s_{2,6} y=0  \end{array} \right.
	\end{split}
\end{align}
We use those two classes to construct the images under the generalized Shioda map 
\begin{align}
\sigma_{\mathbb{Z}_2}(s_0^{(2)}) = [e_2] \, , \quad \sigma (s_1^{(2)}) = [z]-[e_2]  \, ,
\end{align}
where we dropped possible contributions from base divisors.
The intersection of a fibral curve with $\sigma_{\mathbb{Z}_2}(s_0^{(2)})$ and $\sigma(s_1^{(2)})$ respectively
determines the $\mathbb{Z}_2$ and $U(1)$ charge of the corresponding hypermultiplet.
We have again used GV-spectroscopy to determine the matter loci and they are summarized in Table~\ref{tab:nef61Spectrum}.
 \subsubsection*{Analysis of Higgs loci}
In the following, we will consider one singlet locus in more detail as it will lead to the higgs multiplet, which enables the transition from $U(1) \times \mathbb{Z}_2$ to $\mathbb{Z}_4$.
On that locus the two-section $s_0^{(2)}$ degenerates.
This happens, when $\hat{p}_1$ and $\hat{p}_2$ in Equation~\eqref{eq:2sections} develop a common factor that is linear in $(w,y)$, i.e. 
\begin{align}
\hat{p}_1  \rightarrow q_1(w,y) \widehat{q}_1(w,y) \, , \quad 
\hat{p}_2 \rightarrow q_1(w,y) \widehat{q}_2(x,z) \, .
\end{align}
To find this locus we first note that factorization of $\hat{p}_2$ requires that
\begin{align}
s_{2,2} s_{2,9}-s_{2,4} s_{2,6}=0\,.
\end{align}
One can then take the resultant of $q_1(w,y)=s_{2,4}w+s_{2,9}y$ and $\hat{p}_1$.
We find that the locus where the two-section degenerates corresponds to the ideal
\begin{align}
I_1 =  \langle s_{2,2} s_{2,9}-s_{2,4} s_{2,6},\, s_{1,8} s_{2,4}^2 - s_{1,3} s_{2,4} s_{2,9} + s_{1,1} s_{2,9}^2 \rangle\, .
\end{align}
One component of the reducible fiber is then $\mathcal{C}_{1 }=[e_2][q_1(w,y)]$ and intersects the two-sections as $\mathcal{C}_{1} \cdot (s_0^{(2)},s_1^{(2)})= (-1,1)$.
Using the respective Shioda maps we find the charges $q=(2,-)$ under $U(1) \times \mathbb{Z}_2$.

It is now easy to be mislead and to assume that the class of the matter locus is $([s_{2,2}]+[s_{2,9}])([s_{1,8}]+2[s_{2,4}])$.
However, note that the ideal $I_1$ is contained in the simpler ideal $\langle s_{2,4},s_{2,9}\rangle$. A resultant computation reveals the corresponding locus to be included in $V(I_1)$ two times. But this simpler locus does not lead to the correct factorization of the CICY equations. Therefore we have to subtract this locus with multiplicity two. This leads to the multiplicity
\begin{align}
	n_{(2,-)}&=([s_{2,4}]+[s_{2,6}])([s_{1,8}] +2 [s_{2,4}])-2 [s_{2,4}][s_{2,9}] \nonumber \\ & = 4 c_1^2 + \mathcal{S}_2 ( \mathcal{S}_7 + \mathcal{S}_9-\mathcal{S}_6 ) + \mathcal{S}_9 (\mathcal{S}_6 + \mathcal{S}_7 + \mathcal{S}_9) - 
 2 c_1 (\mathcal{S}_2 +\mathcal{S}_7 + 2 \mathcal{S}_9) \, .
\end{align}
The multiplicity and the charges agree with the result from GV-spectroscopy that moreover allows to compute the full spectrum given in Table~\ref{tab:nef61Spectrum}.
The neutral singlets are obtained from the Euler number and a summary of the latter is given in Appendix~\ref{app:eulers}.

Let us stress that to determine the multiplicity by directly analyzing the fiber we had to make use of the special property that the section degenerates and even then determining the correct class is subtle.
None of these problems arise when the matter loci are determined via base independent calculation of fiber GV invariants which always follows the same generic recipe.
\begin{table} 
\begin{center}
	\begin{tabular}{|c|c|c|} \hline
	Representation & Multiplicity&  {\centering   Fiber  }   \\ \hline   
		$\mathbf{1}_{(2,-)}$ & $ \begin{array}{l}4 c_1^2 - 2 c_1 \mathcal{S}_2 - \mathcal{S}_2 \mathcal{S}_6 - 2 c_1 \mathcal{S}_7 + \mathcal{S}_2 \mathcal{S}_7 \\ - 4 c_1 \mathcal{S}_9 + \mathcal{S}_2 \mathcal{S}_9 + \mathcal{S}_6 \mathcal{S}_9 + \mathcal{S}_7 \mathcal{S}_9 + \mathcal{S}_9^2 \end{array} $ &   \rule{0pt}{1.2cm}\parbox[c]{1.8cm}{ \includegraphics[scale=0.25]{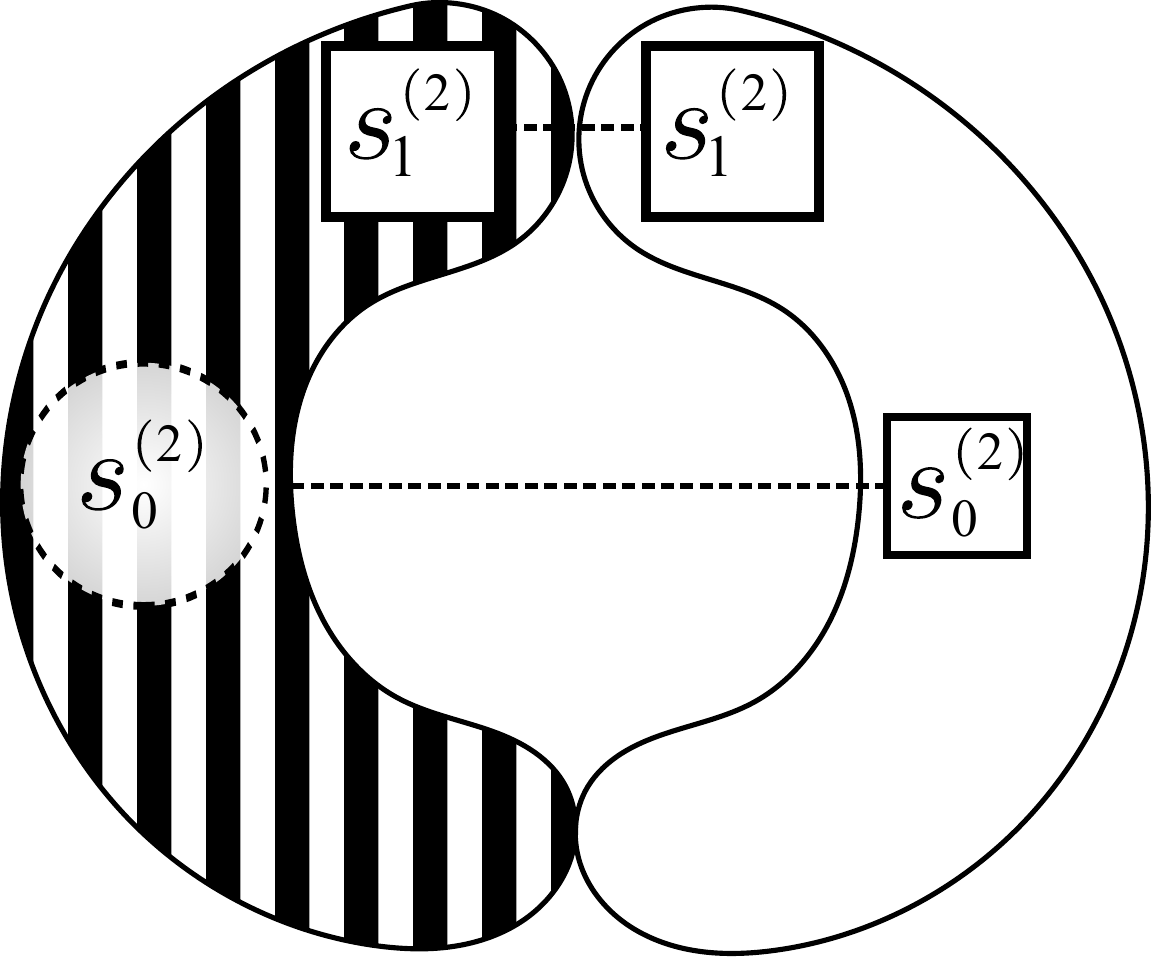}} \hspace{1cm}  \\ \hline 
		
		$\mathbf{1}_{(2,+)} $& $\begin{array}{l}2 c_1^2 - 3 c_1 \mathcal{S}_2 + \mathcal{S}_2^2 + \mathcal{S}_2 \mathcal{S}_6 + 2 c_1 \mathcal{S}_7 \\ - \mathcal{S}_2 \mathcal{S}_7  - c_1 \mathcal{S}_9 + \mathcal{S}_2 \mathcal{S}_9 - \mathcal{S}_6 \mathcal{S}_9 - \mathcal{S}_7 \mathcal{S}_9  \end{array}$ & \rule{0pt}{1.2cm}\parbox[c]{1.8cm}{ \includegraphics[scale=0.25]{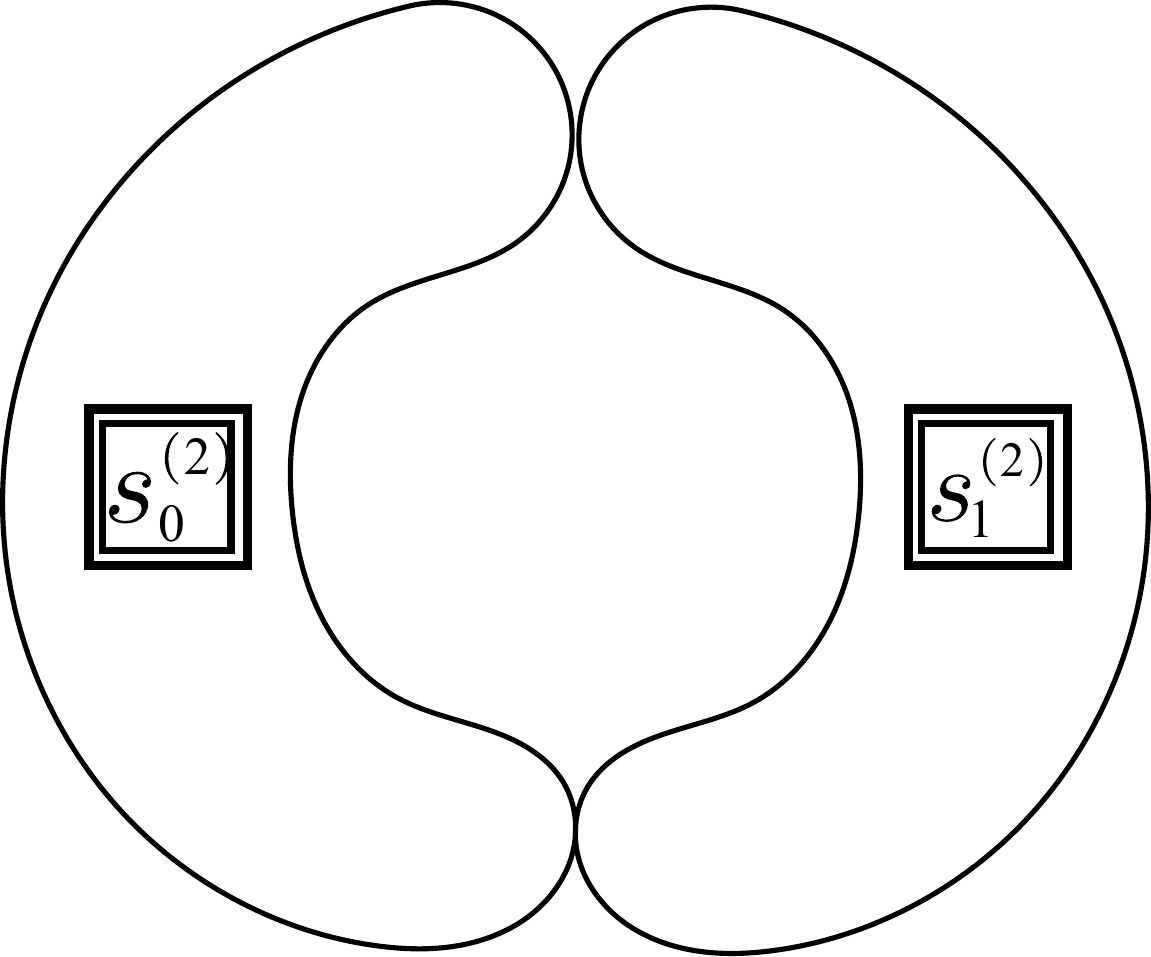}} \hspace{1cm}  \\ \hline 
		
	$	\mathbf{1}_{(1,-)}$  & $ \begin{array}{l}2 (4 c_1^2 - 2 c_1 \mathcal{S}_6 + 2 \mathcal{S}_2 \mathcal{S}_6 + \mathcal{S}_6^2 \\ - \mathcal{S}_7^2  + 2 c_1 \mathcal{S}_9 - 2 \mathcal{S}_2 \mathcal{S}_9 - \mathcal{S}_9^2)\end{array} $ &  \rule{0pt}{1.2cm}\parbox[c]{1.8cm}{ \includegraphics[scale=0.25]{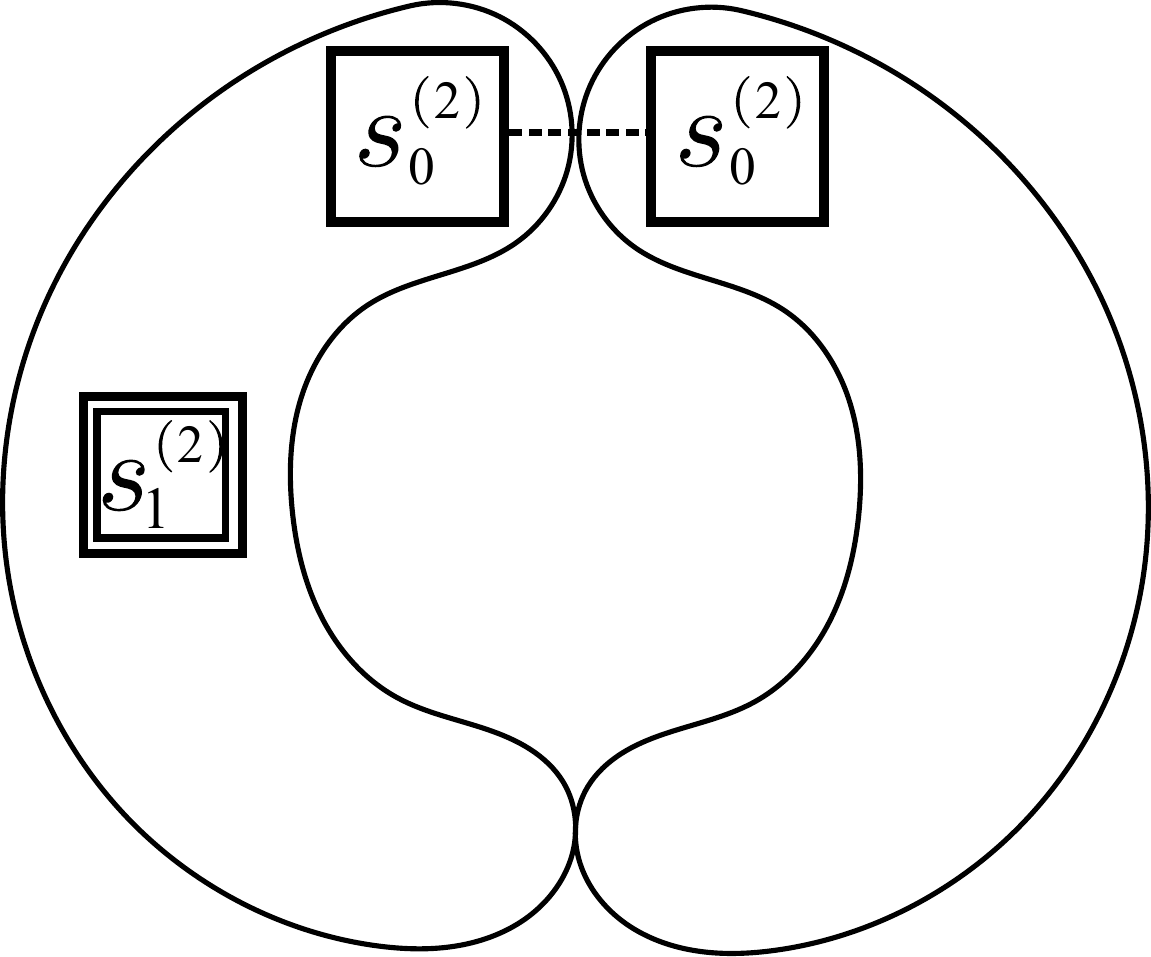}} \hspace{1cm}  \\ \hline
	
		$\mathbf{1}_{(1,+)}$ & $ \begin{array}{l}2  (2 c_1^2 + 4 c_1 \mathcal{S}_2 - 2 \mathcal{S}_2^2 + 2 c_1 \mathcal{S}_6  - 2 \mathcal{S}_2 \mathcal{S}_6 \\  \quad  - \mathcal{S}_6^2 + \mathcal{S}_7^2 + 2 c_1 \mathcal{S}_9 - 2 \mathcal{S}_2 \mathcal{S}_9 - \mathcal{S}_9^2 ) \end{array}$ &  \rule{0pt}{1.2cm}\parbox[c]{1.8cm}{ \includegraphics[scale=0.25]{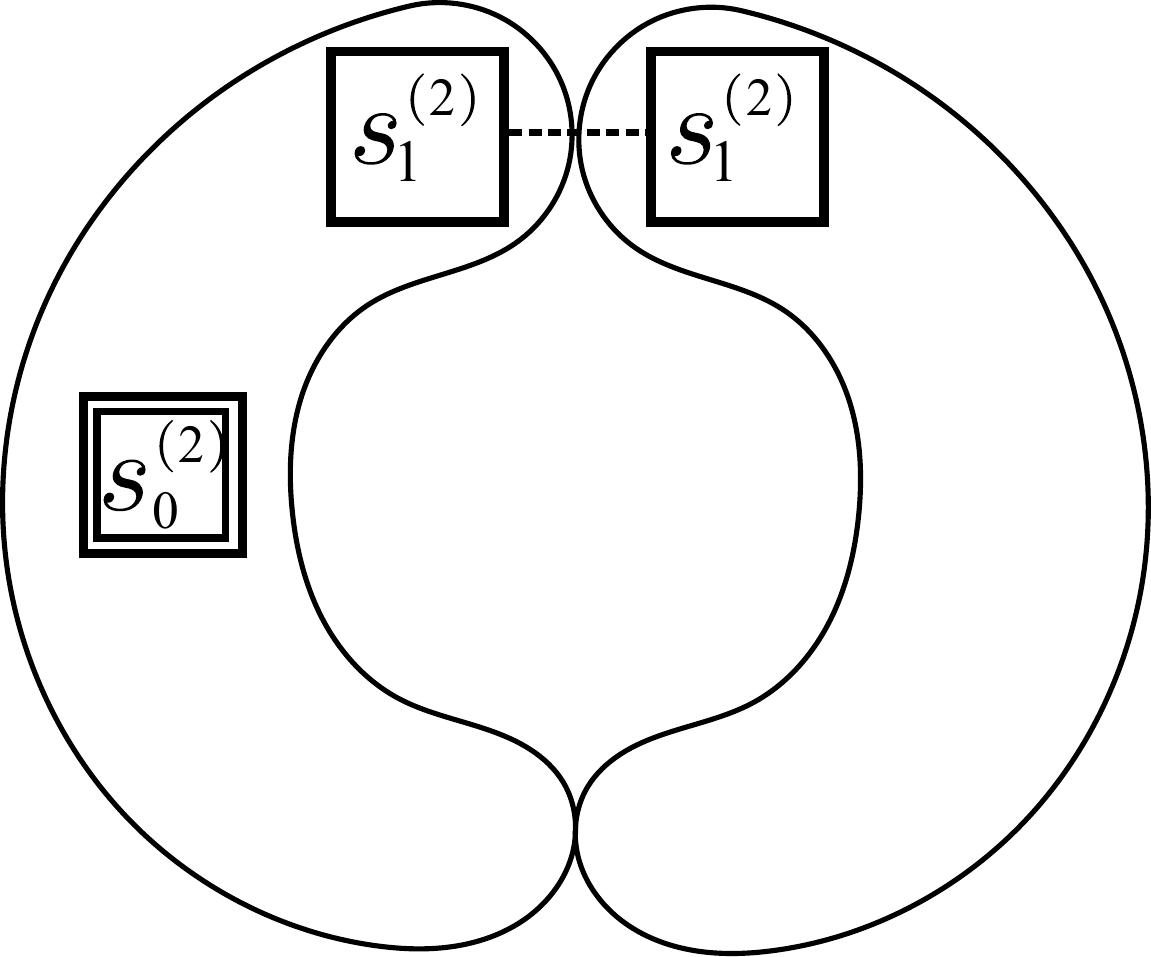}} \hspace{1cm}  \\  \hline  
		
		$\mathbf{1}_{(0,-)}$ &  $ \begin{array}{l} 4 c_1^2 + 2 c_1 \mathcal{S}_2 + 4 c_1 \mathcal{S}_6 - 3 \mathcal{S}_2 \mathcal{S}_6 \\   
		 - \mathcal{S}_2 \mathcal{S}_7 - 2 \mathcal{S}_7^2 + 3 \mathcal{S}_2 \mathcal{S}_9  - 2 \mathcal{S}_6^2 \\ - \mathcal{S}_6 \mathcal{S}_9 - \mathcal{S}_7 \mathcal{S}_9 + \mathcal{S}_9^2 + 2 c_1 \mathcal{S}_7  \end{array} $ &    \rule{0pt}{1.2cm}\parbox[c]{1.8cm}{ \includegraphics[scale=0.25]{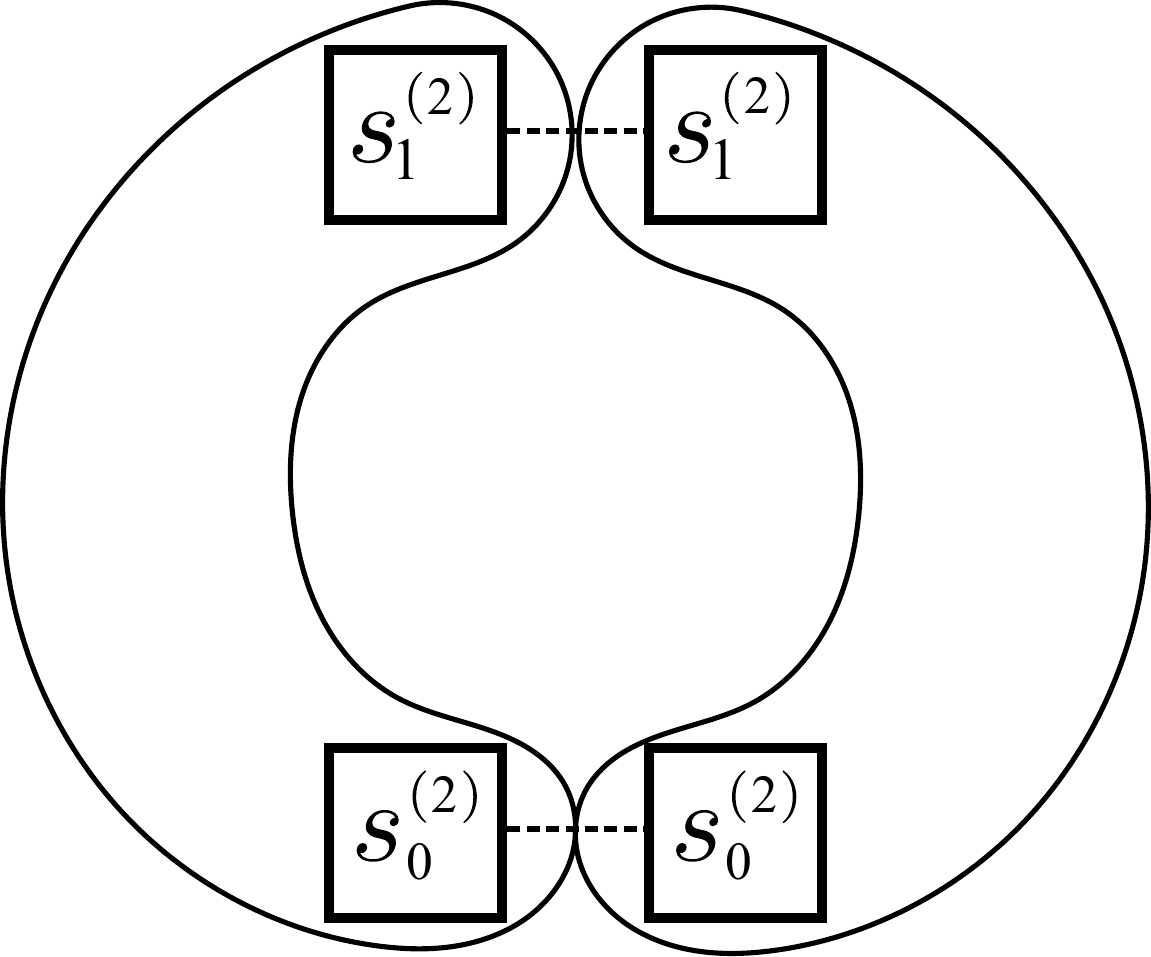}}\hspace{1cm}  \\ \hline 
		 
			$\mathbf{1}_{(0,+)} $&   $\begin{array}{l} 13 + 7 c_1^2 + 3  \mathcal{S}_2^2 + 2 ( \mathcal{S}_6^2 +  \mathcal{S}_7^2) +   \mathcal{S}_6   \mathcal{S}_9  \\+ \mathcal{S}_7   \mathcal{S}_9 + 2  \mathcal{S}_9^2  + 
  \mathcal{S}_2 (3  \mathcal{S}_6 +  \mathcal{S}_7 + 3  \mathcal{S}_9)\\ - c_1 (5  \mathcal{S}_2 + 4  \mathcal{S}_6 + 2  \mathcal{S}_7 + 3  \mathcal{S}_9)   \end{array}$  &  \\ \hline \hline
  Anomaly coefficient & \multicolumn{2}{|c|}{ $b=2 (3 c_1 - \mathcal{S}_2 - \mathcal{S}_9)$ } \\ \hline
	\end{tabular} 
\caption{\label{tab:nef61Spectrum} {Summary of the charged matter representations and multiplicities under $U(1) \times \mathbb{Z}_2$ and codimension two fibers of nef $(6,1)$.                  }}
	\end{center}
\end{table}

\subsection{nef (5,1): $G=U(1)$}
\label{sec:nef51}
\begin{figure}[t!] 
\begin{picture}(0,200)
\put(0,0){
\includegraphics[scale=0.4]{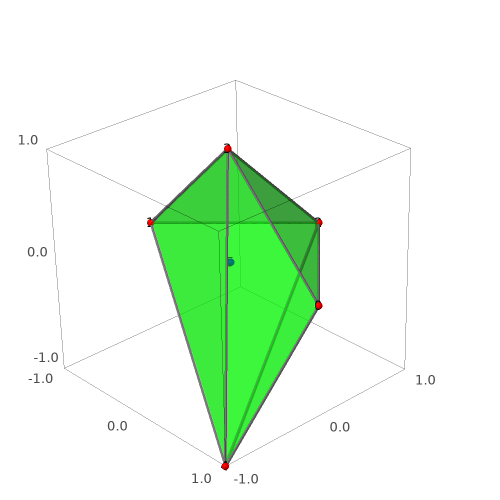}} 
\put(180,100){
{\footnotesize
\begin{tabular}{|c|c|c| }\hline
vertex & coord. & divisor  \\ \hline
(1,-1,-1)&w& $H-E_1 $ \\ 
 (-1,0,0)&x& $H +2 c_1-(\mathcal{S}_2+\mathcal{S}_6 + \mathcal{S}_7+\mathcal{S}_9)$ \\
(0,1,0)&y& $H-E_1+c_1-(\mathcal{S}_6+\mathcal{S}_9)$ \\
(0,0,1)&z& $H-E_1+c_1 -(\mathcal{S}_7+\mathcal{S}_9)  $ \\ 
(1,0,0)&$e_1$& $E_1   $ \\ \hline
Intersections& \multicolumn{2}{|c|}{$H^3=1\, , \, E_1^3=1 \, , \,  E_1 H^2=0\, , \, E_1^2 H = 0 $} \\ \hline
\end{tabular}
}
}
\end{picture}
\caption{\label{fig:Nef51} {The polytope and data associated to the ambient space Bl$_1 \mathbb{P}^3$ of nef (5,1). The latter generically leads to a rank one Mordell-Weil group. }
 }
\end{figure}
The second codimension two nef partition of the three dimensional reflexive polytope with PALP index 5 has been shown in \cite{Braun:2014qka} to generically
lead to an elliptic fibration with Mordell-Weil rank one and no fibral divisors~\footnote{\url{http://wwwth.mpp.mpg.de/members/jkeitel/weierstrass/data/5_1.txt}}.
The defining equations can be obtained from Equation~\eqref{eq:bibiquadric} by setting the two sections $s_{j,5}$ to zero.
They can be written in a suggestive way as
\begin{align}
	\begin{split}
	p_1=&e_1 (s_{1,1} w^2 + s_{1,3} w y + s_{1,8} y^2 + s_{1,4} w z + s_{1,9} y z + 
   s_{1,10} z^2)\\
		&+ x (s_{1,2} w + s_{1,6} y + s_{1,7} z)\, ,  \\
		p_2=&e_1 (s_{2,1} w^2 + s_{2,3} w y + s_{2,8} y^2 + s_{2,4} w z + s_{2,9} y z + 
   s_{2,10} z^2)\\
		&+ x (s_{2,2} w + s_{2,6} y + s_{2,7} z)   \,,
	\end{split}
\end{align}
where the correspondence between homogeneous coordinates $w,x,y,z,e_1$ and vertices of the polytope is given in Figure~\ref{fig:Nef51}.
The tetrahedron of $\mathbb{P}^3$ can be obtained by dropping the point that corresponds to $e_1$ and applying the
lattice automorphism
\begin{align}
	M_{5}\rightarrow M_{0}:\quad\vec{p}\mapsto\left(\begin{array}{ccc}
		-1&0&0\\
		0&1&0\\
		0&0&1
	\end{array}\right)\vec{p}\,,
	\label{eqn:auto5to0}
\end{align}
to the remaining rays.

The polytope admits a unique regular fine star triangulation which results in the Stanley-Reisner ideal
\begin{align}
	\mathcal{SRI}= \langle e_1 x, w y z  \rangle \, .
\end{align}
Calculating the intersections with the generic fiber we find that $[e_1]$ restricts to a section whereas $[x]$ is a four-section and $[w],[y],[z]$ are three-sections.
The homogeneous coordinates $[e_1:x:y:z:w]$ of the intersection of $\{e_1=0\}$ with the fiber are
\begin{align}
s_0 : \quad [ 0:\, 1:\, s_{1,7} s_{2,2} - s_{1,2} s_{2,7}:  s_{1,2} s_{2,6}-s_{1,6} s_{2,2} : s_{1,6} s_{2,7}-s_{1,7} s_{2,6} ] \, .
\end{align}
Using the ``tangent-trick'', see e.g. Section 3.3.1 of~\cite{Klevers:2014bqa}, we can construct a second section $s_1$ and the associated divisor class is
\begin{align}
	[s_1]\sim  [w] -2 [e_1] \, .
\end{align}
The corresponding image under the Shioda map is
\begin{align}
\sigma(s_1)= [s_1]-[s_0] \, ,
\end{align} 
where we have again dropped vertical divisors.
Base independent expressions for the matter loci can be obtained using GV-spectroscopy and are listed in Table~\ref{tab:nef51}.

\subsubsection*{Map into a restricted cubic}
The family can be brought into the form of a restricted cubic hypersurfaces in $\mathbb{P}^2$.
To this end let us first rewrite the defining equations as
\begin{align}
p_1 = e_1 A_{2} + x B_1\,,\qquad 
p_2 = e_1 C_2 + x D_1 \, ,
\end{align}
where $A_2$ and $C_2$ are polynomials of degree two in $w,y,z$ while $B_1$ and $D_1$ are linear.
Away from $e_1=0$ this allows us to map the curve into a cubic hypersurface
\begin{align}
	p_{\text{cubic}}=A_2 D_1 - B_2 C_1 \, ,
\end{align}
inside $\mathbb{P}^2$ with homogeneous coordinates $[w:y:z]$.
Shifting the coordinates $y \rightarrow a y + b z + c w$ to eliminate $y$ and $z$ from $B_1$, and then shifting $z \rightarrow d z + e w + f y$ to eliminate $w$ from $D_1$ allows a direct match with the cubic obtained in~\cite{Raghuram:2017qut} for a model with charge $4$ matter.

\subsubsection*{Analysis of Higgs loci}
We are now going to analyze the matter locus that leads to a hypermultiplet with charge $q=4$. 
This will be the Higgs multiplet which allows us to break the $U(1)$ gauge symmetry to $\mathbb{Z}_4$.
The charge four matter can be found at the locus where $s_0$ degenerates.
This happens when $w,y,z$ vanish simultaneously and this determines the ideal
\begin{align}
\label{eq:I4locus}
	I_1 = \langle s_{1,7} s_{2,2} - s_{1,2} s_{2,7},\,s_{1,2} s_{2,6}-s_{1,6} s_{2,2},\,s_{1,6} s_{2,7}-s_{1,7} s_{2,6}  \rangle \, .
\end{align}
Note that away from $s_{2,2}=0$ this is equivalent to the ideal
\begin{align}
\label{eq:I4codim2}
I_1'= \langle s_{1,7} s_{2,2} - s_{1,2} s_{2,7},\,s_{1,2} s_{2,6}-s_{1,6} s_{2,2}\rangle \,,
\end{align} 
which is in turn contained in $\langle s_{1,2},\,s_{2,2}\rangle$.
Over $V(I_1)$ the CICY equations obtain the form
\begin{align}
\begin{split}
p_1 =& e_1 A_2 + x B_1 \,,\quad p_2 = e_1 C_2 + x \lambda B_1 \, ,
\end{split}
\end{align}
 with $\lambda$ being some non-zero constant.
 Hence the component wrapped by $s_0$ is in the class
 \begin{align}
 \mathcal{C}_{1}=[e_1]\cdot[B_1]\,.
 \end{align}
 The intersection numbers of this fibral curve with the sections are $\mathcal{C}_{1}\cdot s_i=(-1,3)$.
 This implies that the corresponding matter has charge $q=4$.

We note that the charge 4 locus has some remarkable properties.
First, we find that both rational sections degenerate and each wraps a fibral curve.
Moreover, we find that the section $s_1$ wraps the fibral curve $\mathcal{C}_{2}$ two times.
Since $s_1$ is a section it therefore must also intersect the second curve $\mathcal{C}_1$ one additional time~\footnote{Such a behavior was first noted to be possible in \cite{Lawrie:2015hia}.}.
To obtain the multiplicity of the charge four locus defined in~\eqref{eq:I4codim2} we take again the class of the complete intersection and keep in mind that we have to substract the class of $\langle s_{2,2},s_{1,2}\rangle$.
This leads to
\begin{align}
n_{1_4}= [s_{1,6} +s_{2,7}] [s_{2,7} +s_{1,2}]-[s_{1,2}][ s_{2,2}] =   (\mathcal{S}_2 + \mathcal{S}_6) (\mathcal{S}_2 + \mathcal{S}_7) +  \mathcal{S}_2 \mathcal{S}_9 -c_1  \mathcal{S}_2 \, ,
 \end{align} 
which agrees with the result that we obtained from the base independent calculation of fiber GV invariants, see Table~\ref{tab:nef51}. The Euler number is again listed in Appendix~\ref{app:eulers}.
 
\begin{table} 
	\begin{tabular}{|c|c|c|} \hline
	Representation & Multiplicity& Fiber \\ \hline

		$\mathbf{1}_{4}$ & $ (\mathcal{S}_2 + \mathcal{S}_6) (\mathcal{S}_2 + \mathcal{S}_7) + \mathcal{S}_2 \mathcal{S}_9-c_1 \mathcal{S}_2  $ &  \rule{0pt}{1.2cm}\parbox[c]{1.8cm}{ \includegraphics[scale=0.25]{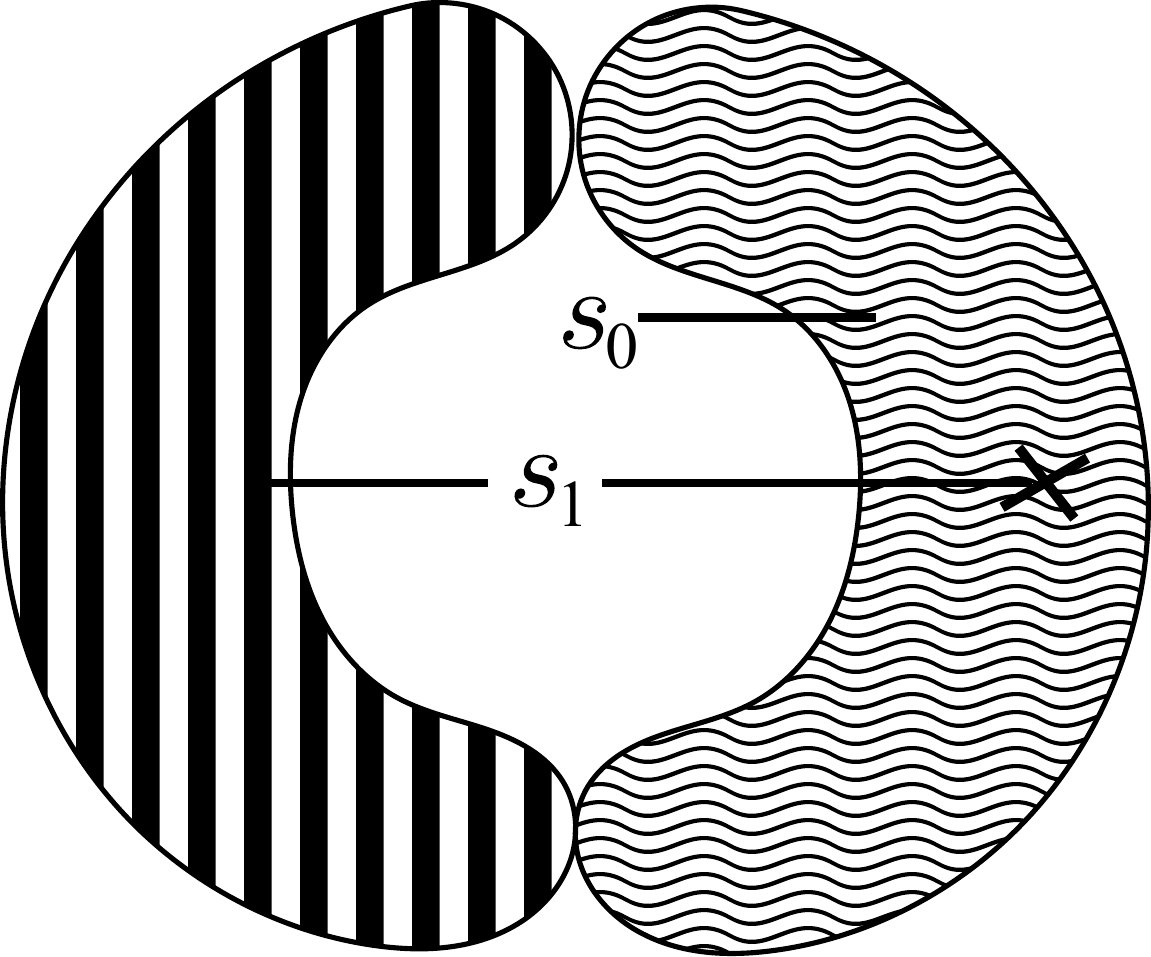}} \hspace{1cm} \\ \hline 

				$\mathbf{1}_{3}$ & $\begin{array}{l} (\mathcal{S}_6 - \mathcal{S}_7)^2 + (\mathcal{S}_6 + \mathcal{S}_7) \mathcal{S}_9  \\+ c_1 (6 \mathcal{S}_2 + \mathcal{S}_6 + \mathcal{S}_7 +\mathcal{S}_9)\\ - 
 \mathcal{S}_2 (\mathcal{S}_6 + \mathcal{S}_7 + 3 \mathcal{S}_9) -2 \mathcal{S}_2^2  \\ \end{array} $ &\rule{0pt}{1.2cm}\parbox[c]{1.8cm}{ \includegraphics[scale=0.25]{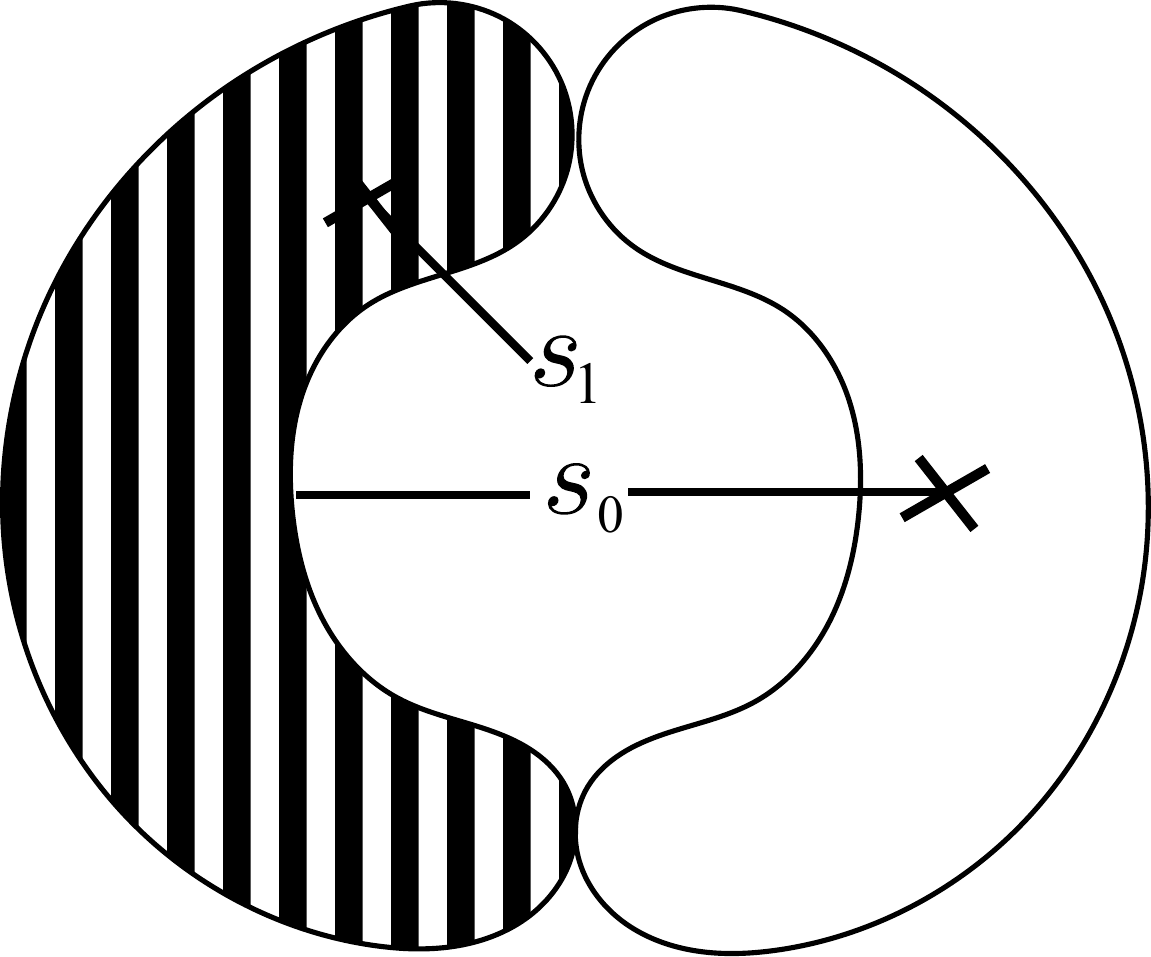}} \hspace{1cm} \\ \hline

	$	\mathbf{1}_{2}$  & $  \begin{array}{l} 6  c_1^2 - c_1 \mathcal{S}_2 + \mathcal{S}_2^2 + 4 c_1 \mathcal{S}_6 - 2 \mathcal{S}_2 \mathcal{S}_6 \\  - 2 \mathcal{S}_6^2 + 4 c_1 \mathcal{S}_7   - 2 \mathcal{S}_2 \mathcal{S}_7
		 - 2 \mathcal{S}_7^2+ \mathcal{S}_9^2 \\ - c_1 \mathcal{S}_9 + 4 \mathcal{S}_2 \mathcal{S}_9 - 2 \mathcal{S}_6 \mathcal{S}_9 - 2 \mathcal{S}_7 \mathcal{S}_9  \end{array}$  & \rule{0pt}{1.2cm}\parbox[c]{1.8cm}{ \includegraphics[scale=0.25]{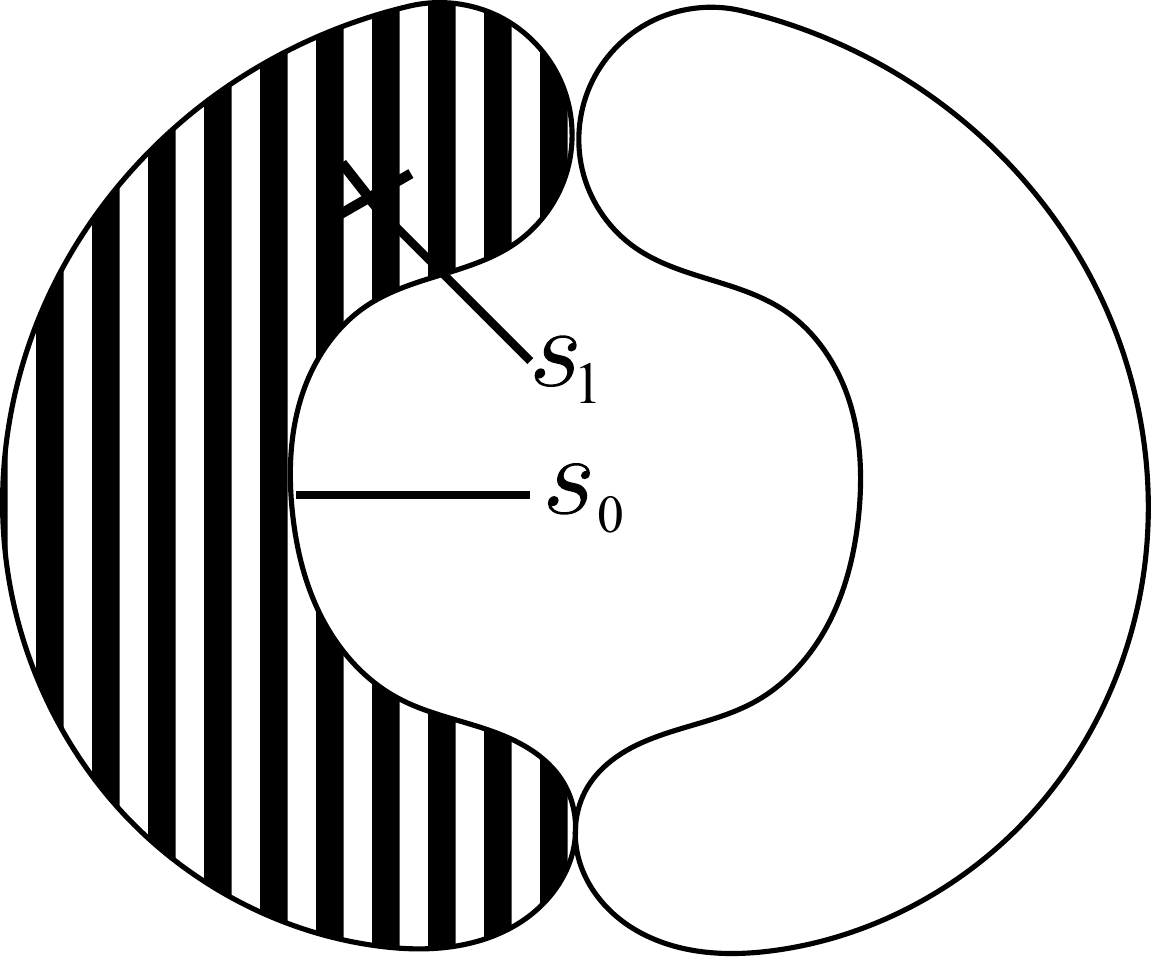}} \hspace{1cm} \\ \hline

				$\mathbf{1}_{1}$ &  $ \begin{array}{l}12 c_1^2 + 2 c_1 \mathcal{S}_2 - 2 \mathcal{S}_2^2 - c_1 \mathcal{S}_6 + \mathcal{S}_2 \mathcal{S}_6 \\  - \mathcal{S}_6^2 - c_1 \mathcal{S}_7 + \mathcal{S}_2 \mathcal{S}_7 
	+ 2 \mathcal{S}_6 \mathcal{S}_7 - \mathcal{S}_7^2 	\\ + 7 c_1 \mathcal{S}_9 - 5 \mathcal{S}_2 \mathcal{S}_9 - \mathcal{S}_6 \mathcal{S}_9 - \mathcal{S}_7 \mathcal{S}_9 - 4 \mathcal{S}_9^2 \end{array}$ & \rule{0pt}{1.2cm}\parbox[c]{1.8cm}{ \includegraphics[scale=0.25]{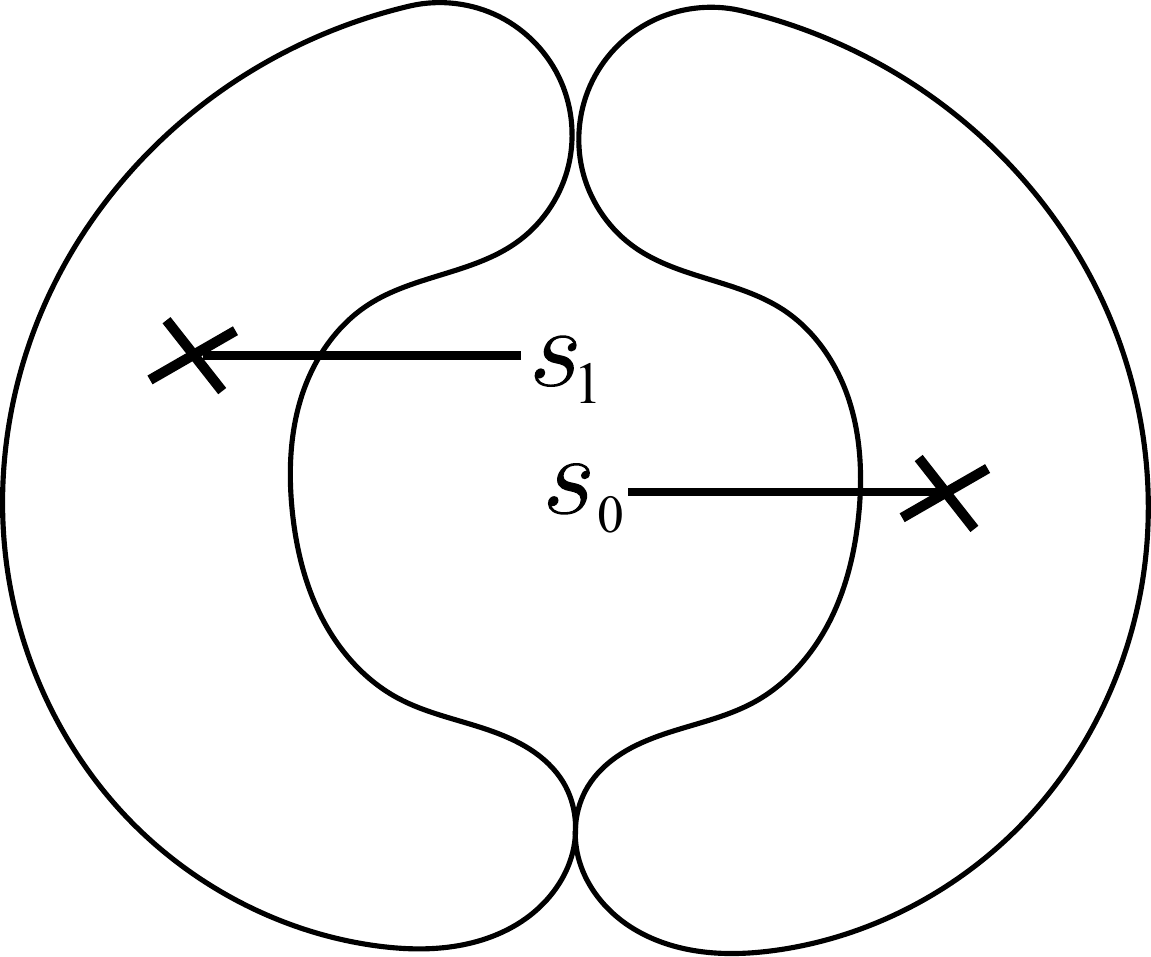}} \hspace{1cm} \\ \hline  
		
		$\mathbf{1}_0$ & $\begin{array}{l} 13 + 11 c_1^2 + 2 \mathcal{S}_2^2 + 2 \mathcal{S}_6^2 - \mathcal{S}_6 \mathcal{S}_7 + 2 \mathcal{S}_7^2   \\ + 2 (\mathcal{S}_6 + \mathcal{S}_7) \mathcal{S}_9   + \mathcal{S}_2 (\mathcal{S}_6 + \mathcal{S}_7 + 3 \mathcal{S}_9)\\ + 
 3 \mathcal{S}_9^2- c_1 (6 \mathcal{S}_2 + 4 (\mathcal{S}_6 + \mathcal{S}_7) + 7 \mathcal{S}_9) \end{array}$ &  \\ \hline \hline
Anomaly coefficient & \multicolumn{2}{|c|}{$ b=  2( 3 c_1 + 3  \mathcal{S}_2 + 2 \mathcal{S}_6 + 2 \mathcal{S}_7 + \mathcal{S}_9)$} \\ \hline  		
	\end{tabular} 
	\caption{ \label{tab:nef51}{Summary of the charged matter representations and multiplicities under $U(1) $ and codimension two fibers of nef $(5,1)$ and its anomaly coefficient.               }}
\end{table} 

\subsection{nef (21,0): $G=U(1)^2$}
\begin{figure}[t!] 
\begin{picture}(0,200)
\put(-10,0){
\includegraphics[scale=0.4]{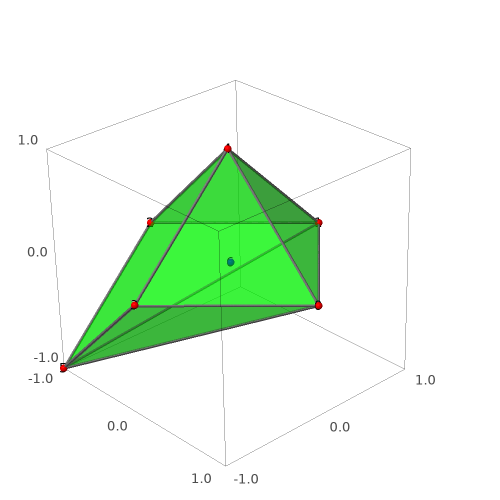}} 
\put(180,110){
 {\footnotesize
 \begin{tabular}{|c|c|c|}\hline
 vertex & coord. & divisor \\ \hline
 (1,0,0)& z& $ H-E_1+c_1 -(\mathcal{S}_7+\mathcal{S}_9) $  \\ \hline 
 (0,1,0)& x &$H-E_2+ 2 c_1-(\mathcal{S}_2+\mathcal{S}_6 + \mathcal{S}_7+\mathcal{S}_9) $\\ \hline 
 (-1,0,0)& e$_2$&$ E_2$\\ \hline 
 (0,-1,0) & e$_1$&$E_1 $\\ \hline 
 (0,0,1)& y& $H-E_1 -E_2+c_1-(\mathcal{S}_6+\mathcal{S}_9) $\\ \hline 
 (-1,-1,-1)&w & $H -E_1-E_2$  \\ \hline 
 \multicolumn{2}{|c|}{Intersections:} & $ \begin{array}{l}H^3=E_1^3=E_2^3=1\, , \\ E_1 E_2=H E_1 = H E_2 =0\end{array}$ \\ \hline 
 \end{tabular}  
 }
}
\end{picture}
\caption{\label{fig:nef21} {The polytope and data associated to the ambient space Bl$_2\mathbb{P}^3$ of nef (21,0). The latter generically leads to a rank two Mordell-Weil group.}  }
\end{figure}

\begin{table}[ht!]
\vspace{-1.5cm}
\begin{center}
\begin{tabular}{|c|l|c|} \hline
Representation & Multiplicity& Fiber  \\ \hline
    
$\mathbf{1}_{(-1,1)}$ &$ \mathcal{S}_2^2 -c_1 \mathcal{S}_2  + \mathcal{S}_2 \mathcal{S}_6 + \mathcal{S}_2 \mathcal{S}_7 + \mathcal{S}_6 \mathcal{S}_7 + \mathcal{S}_2 \mathcal{S}_9$&   \rule{0pt}{1.2cm}\parbox[c]{1.8cm}{  \includegraphics[scale=0.25]{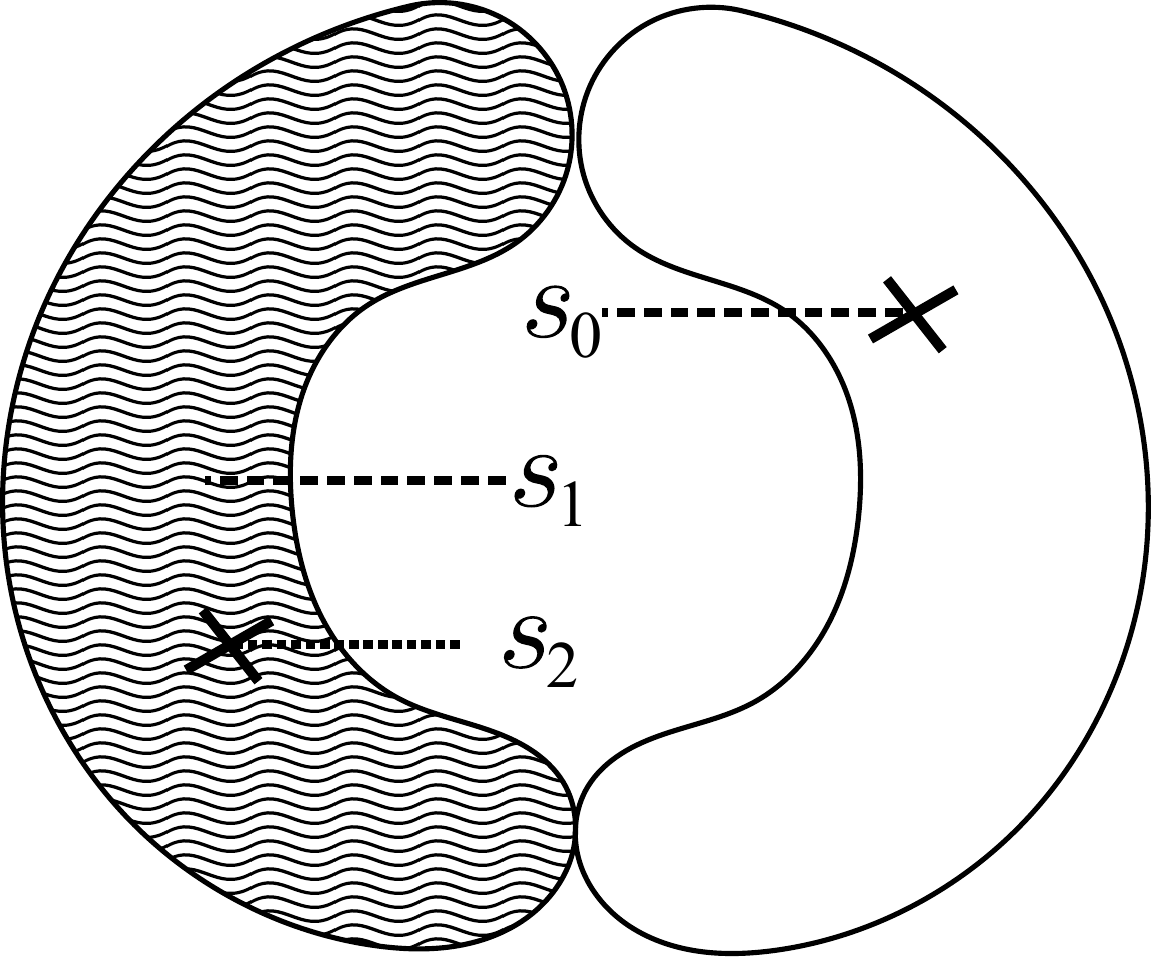}}  \hspace{1cm}      \\ \hline

$\mathbf{1}_{(1,3)} $&$ 
2 c_1 \mathcal{S}_7 - \mathcal{S}_2 \mathcal{S}_7 - \mathcal{S}_6 \mathcal{S}_7 + c_1 \mathcal{S}_9 - \mathcal{S}_6 \mathcal{S}_9
$&  \rule{0pt}{1.2cm}\parbox[c]{1.8cm}{ \includegraphics[scale=0.25]{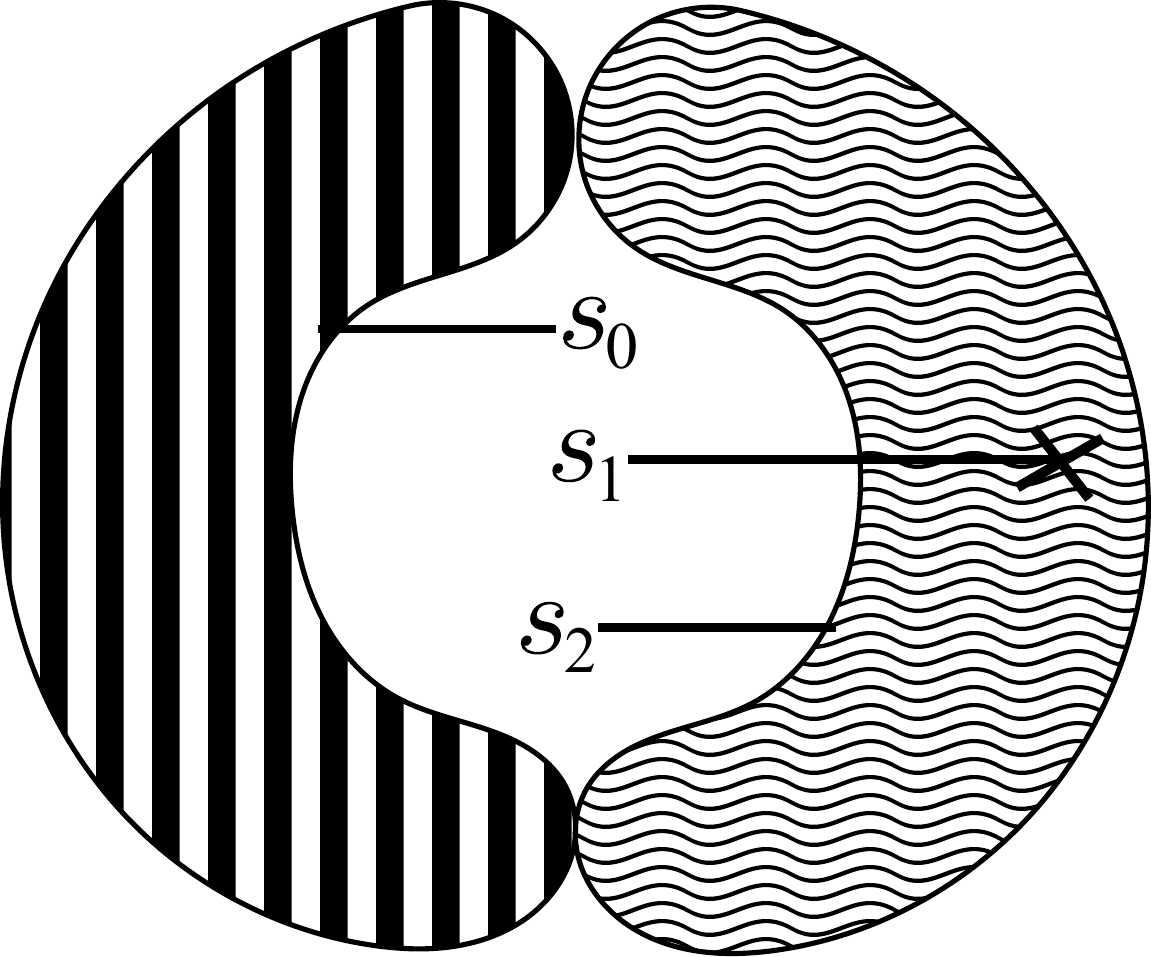}} \hspace{1cm} \\ \hline

$\mathbf{1}_{(0,3)} $ & $-\mathcal{S}_7 (c_1 - \mathcal{S}_2 - \mathcal{S}_7 - \mathcal{S}_9)$& \rule{0pt}{1.2cm}\parbox[c]{1.8cm}{ \includegraphics[scale=0.25]{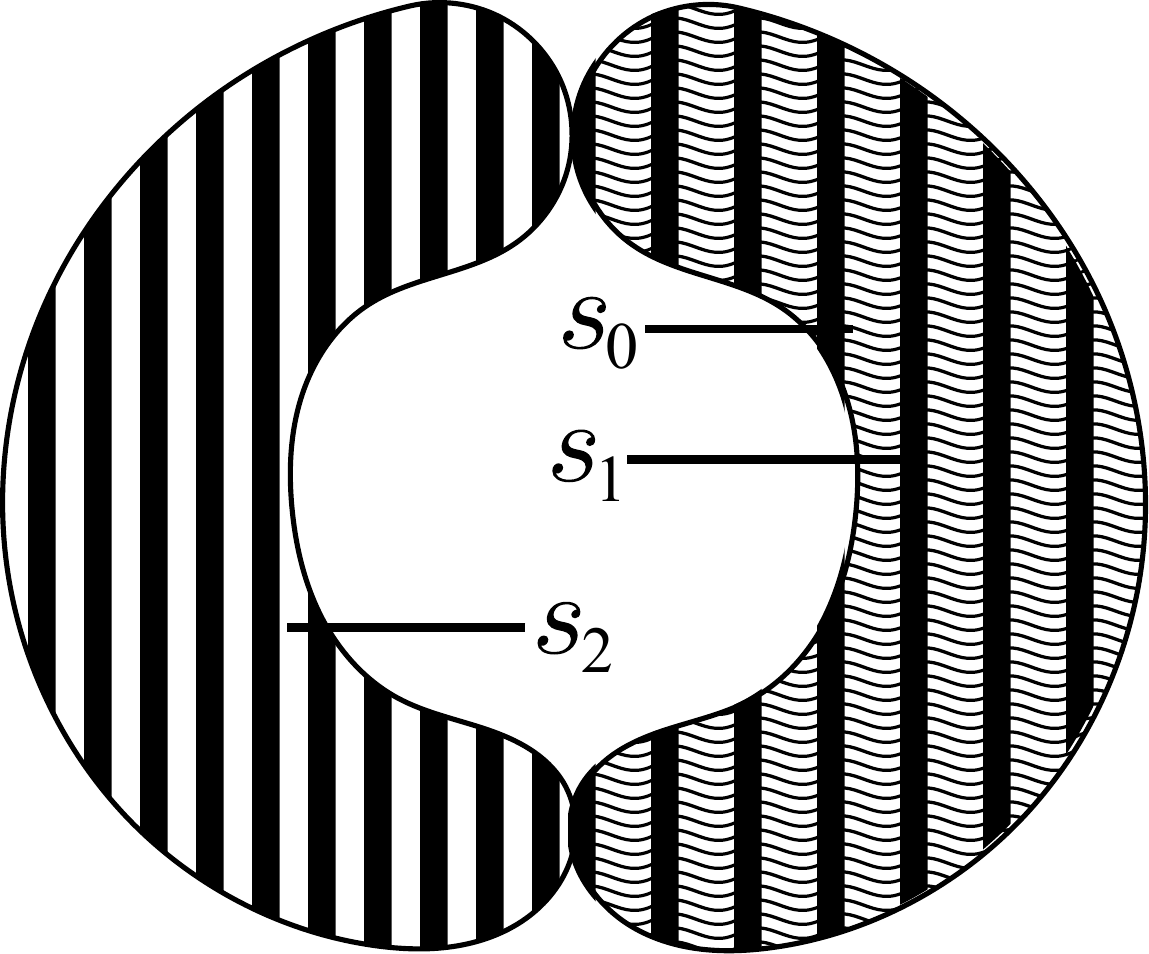}} \hspace{1cm} \\ \hline

$\mathbf{1}_{(1,2)}$&$\begin{array}{l}6 c_1^2 + (\mathcal{S}_2 + \mathcal{S}_6) (\mathcal{S}_6 + 2 \mathcal{S}_7) - \mathcal{S}_2  \mathcal{S}_9\\   
 +  3 \mathcal{S}_6\mathcal{S}_9+ c_1 (2 \mathcal{S}_2 + 5 \mathcal{S}_6 + 2 \mathcal{S}_7 + \mathcal{S}_9) \end{array}$ & \hspace{-1.3cm}  \rule{0pt}{1.2cm}\parbox[c]{1.8cm}{ \includegraphics[scale=0.25]{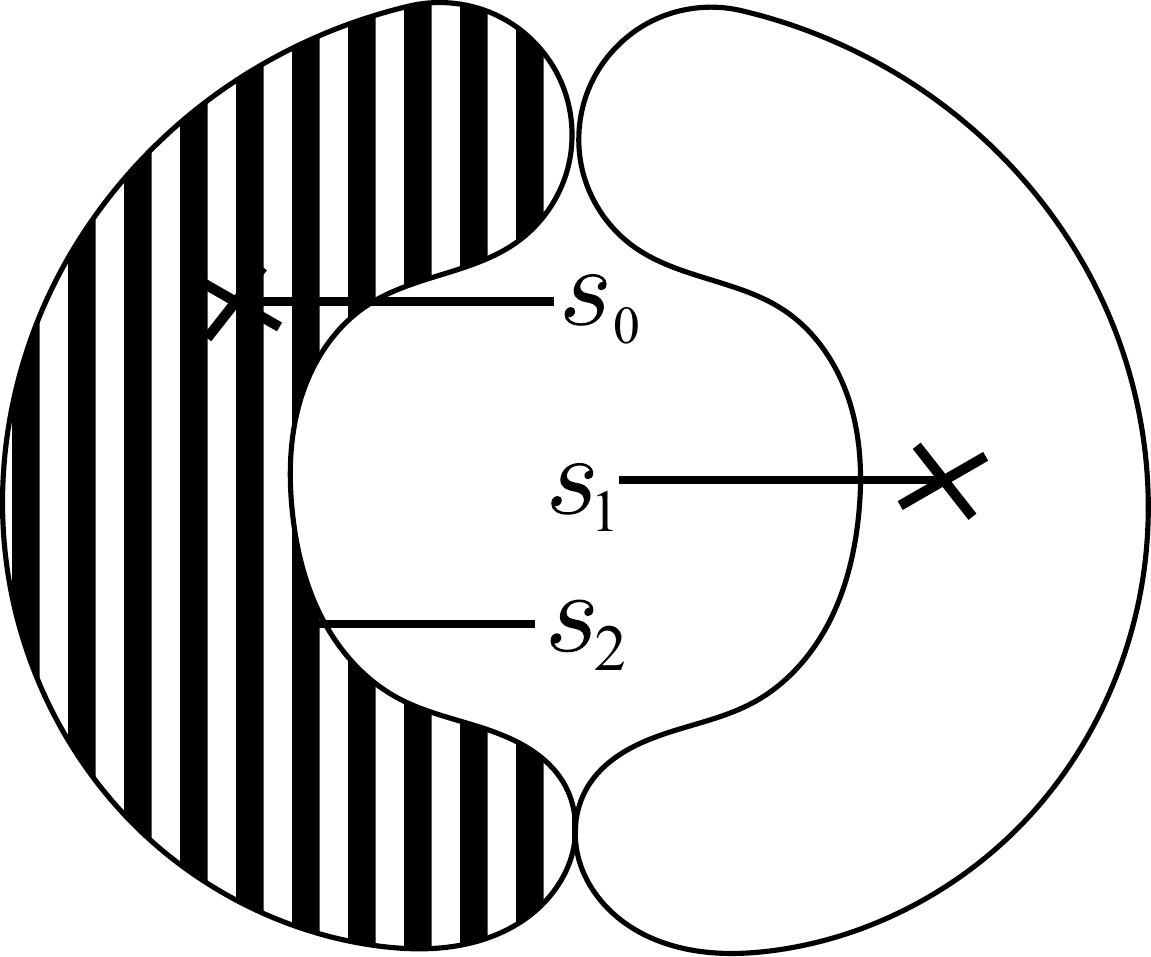}} \\ \hline

$\mathbf{1}_{(0,2)}$ & $\begin{array}{l}-\mathcal{S}_2 \mathcal{S}_6 - 2 \mathcal{S}_2 \mathcal{S}_7 - 2 \mathcal{S}_7^2 + 
 2 c_1 (\mathcal{S}_2 + 3 \mathcal{S}_7)\\ + (\mathcal{S}_2 + \mathcal{S}_6 - 2 \mathcal{S}_7) \mathcal{S}_9 + \mathcal{S}_9^2 \end{array}$ & \hspace{-1.3cm} \rule{0pt}{1.2cm}\parbox[c]{1.8cm}{ \includegraphics[scale=0.25]{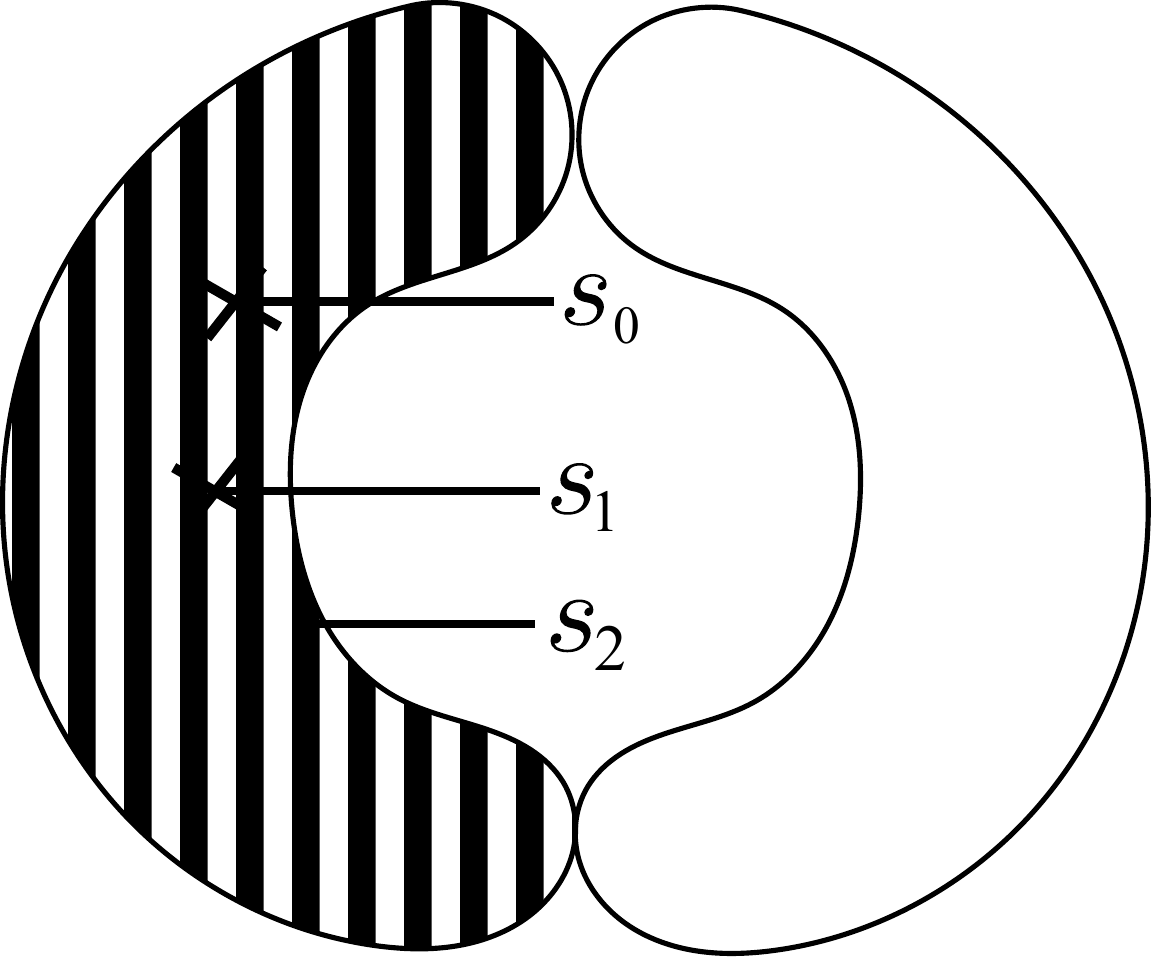}}\\ \hline

$\mathbf{1}_{(1,1)}$&  $\begin{array}{l}6 c_1^2 + \mathcal{S}_2^2 - \mathcal{S}_2 \mathcal{S}_6 - 2 \mathcal{S}_6^2 + 3 \mathcal{S}_2 \mathcal{S}_9\\ - 3 \mathcal{S}_6 \mathcal{S}_9 - 
 c_1 (3 \mathcal{S}_2 - 4 \mathcal{S}_6 + 2 \mathcal{S}_7 + \mathcal{S}_9) \end{array}$ &   \hspace{-1.3cm} \rule{0pt}{1.2cm}\parbox[c]{1.8cm}{ \includegraphics[scale=0.25]{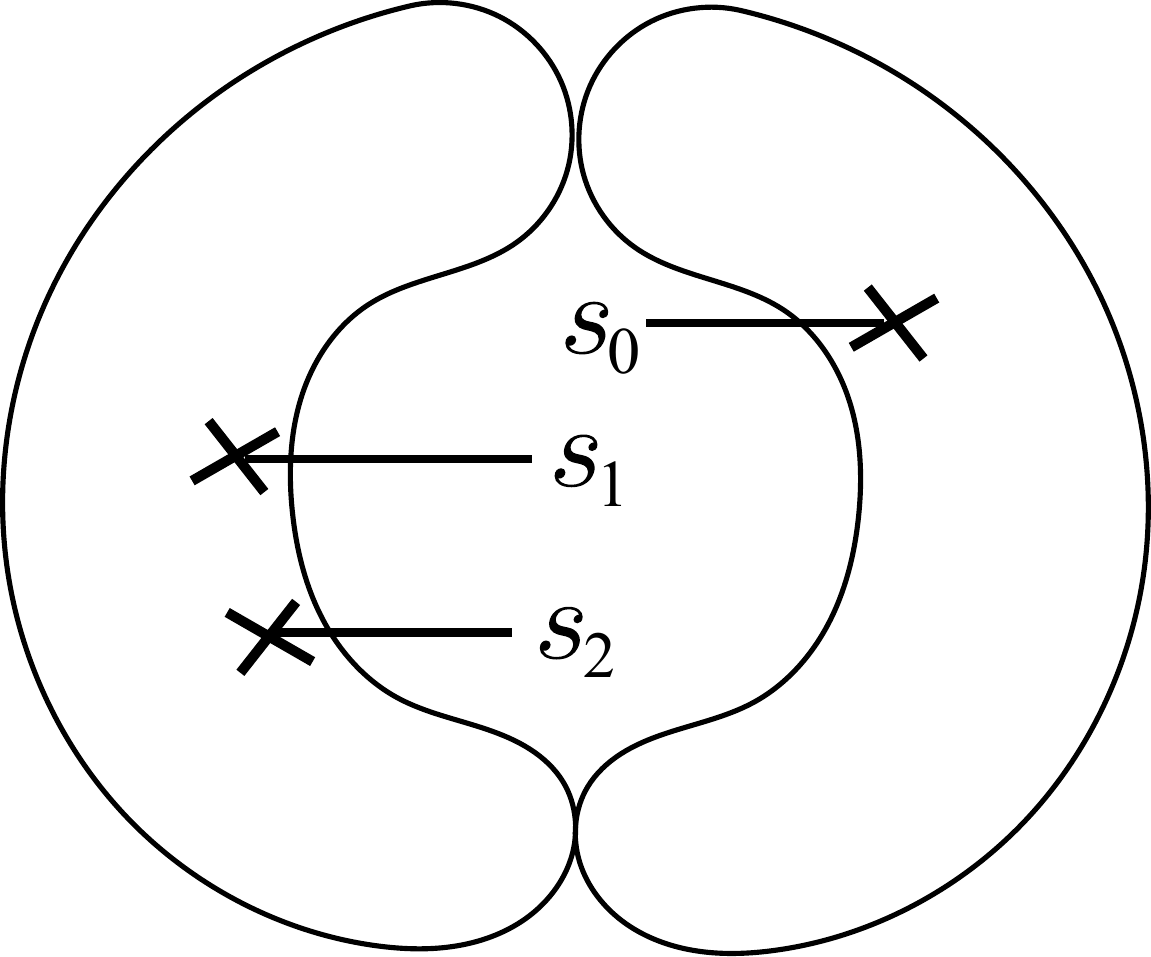}}\\ \hline

$\mathbf{1}_{(1,0)}$ &$\begin{array}{l}-(\mathcal{S}_2 + \mathcal{S}_6) (2 \mathcal{S}_2 - \mathcal{S}_6 + 2 \mathcal{S}_7)  -3 \mathcal{S}_2   \mathcal{S}_9\\  
 +  \mathcal{S}_6  \mathcal{S}_9c_1 (6 \mathcal{S}_2 + \mathcal{S}_6 + 2 \mathcal{S}_7 + \mathcal{S}_9)\end{array}$&\hspace{-1.3cm}  \rule{0pt}{1.2cm}\parbox[c]{1.8cm}{ \includegraphics[scale=0.25]{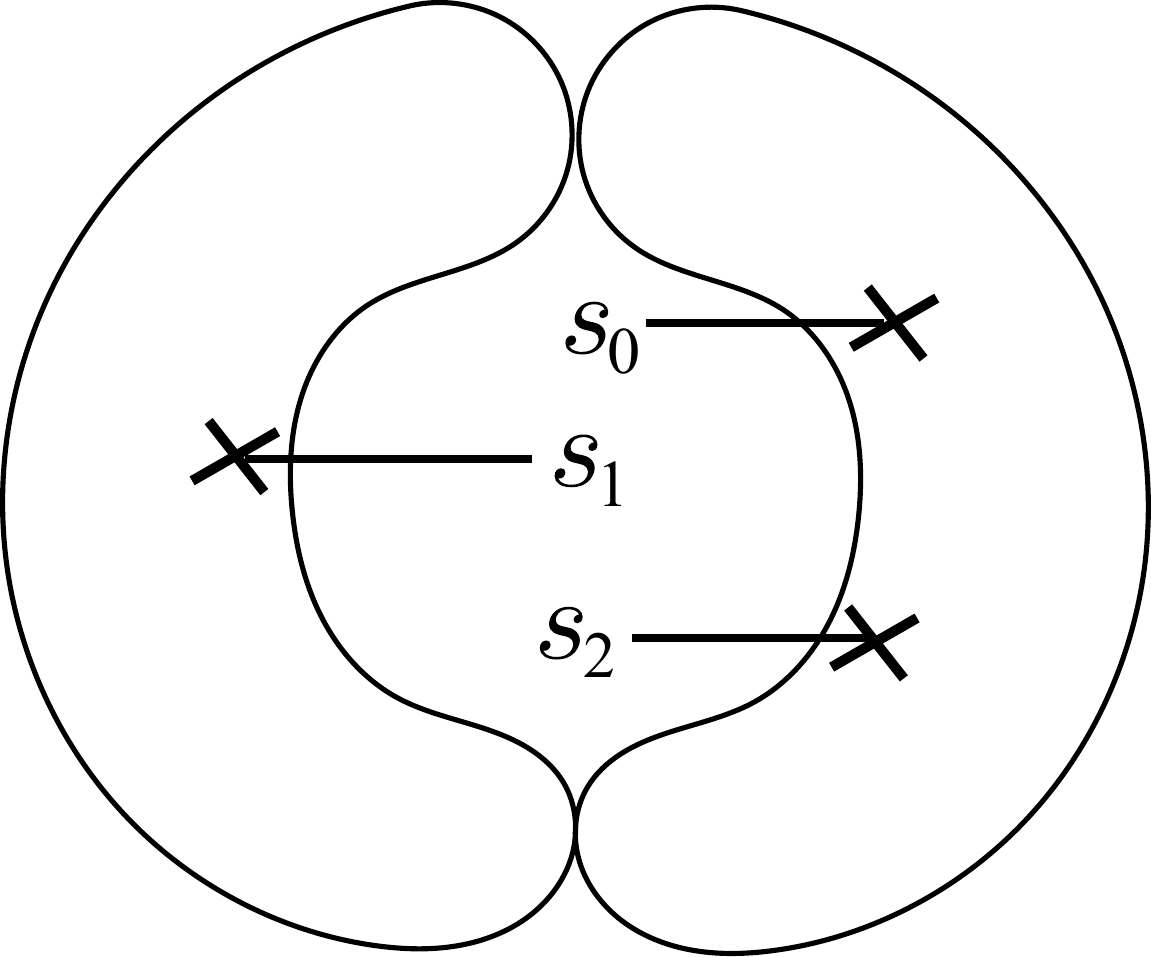}}  \\ \hline
 
$\mathbf{1}_{(0,1)}$&   $\begin{array}{l} 6 c_1^2 - 2 \mathcal{S}_2^2 - 2 \mathcal{S}_6^2 - \mathcal{S}_2 \mathcal{S}_7 - \mathcal{S}_7^2  \\ - (4 (\mathcal{S}_2 + \mathcal{S}_6)  +\mathcal{S}_7 ) \mathcal{S}_9  - 
 4 \mathcal{S}_9^2 \\ + c_1 (4 \mathcal{S}_2 + 4 \mathcal{S}_6 + \mathcal{S}_7 + 8 \mathcal{S}_9)\end{array} $ & \hspace{-1.3cm}  \rule{0pt}{1.2cm}\parbox[c]{1.8cm}{ \includegraphics[scale=0.25]{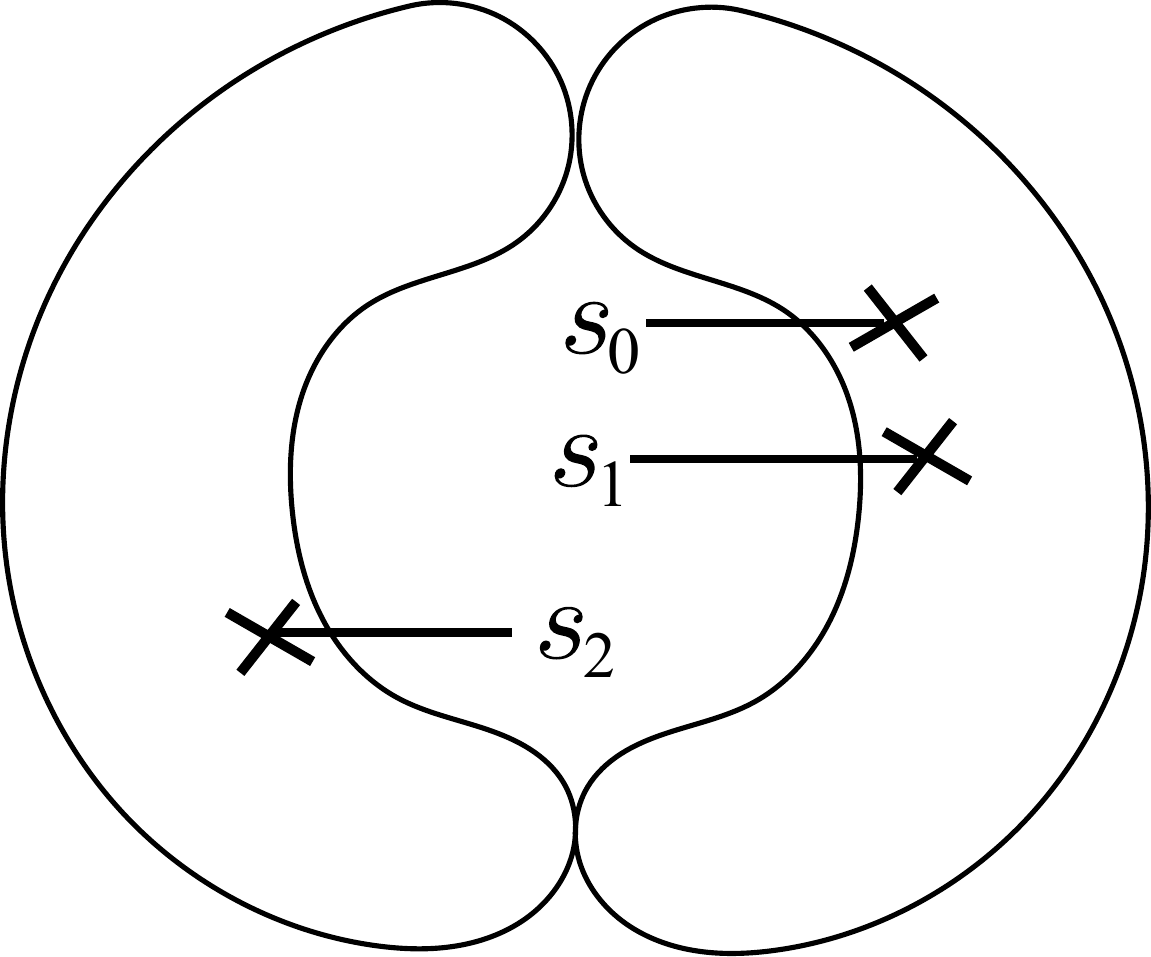}}  \\ \hline 
 
$\mathbf{1}_{(0, 0 )}$ &  $\begin{array}{l} 11 c_1^2 + 2 \mathcal{S}_2^2 + 2 (7 + \mathcal{S}_6^2 + \mathcal{S}_7^2)  + 3 \mathcal{S}_9^2 \\ + (3 \mathcal{S}_6 + 2 \mathcal{S}_7) \mathcal{S}_9+ 
 \mathcal{S}_2 (\mathcal{S}_6 + 2 \mathcal{S}_7 + 3 \mathcal{S}_9) \\  - 2 c_1 (3 \mathcal{S}_2 + 2 \mathcal{S}_6 + 3 \mathcal{S}_7 + 4 \mathcal{S}_9)  \end{array} $ & \\ \hline \hline
 Anomaly Coefficient: & \multicolumn{2}{|c|}{ $\begin{array}{ll} b_{1,1}= 2 c_1 \quad &b_{1,2} =  3 c_1 - \mathcal{S}_2 - \mathcal{S}_6  \,    \\ \multicolumn{2}{l}{b_{2,2}=  2 (3 c_1 - \mathcal{S}_6 + 2 \mathcal{S}_7 + \mathcal{S}_9)} \end{array} $} \\ \hline  
\end{tabular}\vspace{-0.6cm}
\caption{\label{tab:U12Model}{Summary of the charged matter representations and multiplicities under $U(1)^2 $ and codimension two fibers of nef $(21,0)$ and its anomaly coefficients.  }}
\end{center}
\end{table}

The first codimension two nef partition of the three dimensional reflexive polytope with PALP index 21 has been shown in~\cite{Braun:2014qka} to generically lead to an elliptic fibration
with Mordell-Weil trank two and no fibral divisors~\footnote{\url{http://wwwth.mpp.mpg.de/members/jkeitel/weierstrass/data/21_0.txt}}.
The defining equations are given by
\begin{align}
\begin{split}
\label{eq:nef21}
p_1=&e_1 e_2 (s_{1,1} w^2 + s_{1,3} w y + s_{1,8} y^2)+ e_1 z (s_{1,4} w + s_{1,9} y)\\
	&+ e_2 x (s_{1,2} w + s_{1,6} y)+ s_{1,7} x z	\, ,\\
 p_2=&  e_1 e_2 (s_{2,1} w^2 + s_{2,3} w y + s_{2,8} y^2)+ e_1 z (s_{2,4} w + s_{2,9} y)\\
	&+ e_2 x (s_{2,2} w + s_{2,6} y)+ s_{2,7} x z \, .
\end{split}
\end{align}
and they can also be viewed as a specialization of the bi-quadric by setting the coefficients $s_{i,5}, s_{i,10},\,i\in\{1,2\}$ in Equation~\ref{eq:bibiquadric} to zero.
A picture of the polytope, the homogeneous coordinates $w,x,y,z,e_1,e_2$ that correspond to the vertices and the corresponding divisor classes can be found in Figure~\ref{fig:nef21}.
The tetrahedron of $\mathbb{P}^3$ can be obtained by dropping the points that correspond to $e_1,e_2$ and applying the
lattice automorphism
\begin{align}
	M_{21}\rightarrow M_{5}:\quad\vec{p}\mapsto\left(\begin{array}{ccc}
		0&-1&0\\
		1&0&0\\
		0&0&1
	\end{array}\right)\vec{p}\,,
	\label{eqn:auto21to5}
\end{align}
followed by~\eqref{eqn:auto5to0} to the remaining rays.
Using a triangulation of the ambient space we obtain the Stanley-Reisner ideal
\begin{align}
\mathcal{SRI}=\langle  e_2 z, e_1 e_2, e_1 x, w y z, w x yz\rangle \, .
\end{align}
The divisors $[e_i],\,i\in\{1,2\}$ restrict to sections and intersect the generic fiber once.
On the other hand, $[w],[y]$ intersect the generic fiber twice and $[x],[z]$ intersect it three times.
The homogeneous coordinates $[x:y:z:w:e_1:e_2]$ of the sections are
\begin{align}
	\begin{split}
s_0:\,& [s_{1,4} s_{2,9}-s_{1,9} s_{2,4} \,:\, s_{1,7} s_{2,4} - s_{1,4}s_{2,7}\,:\, 1\,:\,    s_{1,9} s_{2,7} - s_{1,7} s_{2,9}\,:\, 1\,:\,0 ]\,, \\
s_1:\,& [ 1 \,:\,  s_{1,7} s_{2,2} -  s_{1,2} s_{2,7} \,:\,    s_{1,2} s_{2,6}-s_{1,6} s_{2,2}  \,:\, s_{1,6} s_{2,7}-s_{1,7} s_{2,6}  \,:\,0\,:\,1]\,.
	\end{split}
\end{align}
The class of a third section is given by
\begin{align}
	[s_2] \sim [x] -  [e_1]-[e_2] \, .
\end{align}
Later we will again consider a map into a restricted cubic, that allows to obtain the coordinates of that section.
The images under the Shioda map are
\begin{align}
\sigma[s_1]= [s_1]-[s_0] \, ,\quad \sigma[s_2]=[s_2]-[s_0]  \, , 
\end{align}
where we have again dropped vertical divisors.
The matter loci can be obtained with GV-spectroscopy and are listed in Table~\ref{tab:U12Model}.

\subsubsection*{Map into a restricted cubic}
Again we can find a map into a restricted family of cubics.
In fact, we can choose to consider either a patch with $e_1=1$ or $e_2=1$ and this leads to two maps
\begin{align}
\label{eq:cubicmap}
	p_{\text{cubic},i}=A_{i,2} D_{i,1}-B_{i,1} C_{i,2} \, .
 \end{align}
Note that in both cases we map one of the sections into a divisor that blows-up the cubic.
This can also be realized torically by considering hypersurfaces associated to the polytope $F_3$.
Using the conventions from~\cite{Klevers:2014bqa} this takes the form
\begin{align}
\label{eq:pf3}
	\begin{split}
	p=&s_1 e_1^2 u^3 + s_2 u^2 v e_1^2 + s_3 u v^2 e_1^2 + s_4 v^3 e_1^2 + s_5 u^2 w e_1\\
	&+ s_6 u v w e_1 + s_7 v^2 w e_1 + s_8 u w^2 + s_9 v w^2 \, ,
	\end{split}
\end{align} 
To obtain the first map we can set $e_1=1$ and solve for $x$ to find
\begin{align}
\begin{split}
A_{1,1}&= (e_2 s_{1,1} w^2 + e_2 s_{1,3} w y + e_2 s_{1,8} y^2 + s_{1,4} w z + s_{1,9} y z)  \, , \\ 
C_{2,1}&= (e_2 s_{2,1} w^2 + e_2 s_{2,3} w y + e_2 s_{2,8} y^2 + s_{2,4} w z + s_{2,9} y z)\,,\\
B_{1,1}&=  (e_2 s_{1,2} w + e_2 s_{1,6} y + s_{1,7} z) \, , \quad D_{2,1}= (e_2 s_{2,2} w + e_2 s_{2,6} y + s_{2,7} z)\, .
\end{split}
\end{align}
We then simply have to relabel the coordinates as $ \{ e_2 \rightarrow e_1, w \rightarrow u, y\rightarrow v, z \rightarrow w \}$  to match the form of Equation~\eqref{eq:pf3} with 
 \begin{align} 
 \label{eq:replacen21}
 \begin{split}
 s_{1}=&   s_{1,1} s_{2,2} -s_{1,2} s_{2,1} \, , \quad s_{2}= s_{1,3} s_{2,2} - s_{1,2} s_{2,3}  -s_{1,6} s_{2,1} + s_{1,1} s_{2,6}\, , \\
 s_{3}=& s_{1,8} s_{2,2} - s_{1,6} s_{2,3} + s_{1,3} s_{2,6} - s_{1,2}s_{2,8}\, , \quad s_{4}= s_{1,8} s_{2,6} - s_{1,6} s_{2,8}\, , \\
 s_{5}=&  s_{1,4}s_{2,2} - s_{1,2} s_{2,4} -s_{1,7} s_{2,1}+ s_{1,1} s_{2,7}\, , \\
 s_{6}=& s_{1,9}s_{2,2} - s_{1,7} s_{2,3} - s_{1,6} s_{2,4 }+ s_{1,4} s_{2,6 }+ s_{1,3} s_{2,7 }- s_{1,2} s_{2,9}\, , \\ 
	 s_{7}=& s_{1,9} s_{2,6}+ s_{1,8} s_{2,7} - s_{1,7} s_{2,8} - s_{1,6} s_{2,9}\, , \\  s_{8}=&  s_{1,4} s_{2,7}-s_{1,7} s_{2,4}\, , \quad  s_{9}= s_{1,9} s_{2,7} - s_{1,7} s_{2,9} \, .
 \end{split}
\end{align}
Similar expressions can be found by setting $e_2=1$.
This map is useful e.g. for finding the Weierstrass description and the coordinates of the non-toric section~\footnote{The $F_3$ coordinate of the non-toric section is given in Section~3.3.1 and the coordinates in Weierstrass form are given in Appendix~B  of \cite{Klevers:2014bqa}.}.
It can also be used to find some of the other singlet loci in a classical analysis.
\subsubsection*{Analysis of Higgs loci}
In the following we will explicitly determine the matter loci that are responsible for the Higgs transition towards the theories associated to nef (5,1).
They can be found by first considering the loci where either $s_0$ or $s_1$ degenerates.
This happens when the linear factors $B_{i,1},D_{i,1}$ become collinear, i.e. $B_{i,1} = \lambda D_{i,1}$ for some $\lambda \neq 0$.
Note that this corresponds to $s_8=s_9=0$ after mapping into the $F_3$ hypersurface.

For $s_1$ this leads to the ideal
\begin{align}
 I_1  = \langle s_{1,6} s_{2,2} -  s_{1,2} s_{2,6}  \, , \,  s_{1,2} s_{2,7} -s_{1,7} s_{2,2} \rangle   \, , 
 \end{align} 
which is contained in $\langle s_{2,2}, s_{1,2}\rangle$.
Computing the intersections with the reducible curve $\mathcal{C}_{1,1}$ that is wrapped by $s_1$ yields $ \mathcal{C}_{1,1} \cdot[s_i]= (0,-1,1)$. Hence this locus leads to singlets of charge $q=(-1,1)$.
The multiplicity of the corresponding hypermultiplets can be obtained by taking the class of $V(I_1)$ and substracting the locus $s_{2,2}=s_{1,2}=0$ once, i.e.
 \begin{align}
	 \begin{split}
	 n_{(1,-1)}=&([s_{1,6}]+[s_{2,2}])([s_{1,2}]+[s_{2,7}])-[s_{2,2}][s_{1,2}]\\
		 =& (\mathcal{S}_2 + \mathcal{S}_6) (\mathcal{S}_2 + \mathcal{S}_7) + \mathcal{S}_2 \mathcal{S}_9 -c_1 \mathcal{S}_2 \, .
	 \end{split}
 \end{align} 

A second locus can be obtained by considering degenerations of $s_0$.
This leads to the ideal
\begin{align}
  I_2 = \langle s_{1,9} s_{2,4}- s_{1,4} s_{2,9}\, ,\, s_{2,7} s_{1,4} - s_{1,7} s_{2,4} \rangle  \, ,
\end{align}
which is contained in $\langle s_{1,4}, s_{2,4}\rangle$.
Taking $\mathcal{C}_{1,2}$ to be the fibral curve wrapped by $s_0$ we calculate the intersections $\mathcal{C}_{1,2}\cdot[s_i] = (-1,0,2)$.
In particular, this locus yields singlets of charge $q=(1,3)$.
The multiplicity of the singlets is given by
\begin{align}
	\begin{split}
		n_{(1,3)}=&([s_{1,9}]+[s_{2,4}])([s_{2,7}]+[s_{1,4}])-[s_{2,4}][s_{1,4}]\\
		=&c_1 (2 \mathcal{S}_7 + \mathcal{S}_9)-\mathcal{S}_2 \mathcal{S}_7 - \mathcal{S}_6 (\mathcal{S}_7 +\mathcal{S}_9) \, .
	\end{split}
\end{align} 
The multiplicities of both loci agree with the results from base independent calculation of fiber GV invariants~\ref{tab:U12Model}.
The singlet sector has been computed using the Euler number and which is again given in Appendix~\ref{app:eulers}.
\subsection{nef (67,0): $G=U(1)^3$}
\begin{figure}[t!]
\begin{center}
\begin{picture}(0,180)
\put(-220,10){
\includegraphics[scale=0.4]{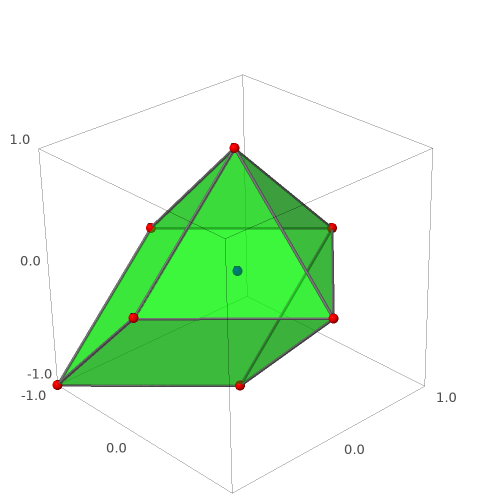}}
\put(- 10,100){
{\footnotesize
\begin{tabular}{|c|c|c|}\hline
vertex & coor. & divisor  \\ \hline
 (1,0,0)& x &  $\begin{array}{c}H-E_1-E_3+ 2 c_1 \\ -(\mathcal{S}_2+\mathcal{S}_6 + \mathcal{S}_7+\mathcal{S}_9) \end{array}$ \\ \hline
 (0,1,0) & y& $H-E_1-E_2+c_1-(\mathcal{S}_6+\mathcal{S}_9)$ \\ \hline
 (0,0,1) &z& $H-E_1-E_3+c_1 -(\mathcal{S}_7+\mathcal{S}_9) $ \\ \hline
 (-1,-1,-1)&w& $H-E_1-E_2-E_3$\\ \hline
 (-1,0,0)&$e_1$& $E_1$ \\ \hline
 (0,0,-1)&$e_2$& $E_2$ \\ \hline 
 (0,-1,0)&$e_3$& $E_3$ \\ \hline \hline
 \multicolumn{2}{|c|}{Intersections:} & 
 
 $\begin{array}{l}
 H^3 = H^2 E_i= H E_i^2=-2 \\
 E_i^3= H E_i E_j = E_i^2 E_j = -1 \\
 E_1 E_2 E_3 = 0 
     \end{array}$ \\ \hline

\end{tabular} 
}
}
\end{picture}
\caption{\label{fig:nef67}{The polytope and data associated to the ambient space Bl$_3\mathbb{P}^3$ of nef (67,0). The latter generically leads to a rank three Mordell-Weil group.}}

\end{center}
\end{figure}
The first codimension two nef partition of the three dimensional reflexive polytope with PALP index 67 has been shown in~\cite{Braun:2014qka} to generically lead to an elliptic fibration
with Mordell-Weil trank three and no fibral divisors~\footnote{\url{http://wwwth.mpp.mpg.de/members/jkeitel/weierstrass/data/67_0.txt}}.
The defining equations are given by
\begin{align}
\begin{split}
	p_1=&s_{1,1}e_1 e_2 e_3 w^2 + s_{1,2}e_2 e_3 w x + s_{1,3}e_1 e_2 w y + s_{1,6}e_2 x y \\
	&+  s_{1,4}e_1 e_3 w z + s_{1,7}e_3 x z + s_{1,9}e_1 y z  \, , \\
	p_2=&s_{2,1}e_1 e_2 e_3 w^2 + s_{2,2}e_2 e_3 w x + s_{2,3}e_1 e_2 w y + s_{2,6}e_2 x y \\
	&+  s_{2,4}e_1 e_3 w z + s_{2,7}e_3 x z + s_{2,9}e_1 y z \, ,
 \end{split}
\end{align} 
and they can also be viewed as a specialization of the bi-quadric by setting the coefficients $s_{i,5}, s_{i,10}, s_{i,8},\,i\in\{1,2\}$ in Equation~\ref{eq:bibiquadric} to zero.
A picture of the polytope, the homogeneous coordinates $w,x,y,z,e_1,e_2,e_3$ that correspond to the vertices and the corresponding divisor classes can be found in Figure~\ref{fig:nef67}.
The tetrahedron of $\mathbb{P}^3$ can be obtained by dropping the points that corresponds to $e_1,e_2,e_3$.
	Using a triangulation of the ambient space we obtain the Stanley-Reisner ideal
\begin{align}
\mathcal{SRI}=\langle e_2 z, w z, e_3 y, e_1 x, w y, w x, e_1 e_2 e_3 \rangle \, .
\end{align}
The divisors $[e_i],\,i\in\{1,2,3\}$ and $[w]$ intersect the generic fiber once and restrict to sections.
In fact, this model has been analyzed in~\cite{Cvetic:2013qsa} and parametrizes the most general fibration with a Mordell-Weil group of rank three.

A convenient choice of generators for the sections and their coordinates in $[x:y:z:w:e_1:e_2:e_3]$ are given by
\begin{align}
 \label{eq:sectionsnef67}
	\begin{split}
s_0:\,& [ s_{1,4} s_{2,9}-s_{1,9} s_{2,4} \,:\,   s_{1,7} s_{2,4}- s_{1,4} s_{2,7}  \,:\, 1\,:\, s_{1,9} s_{2,7} - s_{1,7} s_{2,9}\,:\, 1 \,:\, 0 \,:\, 1]\, , \\
s_1:\,& [1\,:\,  s_{1,2} s_{2,7}-s_{1,7} s_{2,2}  \,:\, s_{1,6} s_{2,2} - s_{1,2} s_{2,6}\,:\, s_{1,7} s_{2,6} - s_{1,6} s_{2,7}\,:\, 0 \,:\, 1 \,:\, 1]\, , \\
s_2:\,& [s_{1,3} s_{2,9}-s_{1,9} s_{2,3} \,:\, 1 \,:\,   s_{1,6} s_{2,3} - s_{1,3}s_{2,6}\,:\, s_{1,9} s_{2,6} - s_{1,6} s_{2,9}\,:\, 1 \,:\, 1 \,:\, 0]\, , \\
s_3:\,& [ 1 \,:\, 1 \,:\, 1 \,:\, 0 \,:\,s_{1,7} s_{2,6} - s_{1,6} s_{2,7}\,:\,   s_{1,9} s_{2,7} - s_{1,7} s_{2,9} \,:\, s_{1,6} s_{2,9}-s_{1,9} s_{2,6}  ] \, .
	\end{split}
\end{align}
The corresponding images under the Shioda map are
\begin{align}
\sigma(s_1)= [e_1]-[e_2] \, ,\quad \sigma(s_2)= [e_3]-[e_2] \, , \quad \sigma(s_3)= [w]-[e_2] \, .
\end{align}

We can use any of the three sections $e_i$ to map nef (67,0) into a family of hypersurfaces inside $dP_2$, which is the most general form for fibrations with Mordell-Weil rank two~\cite{Braun:2013nqa,Borchmann:2013jwa,Borchmann:2013hta,Cvetic:2013jta,Cvetic:2013uta,Cvetic:2013nia, Braun:2013yti}.
This map has been used in~\cite{Cvetic:2013qsa} to find all singlet loci~\footnote{
The conventions in \cite{Cvetic:2013qsa} match ours via the following identifications:
\begin{align*}
\begin{array}{ccc}
\begin{array}{ c|c|c }\hline
\text{\cite{ Cvetic:2013qsa} conv.}&\text{ our conv.}&\text{ our classes }\\ \hline
u & \sim x & 2 c_1 - (\mathcal{S}_2 + \mathcal{S}_6 + \mathcal{S}_7 + \mathcal{S}_9)  \\
v& \sim y &c_1 - (\mathcal{S}_6 + \mathcal{S}_9) \\
w & \sim z & c_1 - (\mathcal{S}_7 + \mathcal{S}_9) \\
t & \sim w & 0 \\
p_{1} & \sim p_{1} & 2c_1-\mathcal{S}_6-\mathcal{S}_7-\mathcal{S}_9 \\ \hline
\end{array}\, .
& 
\begin{array}{ c|c }\hline
   \text{ \cite{Cvetic:2013qsa} classes}  &  \text{our classes} \\ \hline
\mathcal{S}_9 &   3 c_1 - \mathcal{S}_2 - \mathcal{S}_6 - \mathcal{S}_7 - \mathcal{S}_9 \, , \\
 \hat{\mathcal{S}}_7 &  2 c_1 - \mathcal{S}_2 - \mathcal{S}_6\, , \\
  \tilde{\mathcal{S}}_7&   2 c_1 - \mathcal{S}_2 - \mathcal{S}_7\, , \\
   p_2^b &   c_1 - \mathcal{S}_2 \, .\\  \hline
\end{array}
\end{array}
\end{align*} 
}.
However, the spectrum of nef (67,0) as given in~\cite{Cvetic:2013qsa} is neither complete nor do all anomalies cancel.
In Tables~\ref{tab:U13Modela} and~\ref{tab:U13Modelb} we give the complete and fully consistent spectrum obtained via GV-spectroscopy.
Before we conclude this section let us again consider the matter loci that lead to the Higgs fields.
\begin{table}[ht!]
\begin{center}
\begin{tabular}{|c|l|c|} \hline
Representation & Multiplicity & Fiber \\ \hline
$\mathbf{1}_{(1,1,-1)}$&$-\mathcal{S}_6 (c_1 - \mathcal{S}_2 - \mathcal{S}_6 - \mathcal{S}_9) $ & \rule{0pt}{1.2cm}\parbox[c]{1.8cm}{ \includegraphics[scale=0.25]{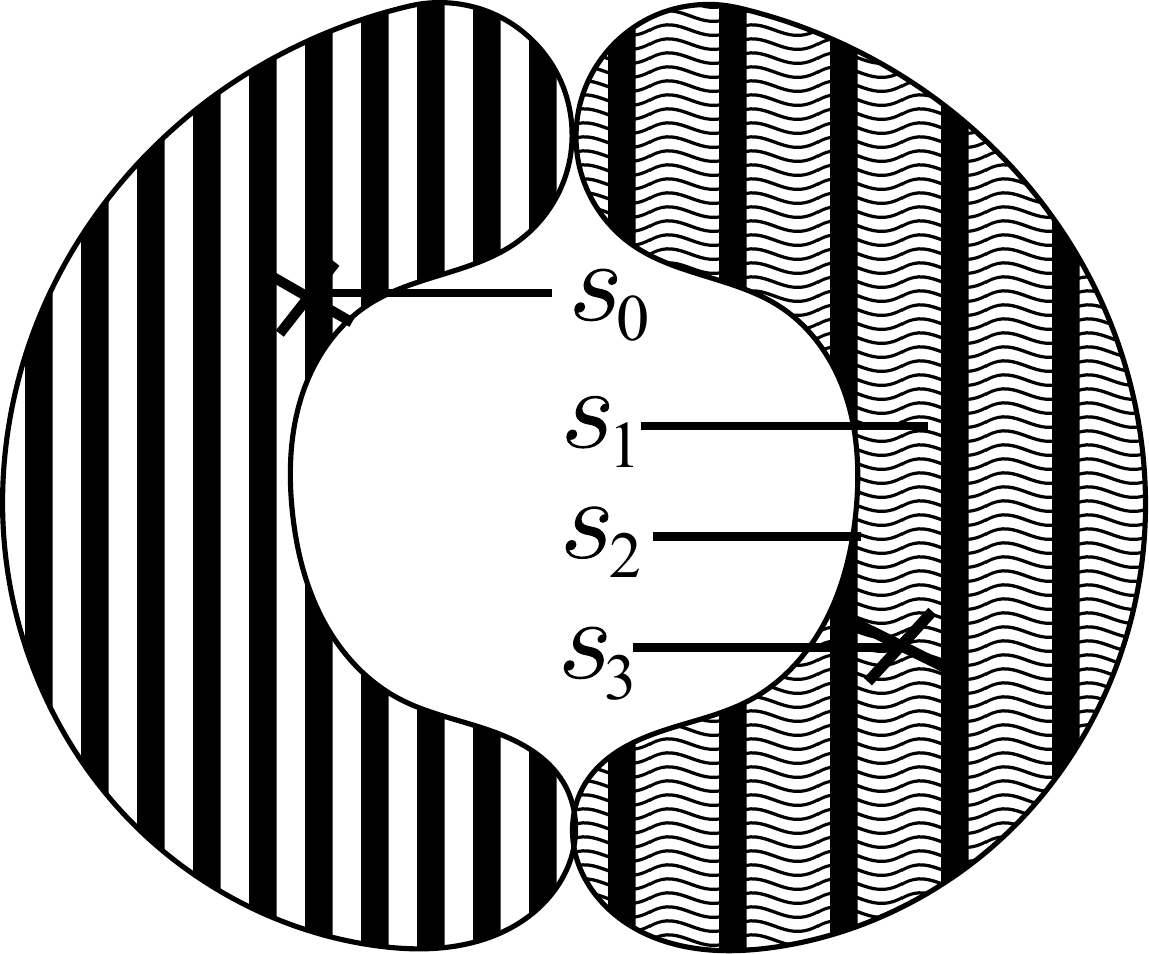}} \hspace{1cm} \\ \hline

$\mathbf{1}_{(0,1,2)}$&$-\mathcal{S}_7 (c_1 - \mathcal{S}_2 - \mathcal{S}_7 - \mathcal{S}_9)$&  \rule{0pt}{1.2cm}\parbox[c]{1.8cm}{ \includegraphics[scale=0.25]{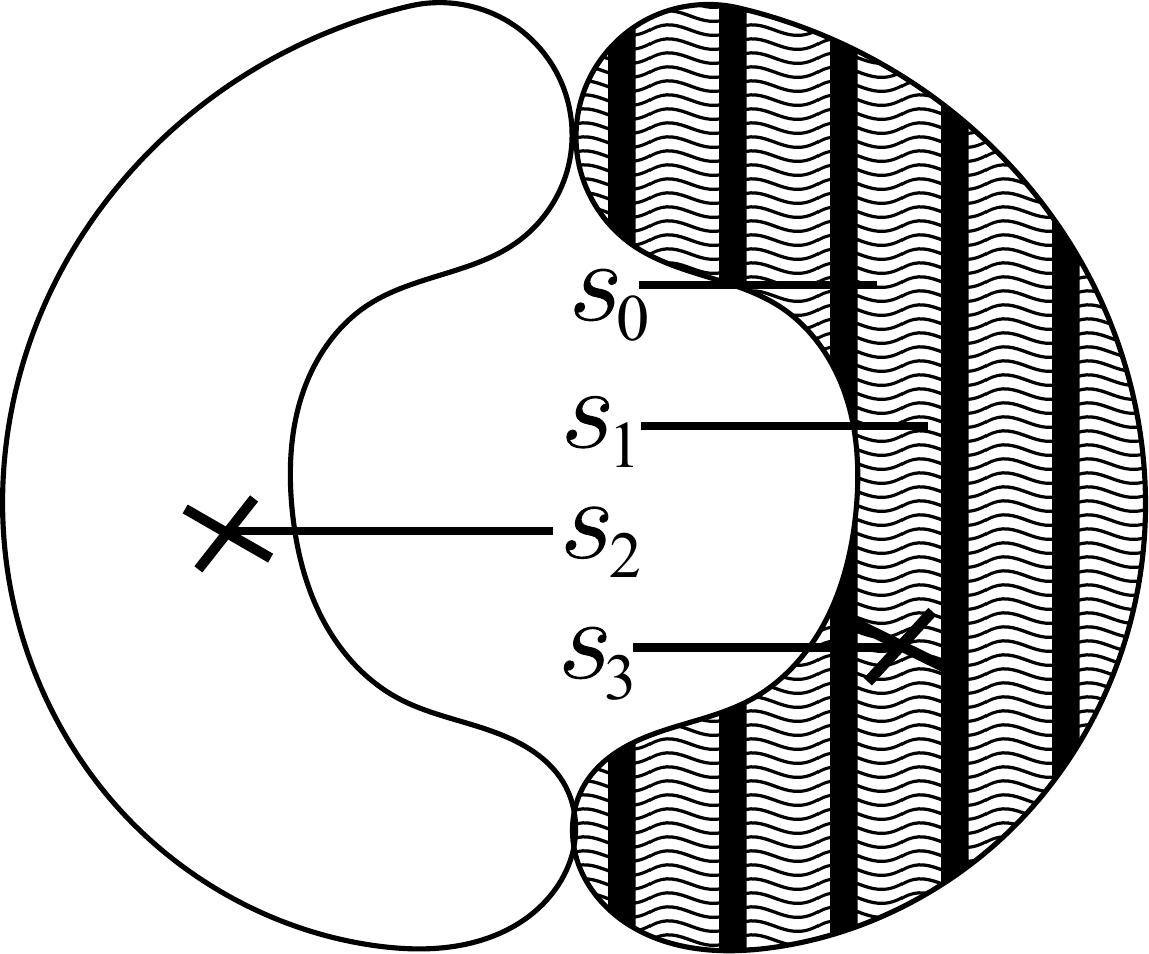}} \hspace{1cm}    \\ \hline

$\mathbf{1}_{(1,0,2)}$&$(c_1 - \mathcal{S}_2) \mathcal{S}_9$ &   \rule{0pt}{1.2cm}\parbox[c]{1.8cm}{ \includegraphics[scale=0.25]{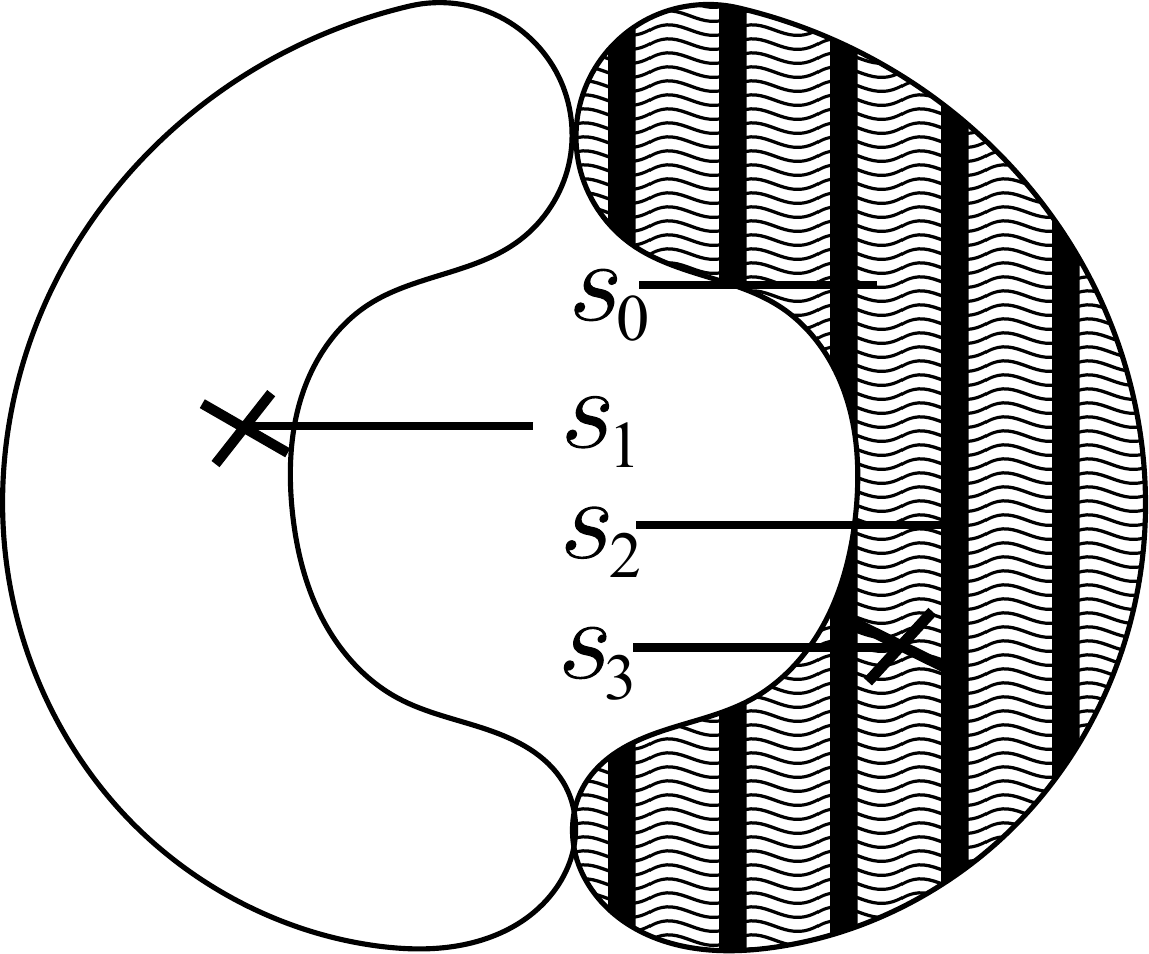}} \hspace{1cm} \\ \hline

$\mathbf{1}_{(-1,0,1)}$ &$-c_1 \mathcal{S}_2 + \mathcal{S}_2^2 + \mathcal{S}_2 \mathcal{S}_6 + \mathcal{S}_2 \mathcal{S}_7 + \mathcal{S}_6 \mathcal{S}_7 + \mathcal{S}_2 \mathcal{S}_9$ &  \rule{0pt}{1.2cm}\parbox[c]{1.8cm}{ \includegraphics[scale=0.25]{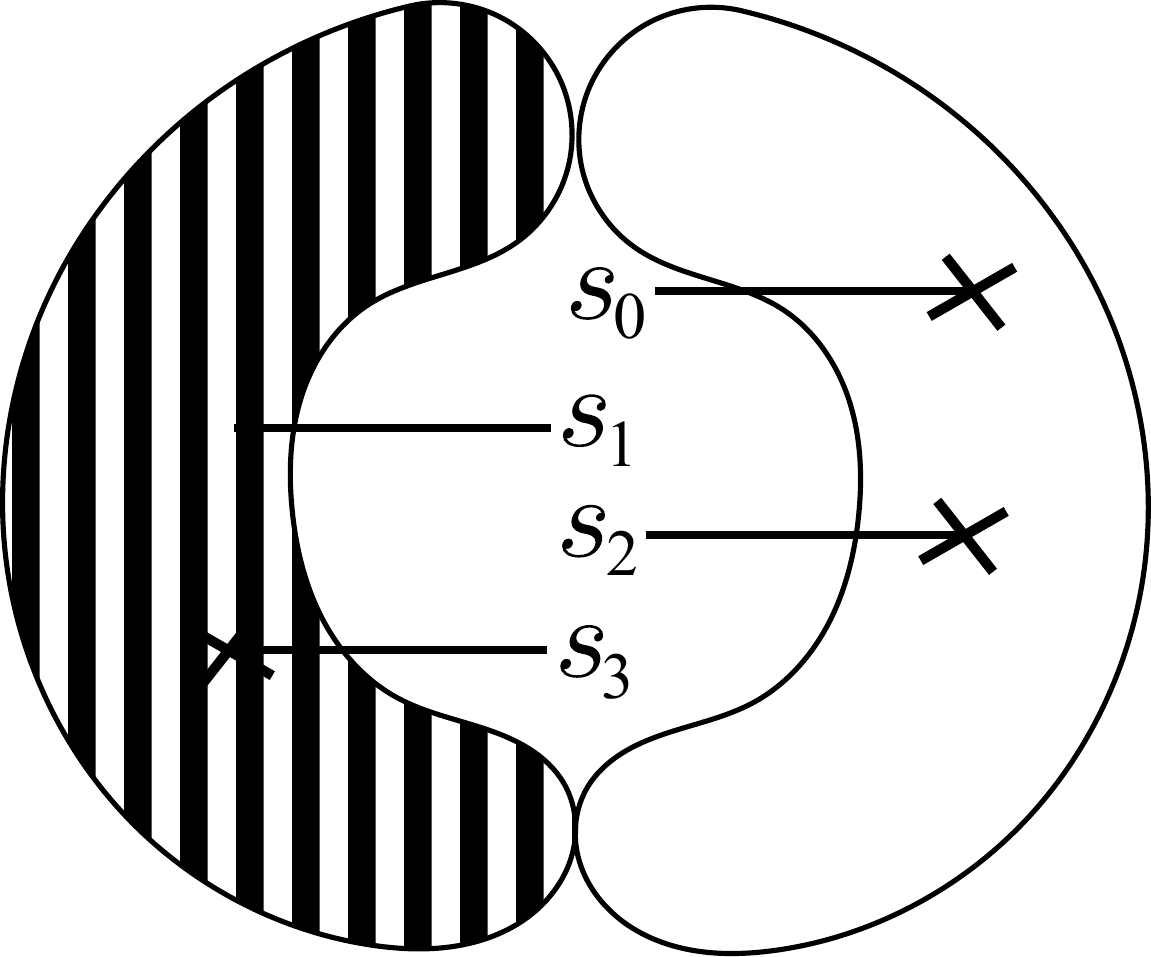}} \hspace{1cm} \\ \hline

$\mathbf{1}_{(0,-1,1)}$&$2 c_1 \mathcal{S}_6 - \mathcal{S}_2 \mathcal{S}_6 - \mathcal{S}_6 \mathcal{S}_7 + c_1 \mathcal{S}_9 - \mathcal{S}_7 \mathcal{S}_9$ &\rule{0pt}{1.2cm}\parbox[c]{1.8cm}{ \includegraphics[scale=0.25]{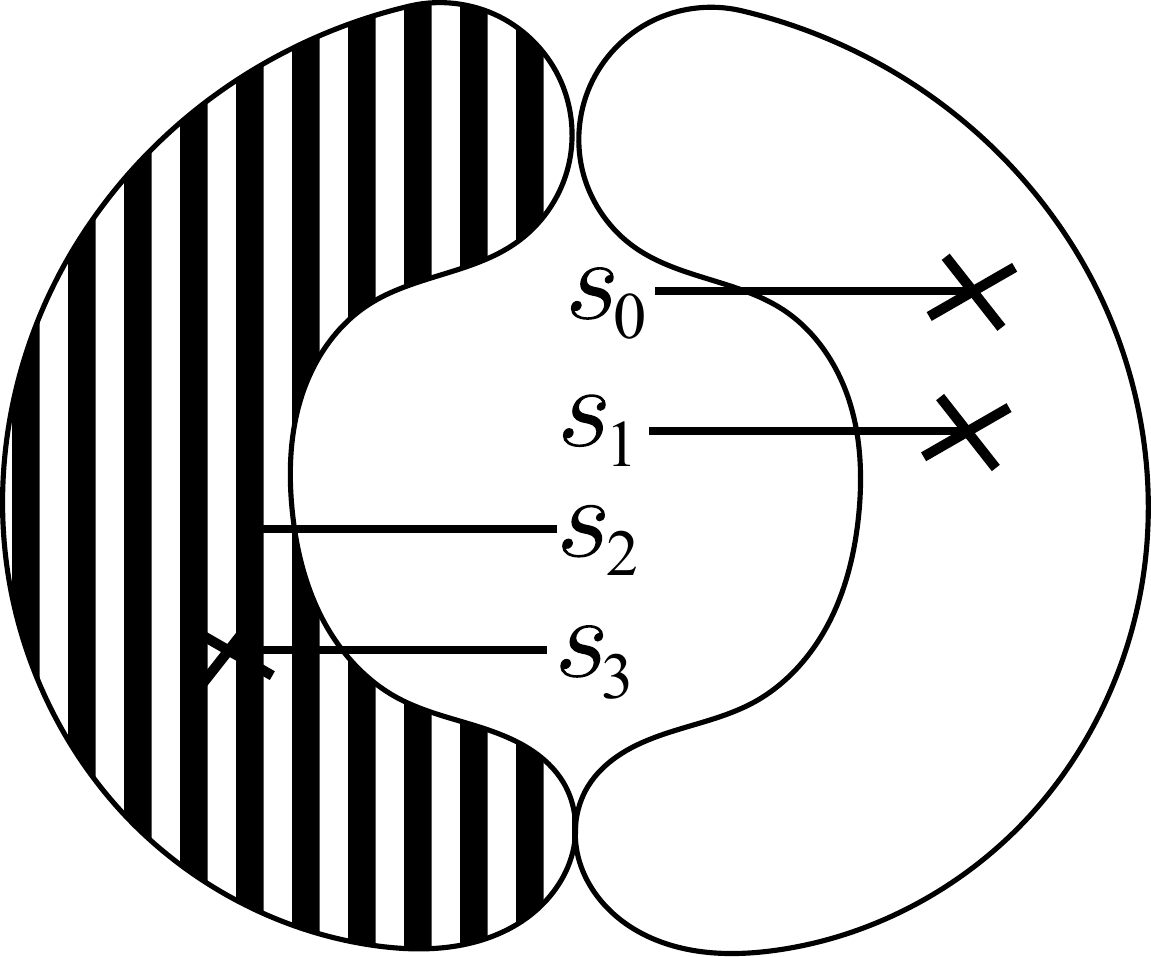}} \hspace{1cm}  \\ \hline

$\mathbf{1}_{(1, 1,2)}$ & $\begin{array}{l}2 c_1 \mathcal{S}_7 - \mathcal{S}_2 \mathcal{S}_7 - \mathcal{S}_6 \mathcal{S}_7 + c_1 \mathcal{S}_9 - \mathcal{S}_6 \mathcal{S}_9 \end{array}$ &\rule{0pt}{1.2cm}\parbox[c]{1.8cm}{ \includegraphics[scale=0.25]{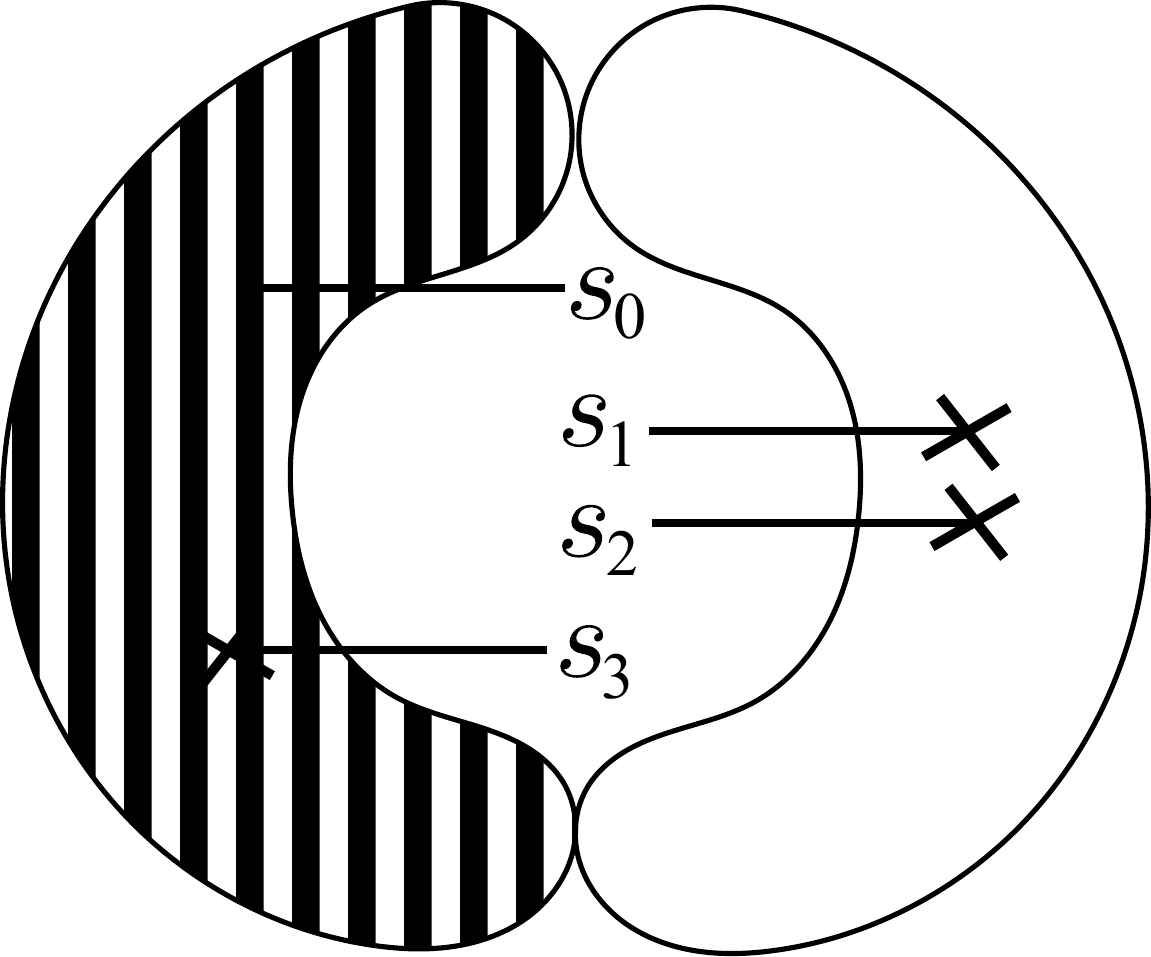}} \hspace{1cm}  \\ \hline

$\mathbf{1}_{(0, 0,2)}$ & $\begin{array}{l}\mathcal{S}_6 \mathcal{S}_7 - c_1 \mathcal{S}_9 + \mathcal{S}_2 \mathcal{S}_9 + \mathcal{S}_6 \mathcal{S}_9 + \mathcal{S}_7 \mathcal{S}_9 + \mathcal{S}_9^2 \end{array}$ & \rule{0pt}{1.2cm}\parbox[c]{1.8cm}{ \includegraphics[scale=0.25]{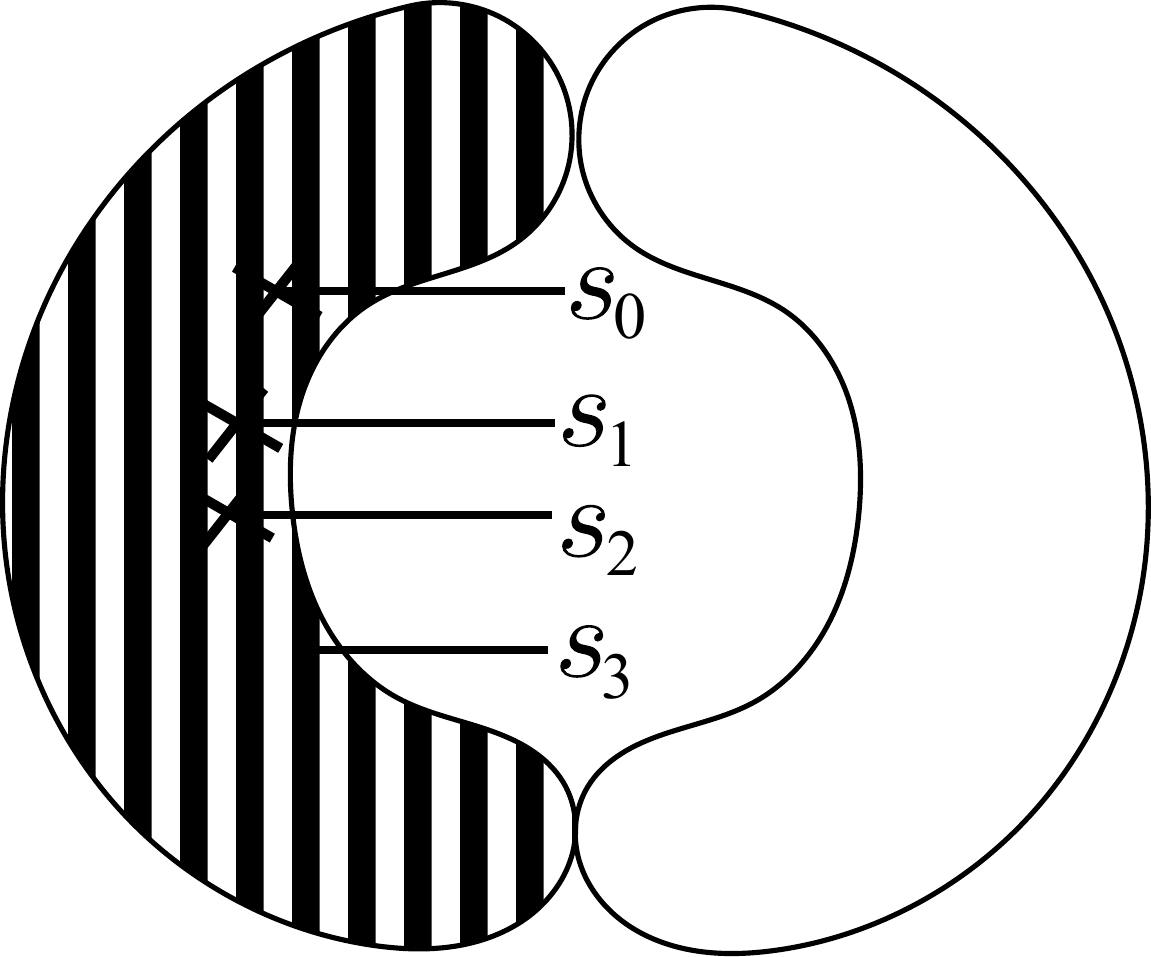}} \hspace{1cm} \\ \hline
 
$\mathbf{1}_{(1, 1,1)}$&  $\begin{array}{l}6 c_1^2 + (\mathcal{S}_2 + \mathcal{S}_6) (\mathcal{S}_6 + 2 \mathcal{S}_7) + 3 \mathcal{S}_6 \mathcal{S}_9 \\ - 
 c_1 (2 \mathcal{S}_2 + 5 \mathcal{S}_6 + 2 (\mathcal{S}_7 + \mathcal{S}_9)) \end{array}$ &  \rule{0pt}{1.2cm}\parbox[c]{1.8cm}{ \includegraphics[scale=0.25]{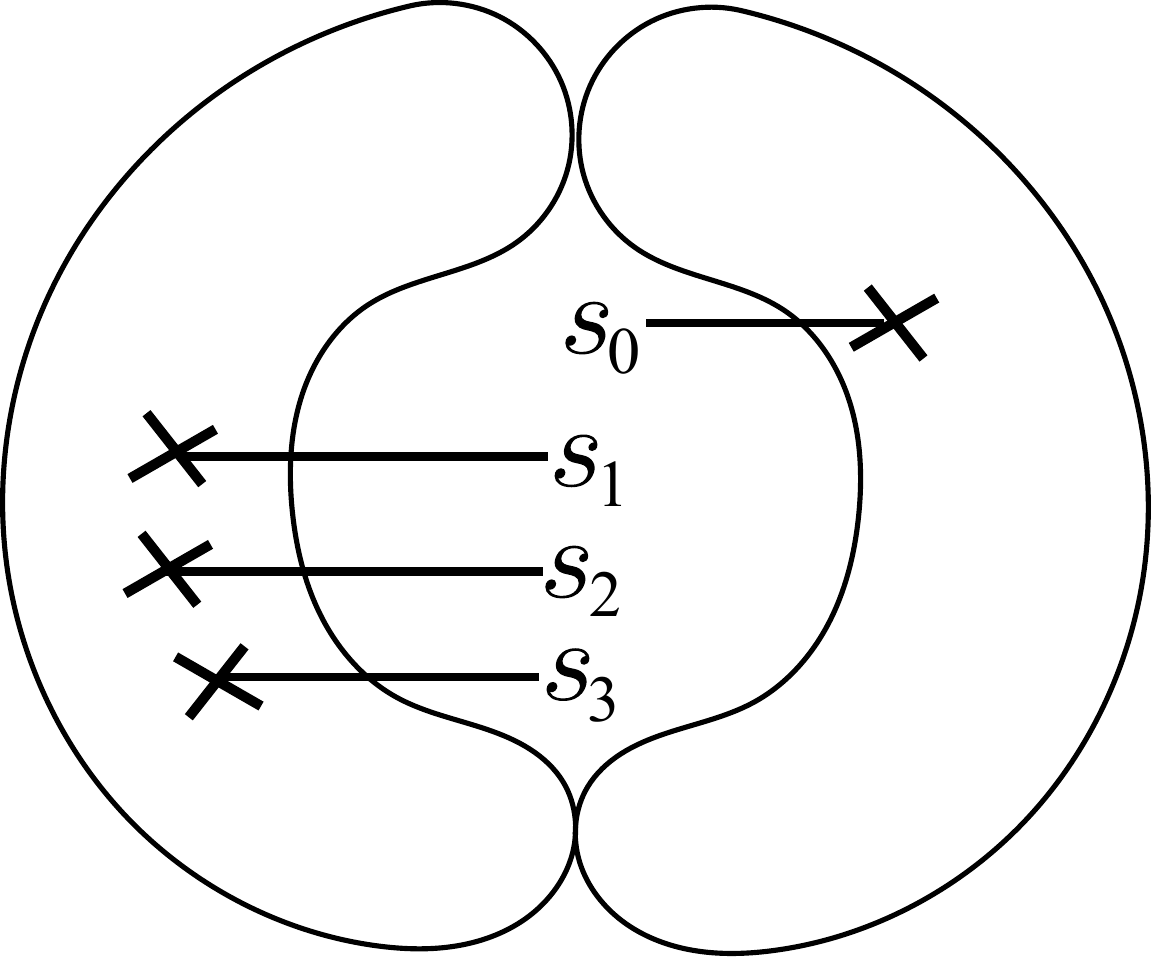}} \hspace{1cm}  \\ \hline
 
\end{tabular} 
\caption{\label{tab:U13Modela}{The first part of the summary of the matter representations and multiplicities from nef (67,0) with $G=U(1)^3$. It is continued in Table~\ref{tab:U13Modelb}. We also depict the intersections of the fiber components with the sections.}}
\end{center}
\end{table}

\begin{table}[ht!]
\begin{center}
\begin{tabular}{|c|l|c|} \hline
Representation & Multiplicity & Fiber \\ \hline 
$\mathbf{1}_{(1, 0,1)}$&   $\begin{array}{l} 6 c_1^2 - 5 c_1 \mathcal{S}_2 + \mathcal{S}_2^2 - 2 c_1 \mathcal{S}_6 + \mathcal{S}_2 \mathcal{S}_6 - 2 c_1 \mathcal{S}_7\\ + \mathcal{S}_2 \mathcal{S}_7 + \mathcal{S}_6 \mathcal{S}_7 - 
 2 c_1 \mathcal{S}_9 + 3 \mathcal{S}_2 \mathcal{S}_9\end{array} $& \rule{0pt}{1.2cm}\parbox[c]{1.8cm}{ \includegraphics[scale=0.25]{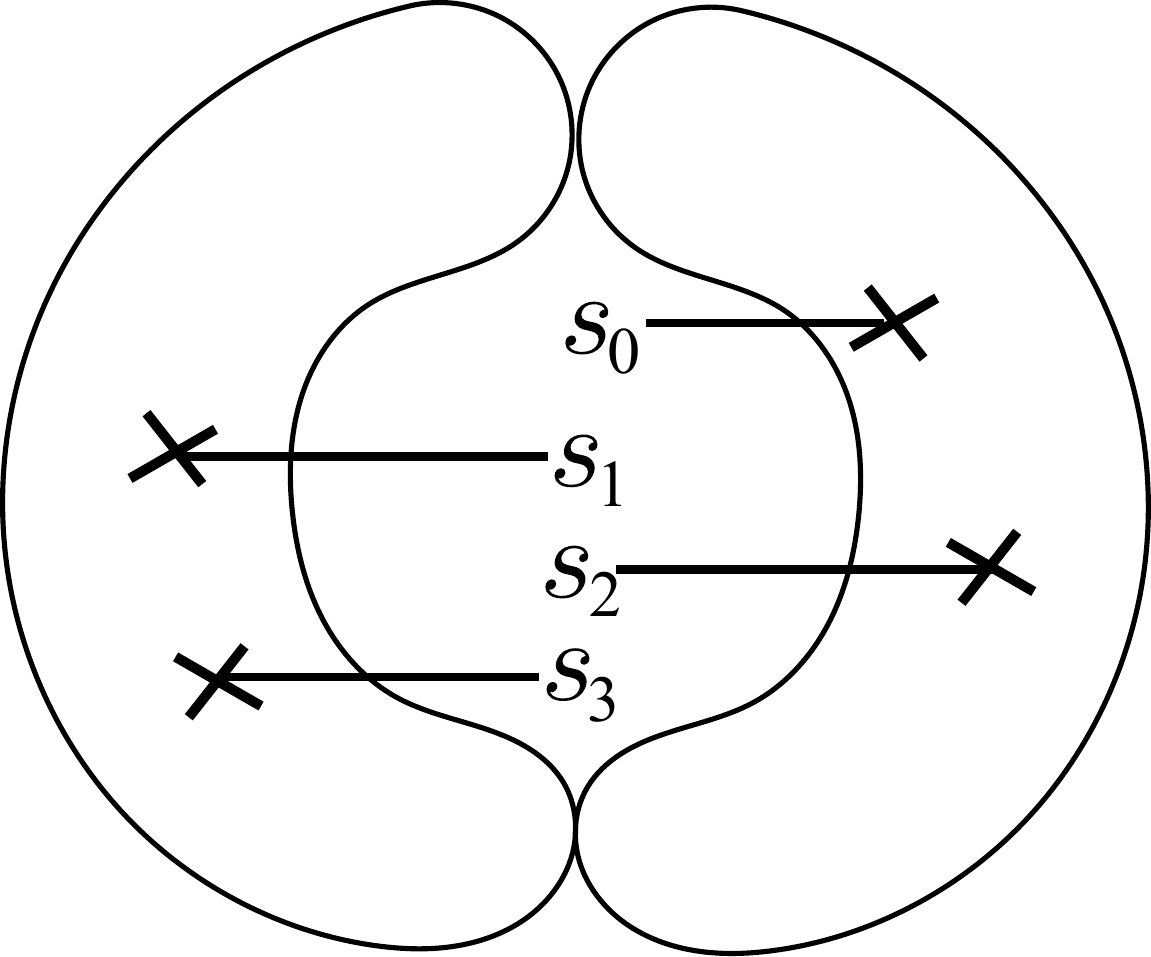}} \hspace{1cm} \\ \hline 

$\mathbf{1}_{(0, 1,1)}$& $ \begin{array}{l}2 c_1 \mathcal{S}_2 - \mathcal{S}_2 \mathcal{S}_6 + 6 c_1 \mathcal{S}_7 - 2 \mathcal{S}_2 \mathcal{S}_7 \\- \mathcal{S}_6 \mathcal{S}_7 - 2 \mathcal{S}_7^2  + c_1 \mathcal{S}_9 - 3 \mathcal{S}_7 \mathcal{S}_9\end{array}$&   \rule{0pt}{1.2cm}\parbox[c]{1.8cm}{ \includegraphics[scale=0.25]{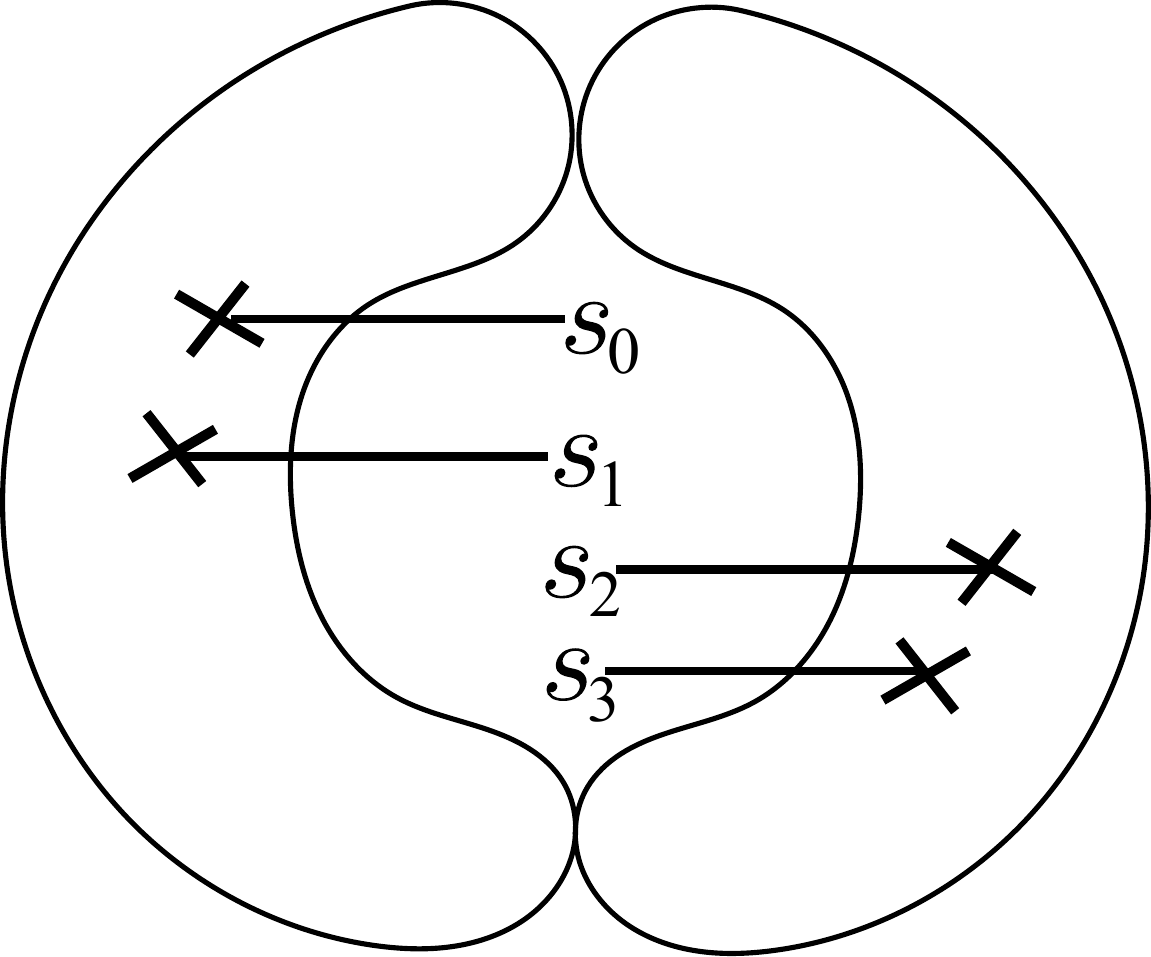}} \hspace{1cm} \\ \hline

$\mathbf{1}_{(1, 1,0)}$&  $\begin{array}{l}2 c_1 \mathcal{S}_2 + 6 c_1 \mathcal{S}_6 - 2 \mathcal{S}_2 \mathcal{S}_6 - 2 \mathcal{S}_6^2 \\ - \mathcal{S}_2 \mathcal{S}_7 - \mathcal{S}_6 \mathcal{S}_7 + c_1 \mathcal{S}_9 - 3 \mathcal{S}_6 \mathcal{S}_9 \end{array}$&   \rule{0pt}{1.2cm}\parbox[c]{1.8cm}{ \includegraphics[scale=0.25]{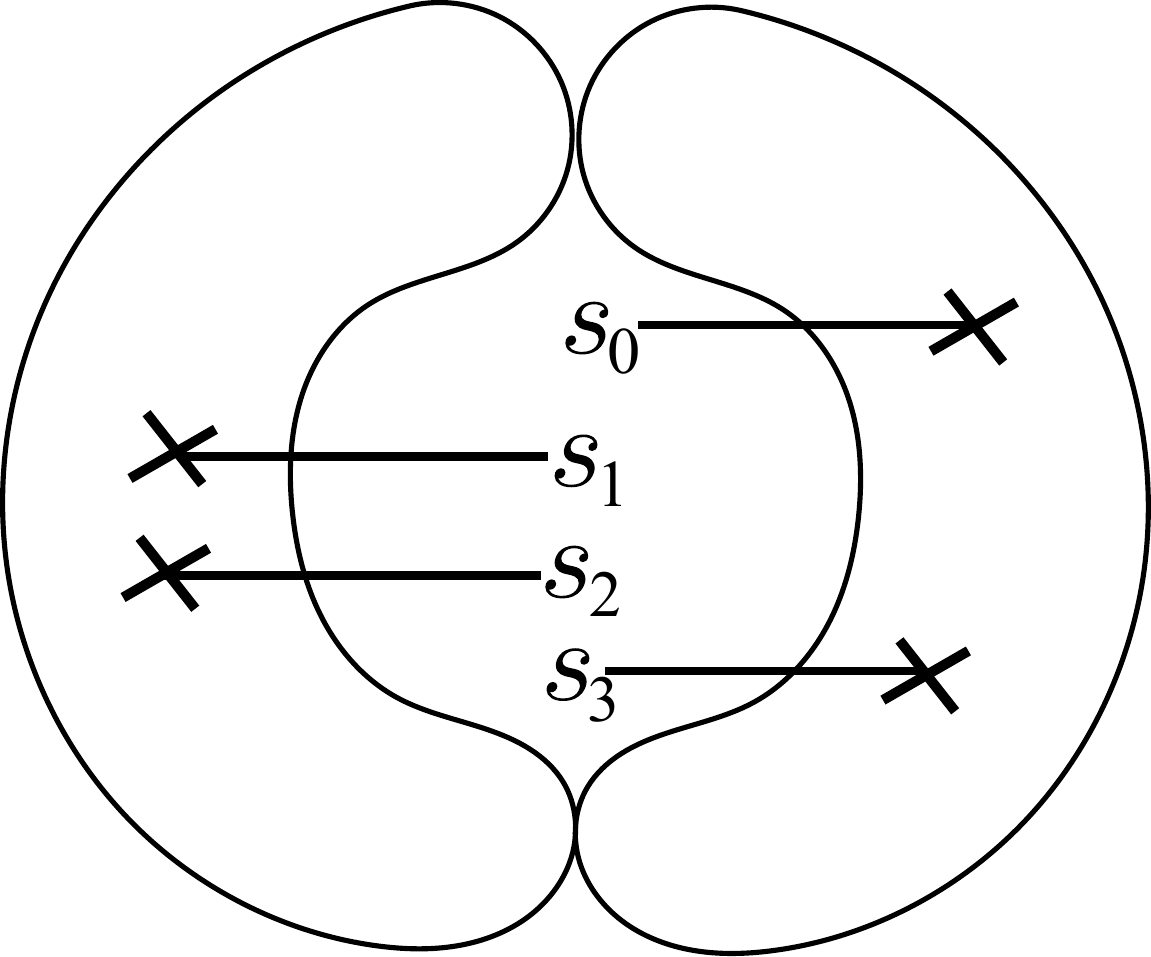}} \hspace{1cm} \\  \hline

$\mathbf{1}_{(1, 0,0)}$& $ \begin{array}{l}6 c_1 \mathcal{S}_2 - 2 \mathcal{S}_2^2 + 2 c_1 \mathcal{S}_6 - 2 \mathcal{S}_2 \mathcal{S}_6 + 2 c_1 \mathcal{S}_7\\ - 2 \mathcal{S}_2 \mathcal{S}_7 - 2 \mathcal{S}_6 \mathcal{S}_7 + 
 c_1 \mathcal{S}_9 - 3 \mathcal{S}_2 \mathcal{S}_9\end{array}$ & \rule{0pt}{1.2cm}\parbox[c]{1.8cm}{ \includegraphics[scale=0.25]{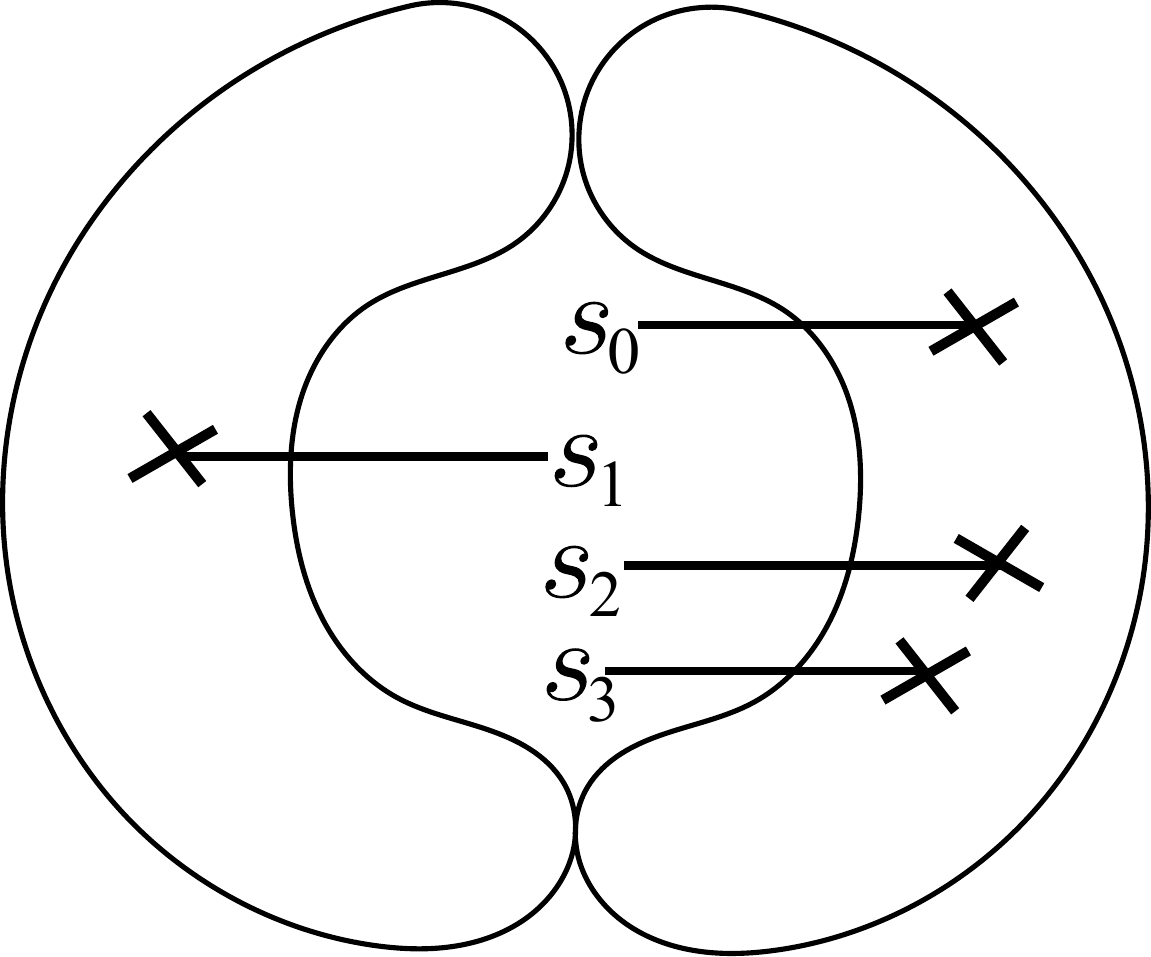}} \hspace{1cm}  \\ \hline
    
$\mathbf{1}_{(0, 1,0)}$&$ \begin{array}{l}6 c_1^2 - 2 c_1 \mathcal{S}_2 - 2 c_1 \mathcal{S}_6 + 2 \mathcal{S}_2 \mathcal{S}_6 - 5 c_1 \mathcal{S}_7 \\+ \mathcal{S}_2 \mathcal{S}_7  + 
 2 \mathcal{S}_6 \mathcal{S}_7 + \mathcal{S}_7^2 - 2 c_1 \mathcal{S}_9 + 3 \mathcal{S}_7 \mathcal{S}_9\end{array}$& \rule{0pt}{1.2cm}\parbox[c]{1.8cm}{ \includegraphics[scale=0.25]{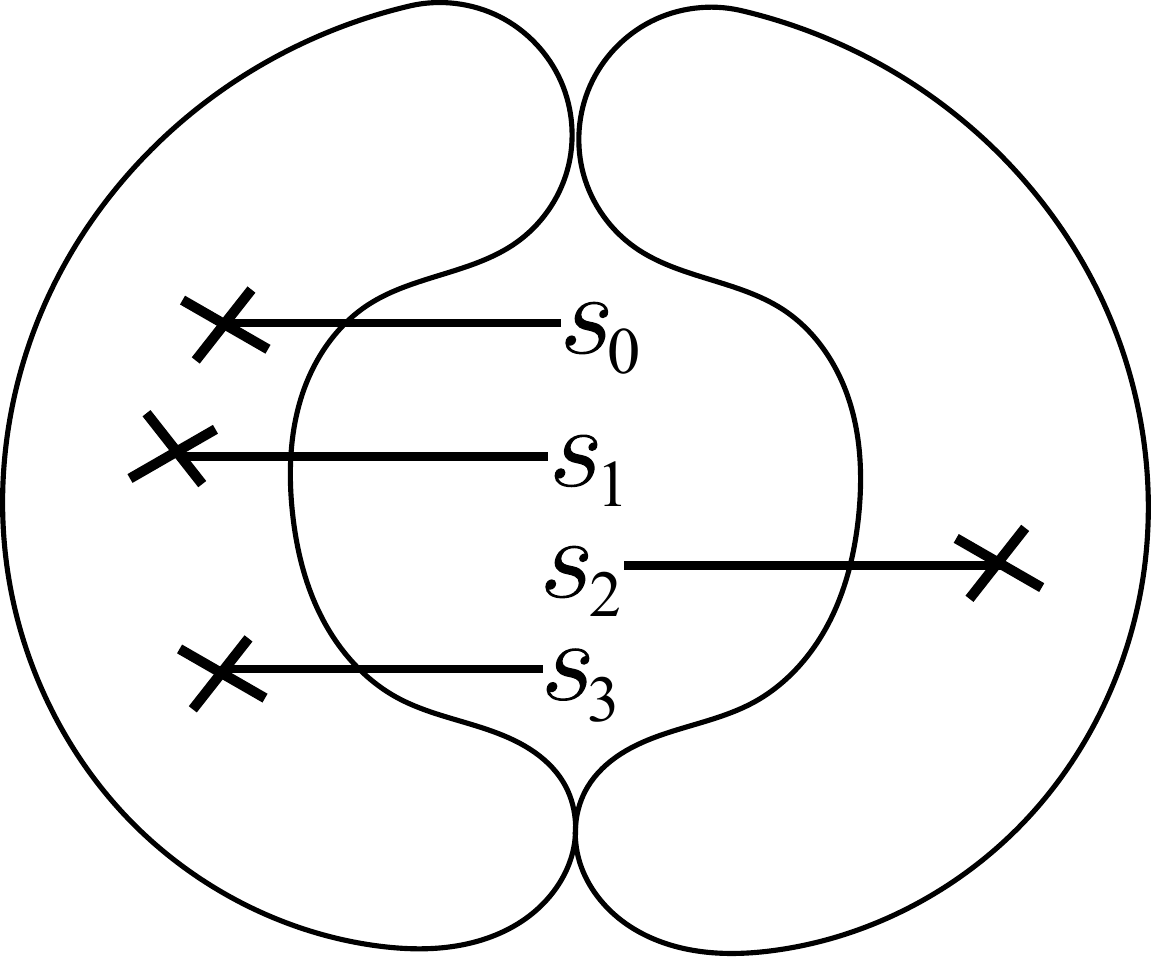}} \hspace{1cm} \\  
  \hline

$\mathbf{1}_{(0, 0,1)}$ & $\begin{array}{l}2 (3 c_1 \mathcal{S}_2 - \mathcal{S}_2^2 + 3 c_1 \mathcal{S}_6 - \mathcal{S}_2 \mathcal{S}_6 - \mathcal{S}_6^2 \\ \quad+ 3 c_1 \mathcal{S}_7 - \mathcal{S}_2 \mathcal{S}_7  - \mathcal{S}_6 \mathcal{S}_7 -
    \mathcal{S}_7^2 + 5 c_1 \mathcal{S}_9 \\ \quad- 2 \mathcal{S}_2 \mathcal{S}_9 - 2 \mathcal{S}_6 \mathcal{S}_9 - 2 \mathcal{S}_7 \mathcal{S}_9 - 2 \mathcal{S}_9^2) \end{array}$ &  \rule{0pt}{1.2cm}\parbox[c]{1.8cm}{ \includegraphics[scale=0.25]{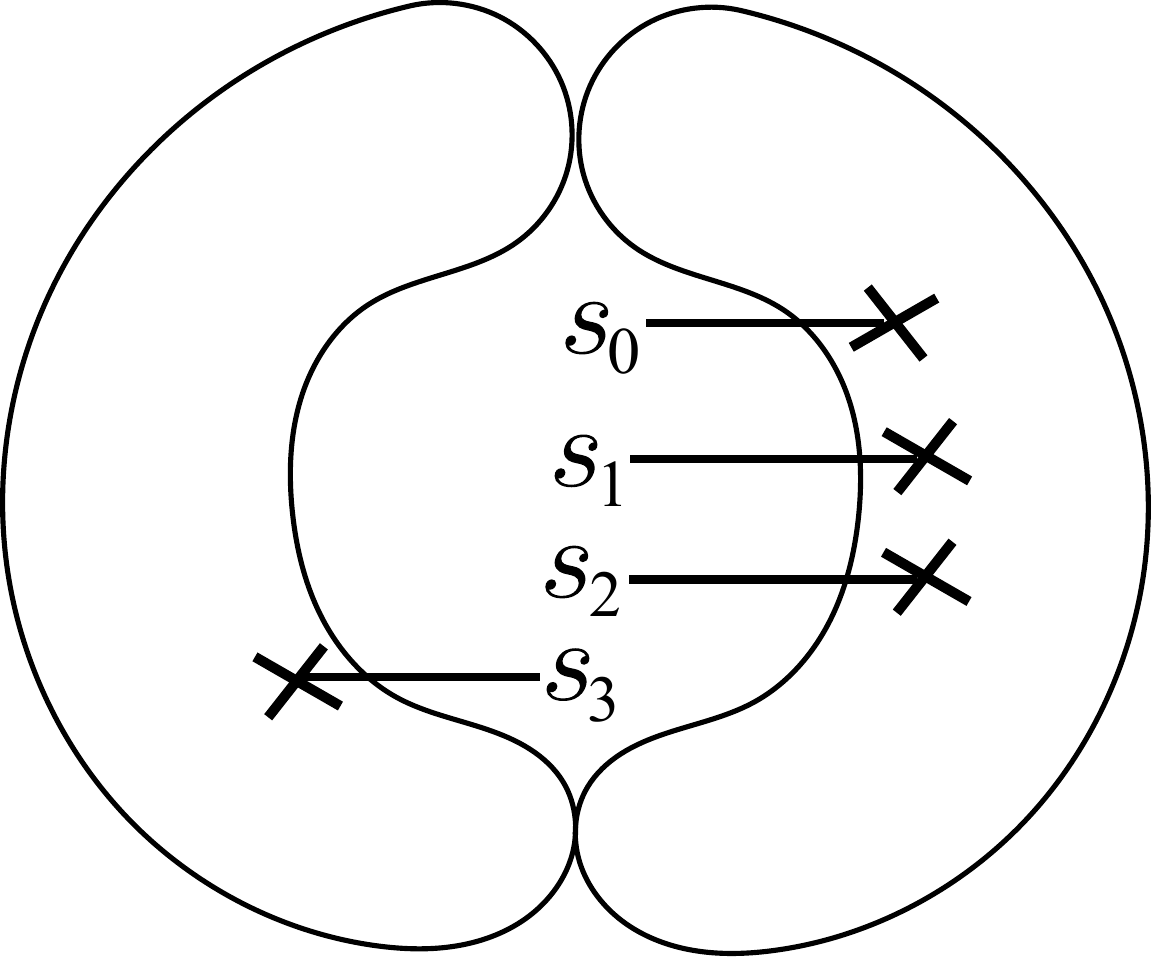}} \hspace{1cm}  \\ \hline
   
$\mathbf{1}_{(0, 0,0)}$ &  $\begin{array}{l}15 + 11 c_1^2 - 6 c_1 \mathcal{S}_2 + 2 \mathcal{S}_2^2 - 6 c_1 \mathcal{S}_6 + 2 \mathcal{S}_2 \mathcal{S}_6 \\ + 2 \mathcal{S}_6^2  - 
 6 c_1 \mathcal{S}_7 + 2 \mathcal{S}_2 \mathcal{S}_7 + \mathcal{S}_6 \mathcal{S}_7 + 2 \mathcal{S}_7^2\\  - 9 c_1 \mathcal{S}_9 + 3 \mathcal{S}_2 \mathcal{S}_9 + 3 \mathcal{S}_6 \mathcal{S}_9 + 
 3 \mathcal{S}_7 \mathcal{S}_9 + 3 \mathcal{S}_9^2 \end{array} $ &  \\ \hline \hline
 \begin{tabular}{c}Anomaly- \\ Coefficients:\end{tabular} &  \multicolumn{2}{|c|}{
$\begin{array}{ll} b_{11}= 2 c_1 \, , & b_{12}=c_1 \, , \\ b_{22} = 2 c_1 \, , & b_{23} = (c_1 - \mathcal{S}_6 + \mathcal{S}_7) \, ,\\
b_{33}=  2 (c_1 + \mathcal{S}_7 + \mathcal{S}_9)\, , & b_{13}=  (2 c_1 - \mathcal{S}_2 - \mathcal{S}_6) \, .  
\end{array}$
}\\ \hline
\end{tabular}
	\caption{\label{tab:U13Modelb}{Continuation of Table~\ref{tab:U13Modela} of matter representations and multiplicities from nef (67,0) with $G=U(1)^3$. We also depict the intersections of the fiber components with the sections as well as the the anomaly coefficients.}}
\end{center}
\end{table}  
\subsubsection*{Analysis of Higgs loci}
All of the toric singlet loci can be found by first considering where the sections $s_i$ degenerates.
The common feature of those loci are that the CICY equations develop a common factor.

For $s_0$ this locus corresponds to the ideal
\begin{align}
 I_1  = \langle s_{2,7} s_{1,9} - s_{1,7} s_{2,9}, s_{1,7} s_{2,4} - s_{1,4} s_{2,7} \rangle\,, 
\end{align}
and a reducible fiber component is in the class
\begin{align}
	\mathcal{C}_{1,1} = [e_2]([e_1]+[y]) \, . 
\end{align}
This results in the singlets with charges $q=(1,1,2)$.
The multiplicity of that locus is given by taking the class of $I_1$ and substracting the class of the sublocus $s_{1,7}=s_{2,7}=0$, leading to
\begin{align}
	\begin{split}
		n_{(1,1,2)}=&([s_{2,7}]+[s_{1,9}])([s_{1,7}]+ [s_{2,4}]) - [s_{1,7}][s_{2,7}]\\
		=&2 c_1 \mathcal{S}_7 - \mathcal{S}_2 \mathcal{S}_7 - \mathcal{S}_6 \mathcal{S}_7 + c1 \mathcal{S}_9 - \mathcal{S}_6 \mathcal{S}_9 \, .
	\end{split}
\end{align}
 Similarly, the locus where $s_1$ degenerates corresponds to the ideal
 \begin{align}
 I_2  = \langle s_{1,7} s_{2,6} -s_{1,6} s_{2,7}, s_{2,2} s_{1,6}- s_{1,2} s_{2,6}  \rangle \, .
\end{align}
Here we find a fibral curve in the class
\begin{align}
	\mathcal{C}_{1,2} = [e_1]([e_3]+[z]) \, ,
\end{align}
 whose intersections with $s_i$ reveal it to yield a $q=(-1,0,1)$ singlet of multiplicity
 \begin{align}
	 \begin{split}
		 n_{(-1,0,1)}=&([s_{1,6}]+[s_{2,7}])([s_{1,2}]+[s_{2,6}]) - [s_{1,6}][s_{2,6}]\\
	 =& \mathcal{S}_2^2-c_1 \mathcal{S}_2 + \mathcal{S}_2 \mathcal{S}_6 + \mathcal{S}_2 \mathcal{S}_7 + \mathcal{S}_6 \mathcal{S}_7 + \mathcal{S}_2 \mathcal{S}_9 \, .
	 \end{split}
 \end{align}
 Similarly one proceeds for $s_2$, which degenerates over the locus that corresponds to the ideal
 \begin{align}
 I_3 = \langle     s_{2,6} s_{1,9}- s_{1,6} s_{2,9}, s_{1,6} s_{2,3} - s_{1,3} s_{2,6}\rangle\, .
 \end{align}
We find a fibral curve in the class
 \begin{align}
	 \mathcal{C}_{1,3}=[e_3]([e_1]+[z]) \, ,
 \end{align}
and the intersections with $s_i$ reveal the associated singlets to have charges $q=(0,-1,1)$ and multiplicity
 \begin{align}
	 \begin{split}
		 n_{(0,-1,1)}=& 2 c_1  \mathcal{S}_6 -  \mathcal{S}_2  \mathcal{S}_6 -  \mathcal{S}_6 \mathcal{S}_7 + c_1  \mathcal{S}_9 -  \mathcal{S}_7  \mathcal{S}_9 \, .
	 \end{split}
 \end{align}
 The final toric locus can be obtained by starting from the ideal
\begin{align}
I_4 = \langle   s_{2,6} s_{1,7}-s_{1,6} s_{2,7} , s_{1,6} s_{2,9} -s_{1,9} s_{2,6} \rangle \, . 
\end{align}
Over this locus, the fiber factors into a quadric polynomial in $w$ and a linear polynomial in the $e_i$.
This results in the class of one of the irreducible components
\begin{align}
\mathcal{C}_{1,4}= [w][e_i] \, .
\end{align}
 The resulting charges are $q=(0,0,-2)$ and the  multiplicity is
 \begin{align}
	 \begin{split}
		 n_{(0,0,2)}=&([s_{1,6}]+[s_{2,7}])([s_{1,9}]+[s_{2,6}]) - [s_{2,6}][ s_{1,6}]\\
		 =&  \mathcal{S}_6  \mathcal{S}_7 - c_1  \mathcal{S}_9 +  \mathcal{S}_2  \mathcal{S}_9 +  \mathcal{S}_6  \mathcal{S}_9 +  \mathcal{S}_7  \mathcal{S}_9 +  \mathcal{S}_9^2 \, .
	 \end{split}
 \end{align}
 Note that the above singlet is the only one that does not Higgs to the $U(1)^2$ theory described by nef (21,0).
 Instead the Higgsing corresponds to a transition to nef (30,2) which engineers $G=U(1)^2 \times \mathbb{Z}_2$ but this is not part of the chain that we consider in the present work.  
%\newpage
\section{The six- and five-dimensional field theory}
\label{sec:FieldTheory}
In this section we study the supergravities that we have geometrically engineered in the previous section.
More precisely, we will focus on the dynamical connections among them via the Higgs mechanism.
All spectra that have been found in the previous sections satisfy the six-dimensional gravitational and $U(1)$ gauge anomaly cancellation conditions for any choice of base und bundles.

The conditions read
\begin{align}
	\begin{split}
		H- V+29 T= 273 \, ,& \quad 9-T= c_1^2 \, ,\\ 
\sum_q x_{q_m, q_n,q_k,q_l} q_m q_n q_k q_l &= b_{\left(m,n \right.} \cdot b_{\left.k,l\right)} \, , \\ 
\frac16 \sum_q x_{q_m, q_n} q_m q_n &= c_1 \cdot b_{m,n}  \, ,
	\end{split}
\end{align}
where $b_{m,n}$ denotes the height pairing of the Shioda maps, see e.g.~\cite{Klevers:2014bqa}, and $x_{q_{i_1},\dots,q_{i_n}}$ are the multiplicities of hypermultiplets with charges $q_{i_1},\dots,q_{i_n}$ under $n$ not necessarily inequivalent $U(1)$ factors.
Let us point out that the height pairings that we provided in the previous section have not been computed geometrically but were deduced from the anomalies.
This is still a highly non-trivial cross-check of the validity of the spectrum.
Two of the models contain discrete symmetries and we comment on their anomalies in Section~\ref{sec:discAnomalies}.
\subsection{Higgs transitions among the vacua}
\label{subsec:higgsing}
  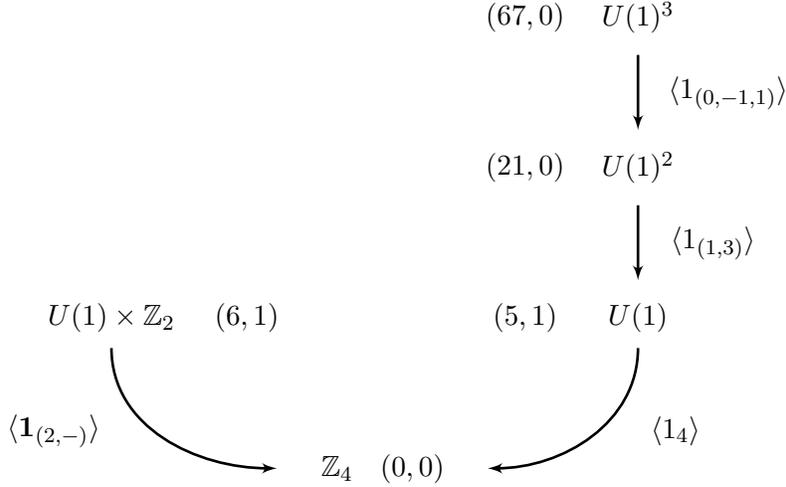
\begin{figure}[h!]
\centering
\begin{tikzpicture}[node distance=4mm, >=latex',block/.style = {draw, rectangle, minimum height=65mm, minimum width=83mm,align=center},]
\begin{scope}[shift={(0,0)},scale=1]
	\node at (1.5,4) {$U(1)\times\mathbb{Z}_2$};
			\node at (0.7,2.5) {$\, \langle \mathbf{1}_{(2,-)} \rangle $};
	\node at (3.3,4) {${(6,1)}$};
	\node at (5.5,2) {${(0,0)}$};
	\node at (4.5,2) {$\mathbb{Z}_4$};
	\node at (7,8) {${(67,0)}$};
	\node at (7,6) {${(21,0)}$};
	\node at (7,4) {${(5,1)}$};
	\node at (8.5,8) {$U(1)^3$};
	\node at (8.5,6) {$U(1)^2$};
	\node at (8.5,4) {$U(1)$};
	\draw [line width=1,->] (8.5,7.5)  -- (8.5,6.5);
	\draw [line width=1,->] (8.5,5.5)  -- (8.5,4.5);

	\node at (9.5,5) {$\langle 1_{(1,3) } \rangle$};

	\node at (9.7,7) {$\langle 1_{(0,-1,1)  }\rangle$};

	\draw [line width=1,->] (8.5,3.6) to[out=270,in=0] (6.5,2);
	\node at (9,2.5) {$ \langle 1_4 \rangle $};

	\draw [line width=1,->] (1.5,3.6) to[out=270,in=180] (3.7,2);
\end{scope}
\end{tikzpicture}
\caption{ Depiction of the two Higgs branches towards the $\mathbb{Z}_4$ genus-one model. Gauge groups and Higgs multiplets are highlighted, together with the nef partition that engineers the vacuum.  }
\label{fig:higgschains}
\end{figure}

Field theoretically we can connect two vacua by giving a vacuum expectation value (vev) to some hypermultiplet.
In general, this hypermultiplet will have a non-trivial representation under some gauge group which therefore gets partially or fully broken by the vev.
The Higgs field then contributes the respective Goldstone mode for the massive vector boson(s).
For purely Abelian theories the unbroken generators are then simply the combinations under which the Higgs field is not charged.

In order to preserve supersymmetry, the vev also needs to satisfy the D-flatness conditions.
Such a vev can always be found when there are two Higgs multiplets with the same U(1) charge~\cite{Honecker:2006qz}.
In the following we will always assume that this constraint is fulfilled~\footnote{In~\cite{Klevers:2014bqa} it was argued that a transition which does not satisfy the D-flatness condition would result in poles in the defining equation of the deformed Calabi-Yau geometry.}.
For each individual transition we then match charges and multiplicities of the hypermultiplets under the unbroken gauge group with our geometric computations.
This matching is particularly easy in six-dimensional theories, due to the absence of a superpotential.
At the bottom of our Higgs chains is the~$\mathbb{Z}_4$ model.
Note that it can be reached via two different Higgs branches depicted in Figure~\ref{fig:higgschains}.
 \subsection*{Branch 1: The $U(1)^3 \rightarrow \mathbb{Z}_4$ chain}
The first branch starts with the $U(1)^3$ theory that is engineered by fibrations of nef (67,0) and the spectrum is summarized in Tables~\ref{tab:U13Modela}-\ref{tab:U13Modelb}.
The unHiggsing from $\mathbb{Z}_4$ to $U(1)^3$ can geometrically be understood as a splitting of the four-section into four sections~\cite{Morrison:2014era}.

The Higgsing of $U(1)^3$ proceeds by assigning a vev to the hypermultiplets in the representation $\mathbf{1}_{(0,1,-1)}$.
Note that there is also the possibility to Higgs with the multiplets $\mathbf{1}_{(1,1,2)}$ or $\mathbf{1}_{(-1,0,1)}$.
The resulting $U(1)^2$ theories are equivalent to each other and the spectra can be geometrically matched by redefining the divisors.
To directly match the conventions that we laid out for nef (21,0), we consider the first option, which results in the unbroken charges
 \begin{align}
\widetilde{Q_1} = Q_1 \, , \quad\widetilde{Q_2} = Q_2+Q_3 \, .
	 \label{eqn:chargesu13}
\end{align}
The Higgs multiplet minus the Goldstone mode is  matched to the additional neutral singlets in the $U(1)^2$ theory.
Those can be computed directly via the difference of the Euler numbers that are summarized in Appendix~\ref{app:eulers}
\begin{align}  
	\Delta n_{(0,0)}+1  = n_{(0,1,-1)} =  \frac12 ( \chi_{(21,0)}-\chi_{(67,0)}) \, . 
 \end{align}
We can also match the remaining charged hypermultiplets with those obtained from the geometric computations summarized in Table~\ref{tab:U12Model}.
The identification of charges~\eqref{eqn:chargesu13} determines
 \begin{align} 
 \begin{split}
	 n_{(1,1)} =&  n_{(1,0,1)} +n_{(1,1,0)} \, ,\quad n_{(0,2)}=    n_{(0,0,2)}+    n_{(0,1,1)} \,,\\
	 n_{(1,1)}=&     n_{(1,0,1)}+    n_{(1,1,0)} \, ,\quad   n_{(0,1)}=     n_{(0,1,0)}+    n_{(0,0,1)} \, .\\
	 n_{(1,2)} =& n_{(1,0,2)}+ n_{(1,1,1)} \, , \quad n_{(-1,1)}  = n_{(-1,0,1)}\,,\\
	 n_{(1,3)}=&  n_{(1,1,2)}\, , \quad n_{(0,3)}  =  n_{(0,1,2)} \,.\\ 
\end{split}
\end{align} 
One observes in the identification of hypermultiplets that the aforementioned Higgs multiplets do not combine with other representations.

The next step in the chain is performed, by assigning a vev either to the $\mathbf{1}_{(1,3)}$ or the $\mathbf{1}_{(-1,1)}$ hypermultiplets.
Again the choices result in equivalent theories.
The former one leads to a straight forward match conventions that we laid out for nef (5,1).
It leaves an unbroken U(1) charge
\begin{align}
\widetilde{Q} = 3 Q_1 - Q_2 \, .
\end{align}
We can again match the difference of singlets
\begin{align}
	\Delta n_{(0)}+1  = n_{(3,1)} =  \frac12 ( \chi_{(5,1)}-\chi_{(21,0)}) \, ,  
\end{align}
as well as the other hypermultiplets
\begin{align}
\begin{array}{ll}
	n_{(4)} =  n_{(-1,1)} \, ,&   n_{(3)}= n_{(1,0)} + n_{(0,3)}\, , \\
	n_{(2)} =  n_{(0,2)}+n_{(1,1)}\, , & n_{(1)} = n_{(1,2)}+n_{(0,1)} \,  ,
\end{array}
\end{align}
where the matter loci of nef (5,1) are summarized in Table~\ref{tab:nef51}.

The last transition in this chain Higgses the U(1) theory to $\mathbb{Z}_4$ via the charge four singlets.
We again check the difference in singlets
\begin{align}
	\Delta n_{(0)}+1 &= n_{(4)} =  \frac12 ( \chi_{(5,1)}-\chi_{(0,0)}) \, ,
\end{align}
and the matter multiplicities can be matched by taking all U(1) charges to be modulo 4.
Note that in six dimensions the hypermultiplets contain two conjugated representations and hence $\mathbf{1}_3$ multiplets are not distinguished from $\mathbf{1}_{-3}$.
Both are identified as $\mathbf{1}_1$ with respect to the $\mathbb{Z}_4$ gauge symmetry.
To conclude, we  are left with only two types of hypermultiplets
\begin{align}
	n_{(1)} =& n_{(1)} + n_{(3)}\, , \qquad n_{(2)} = n_{(2)}\, ,
\end{align}
which matches the geometric computations given in Table~\ref{tab:nef00Spectrum}.

\subsection*{Branch 2: The $U(1) \times \mathbb{Z}_2 \rightarrow \mathbb{Z}_4$ chain}
The second branch starts with a theory that already exhibits a discrete symmetry.
The gauge group is $U(1) \times \mathbb{Z}_2$ and the spectrum is summarized in Table~\ref{tab:nef61Spectrum}.
The Higgsing mixes the $\mathbb{Z}_2$ and the $U(1)$ symmetry such that only a $\mathbb{Z}_4$ subgroup remains unbroken~\footnote{This mechanism was speculated to exist in the works \cite{Kimura:2019bzv,Kimura:2019syr,Kimura:2019qxf}.}.
The unique Higgs candidate that can trigger such a transition is the $\mathbf{1}_{(2,-)}$ hypermultiplet.
After giving a vev to this field we can write down the $\mathbb{Z}_4$ charge
\begin{align}
Q_{\mathbb{Z}_4} = Q_{U(1)} \pm 2 Q_{\mathbb{Z}_2} \, \text{ mod } 4 \, . 
\end{align} 
This allows us to match the difference in the number of singlets via
\begin{align}
	\Delta n_{(0)}+1 & = n_{(2,-)} =  \frac12 ( \chi_{(0,0)}-\chi_{(6,1)}) \, .
\end{align}
Moreover, we can identify the hypermultiplets via
\begin{align}
	n_{(1)} =  n_{(1,+)} + n_{(1,-)}\, , \qquad n_{(2)} = n_{(2,+)}+n_{(0,-)}\, ,
\end{align}
and this matches with the geometric computation of the matter loci for nef (0,0) that is summarized in Table~\ref{tab:nef00Spectrum}.  
\subsection{The structure of the $\mathbb{Z}_4$ Tate-Shafarevich group}
In this section we would like to comment on the structure of the $\mathbb{Z}_4$ Tate-Shafarevich (TS) group.
In the spirit of~\cite{Cvetic:2015moa} we want to investigate this structure via the conifold transition from the vacum with $U(1)$ via the charge $q=4$ singlet.
As shown in Table~\ref{tab:nef51}, this field arises in fibrations of nef (5,1) from M2-branes that wrap one of the two fibral curves $\mathcal{C}_{1,1},\mathcal{C}_{2,1}$ which intersect the sections as follows:
\begin{align}
 \begin{array}{|c|c|c|}\cline{2-3}
\multicolumn{1}{c|}{} & \mathcal{C}_{1,1 } & \mathcal{C}_{2,1 }  \\ \hline
 s_0 & -1 & 2 \\ \hline
 s_1 & 3 & -2 \\ \hline
 \sigma(s_1)& -4 & 4 \\ \hline
 \end{array} 
\label{eq:nef51h}
\end{align}
Indeed, it seems that all non-trivial TS elements can be obtained from shrinking down the two curves above~\footnote{The deformation towards the Jacobian fibration however required a different model with a holomorphic zero-section over the singlet locus, similar as constructed in the $\mathbb{Z}_3$ case in~\cite{Cvetic:2015moa}.}.
One can deduce from the intersections of the other fibral curves in Table~\ref{tab:nef51} that the limit where the volume of $\mathcal{C}_{1,1}$ goes to zero lies on the boundary of the K\"ahler cone.
We can therefore shrink it and identify $s_0$ as the KK-generator which requires the U(1) flux $\xi_1 = \frac14$.
Geometrically this blows down down the exceptional divisor $[e_1]$.
Adding the deformation parameters $s_{i,5}, i\in\{1,2\}$ deforms the singularity and leads us to the generic bi-quadric in $\mathbb{P}^3$.
Similarly, we can swap $s_0$ and $s_1$ as the zero-section and shrink the same curve.
This allows us to deform to the same genus-one fibration but now a flux $\xi_3=-\frac34$ is switched on.
It is the conjugate TS element with flux $\xi_1$.
This is analogous to the situation for the generic cubic~\cite{Cvetic:2015moa} in that two different elements of the TS group correspond to the same geometry but are distinguished by the action of the Jacobian.

We also have the $\mathcal{C}_{1,2}$ curve that admits the right intersections with both $s_0$ and $s_1$ to Higgs to the order two element with fluxes $\xi=\frac12$.
However, the limit in which this curve shrinks does not lie on the boundary of the K\"ahler cone that is torically realized but on the boundary of a different K\"ahler cone.
We therefore expect that another birational model is needed to describe the transition.

A slightly different perspective can be given from the $\mathbb{Z}_2 \times U(1)$ theory of nef (6,1). Here the respective curves that lead to the Higgs field intersect as follows:  
\begin{align}
 \begin{array}{|c|c|c|}\cline{2-3}
\multicolumn{1}{c|}{} & \mathcal{C}_{1,2 } & \mathcal{C}_{2,2 }  \\ \hline
 s^{(2)}_0 & -1 & 3 \\ \hline
 s^{(2)}_1 & 1 & 1 \\ \hline
 \sigma( s^{(2)}_1)& -2 & 2 \\ \hline
 \end{array}
\end{align} 
The hypermultiplets that originate from the curves $\mathcal{C}_{1,2}$ can become massless when a flux $\xi_1= \frac12$ is switched on.
Hence naively only a $\mathbb{Z}_2$  group might be visible.
However note, that the above model is already a genus one fibration and hence in M-theory there is already an order two flux $\tilde{\xi}=\frac12$. In total we expect to have a $\xi=\frac14$ flux from that deformation~\footnote{The order two element of the TS group might be constructed from the resolved Jacobian of nef $(6,1)$ by performing the conifold from there.}.  

In both families (5,1) and (6,1) we find a unique curve that can be shrunken such that a subsequent complex structure deformation leads to a fibration of bi-quadrics in $\mathbb{P}^3$.
However, due to the correspondence of Higgs transitions and compactifications with discrete Wilson lines reviewed in Section~\ref{sec:ftpre}, we expect to be able to find extremal transitions to the order two element and to the Jacobian, possibly from other birational models.
The curve $\mathcal{C}_{2,1}$ in~\eqref{eq:nef51h} would be a candidate that admits the right intersections with the zero section to reach the order two TS element.
But to shrink this curve we have to move outside of the boundaries of the K{\"a}hler cone.
It would be interesting to find the birational geometry that realizes the adjacent K\"ahler cone and to construct the transition to the order two TS element.

In the following we would like to comment on some implications of the field theory perspective for a general TS element $M$ that has an order $K$ which divides the order of the TS group $\mathbb{Z}_N$ with $N=K \cdot L$.
Let us assume that there exists an elliptic fibration $M^\prime$ with Mordell-Weil group of non-vanishing rank that exhibits a shrinkable fibral curve and a corresponding extremal transition that leads to the TS element.
We denote the zero-section by $s_0$ and the additional section, that leads to the $U(1)_F$ which is broken to $\mathbb{Z}_N$ in the transition, by $s_1$.
The fibral curve then must have intersections
\begin{align}
\label{eq:subhiggs}
 \mathcal{C} \cdot (s_0, s_1) =(L a ,L (a-K))  \, ,
\end{align}
possibly with $s_0,s_1$ exchanged.
From an M-theory perspective the massive hypermultiplet obtained from the M2 brane state that wraps the above curve has $U(1)_F \times U(1)_{KK}$ charges $(K \cdot L, a L)$.
In particular, it becomes massless after turning on a discrete Wilson line $\xi=-a/K$. 
The key observation is now that both charges of the state are multiples of $L$.
Therefore giving a vev to this hypermultiplet yields the unbroken symmetry
\begin{align}
\widetilde{U(1)} \times \mathbb{Z}_L \quad \text{ with } \quad \widetilde{U(1)} = a U(1)_{F} - K U(1)_{KK}  \, .
\end{align}
We find the expected mixing of the F-theory $U(1)_F$ and the Kaluza-Klein $U(1)_{KK}$ but also an additional $\mathbb{Z}_L$ discrete factor. 

We are now in the position to relate those field theory properties back to the geometry of $M$ that is obtained after shrinking $\mathcal{C}$ and deforming the resulting singularity.
Recall that per construction $M$ is a genus one fibration that is an order $K$ element of a TS group $\mathbb{Z}_{K\cdot L}$.
From the field theory analysis and the M-theory perspective on the origin of discrete symmetries we expect that
\begin{align}
 \text{Tors}H_2(M, \mathbb{Z})=\mathbb{Z}_L \, .
 \end{align}
Recall that this torsion cycle must be present in order to yield a discrete symmetry in five-dimensions from the three-form reduction, as reviewed in Section~\ref{sec:ftpre}.
We therefore expect that $M$ is a genus one fibration with $K$-sections and $L$-torsional cycles.

\subsection{Absence of discrete anomalies}
\label{sec:discAnomalies}
Finally we want to comment on discrete anomalies within F-theory models obtained from genus one fibrations.
These include the ones that we discussed in Section~\ref{sec:CICYFI} but for completeness we also include the hypersurfaces studied in~\cite{Klevers:2014bqa}.
The potential anomalies have recently been discussed by Monnier and Moore in \cite{Monnier:2018nfs}.
There the authors found an anomaly for discrete groups but no Green-Schwarz mechanism.
The result is a non-trivial condition on the number of hypermultiplets $n_s$ and their charges $s$
\begin{align}
\sum_{s=1}^{k-1} n_{s} \frac{1}{24 k}  (-2 k s + 2 s^2 - k^2 s^2 + 2 k s^3 - s^4) = 0 \text{ mod } 1 \, ,
\end{align}
in order to keep the discrete $\mathbb{Z}_k$ symmetry free from anomalies as

However, the authors argued that these constraints might be too strong and only need to vanish modulo $1/2k$.
This is due to the possibility to add certain torsion classes to the field strengths in the anomaly polynomial.
The base independent results obtained in this work support this interpretation.

In the following we show that our $U(1)\times\mathbb{Z}_2$ and  $\mathbb{Z}_4$ models satisfy the refined integrality condition for any choice of base and bundles.
For $\mathbb{Z}_k,\,k\in\{2,3,4\}$ those conditions read
\begin{align}
	\begin{split}
\label{eq:discAnomaly}
\mathbb{Z}_2 &: \quad \frac{1}{4} n_1 \equiv 0 \text{ mod } 1  \, , \\ 
\mathbb{Z}_3 &: \quad \frac{1}{3} (n_1+
n_2) \equiv 0 \text{ mod } 1  \, , \\ 
\mathbb{Z}_4 &: \quad \frac14 \left(5 n_1 + 8 n_2 + 5 n_3    \right) \equiv 0 \text{ mod } 1 \, .
	\end{split}
\end{align}
Since a hypermultiplet in six dimensions contains two conjugated states, there is the additional identification $n_1 \sim n_{-1} \sim n_{-1+k}$.
For the pure $\mathbb{Z}_4$ theory one therefore finds only the looser constraint
\begin{align}
\mathbb{Z}_4 &:\quad \frac12 n_1\equiv 0   \text{ mod } 1 \, .
\end{align}
Note that charge two singlets do not contribute to this condition at all.

For the discrete symmetries obtained in Section~\ref{sec:CICYFI} we find the above constraints to be satisfied:
\begin{align}
	\begin{split}
U(1) \times \mathbb{Z}_2 :\quad&   n_{(0,-)}+ n_{(1,-)}+ n_{(2,-)}  =4(2 c_1 - \mathcal{S}_7) (2 c_1 +\mathcal{S}_7) \\
\mathbb{Z}_4:\quad& n_1 =2(3 c_1 - \mathcal{S}_2 - \mathcal{S}_9) (c_1 +\mathcal{S}_2 + \mathcal{S}_9)
	\end{split}
\end{align}
Finally, we would also like to comment on a generic class of $\mathbb{Z}_2$ and $\mathbb{Z}_3$ theories that can be constructed from hypersurfaces inside the toric ambient spaces $\mathbb{F}_0$, $\mathbb{P}^2_{1,1,2}$ and $\mathbb{P}^2$~\cite{Klevers:2014bqa}.
The matter loci are given by
\begin{align}
	\begin{split}
U(1) \times \mathbb{Z}_2:\quad&  n_{(1/2\pm 1/2,-)}  =2 (3 c_1^2 \pm 2   c_1 \mathcal{S}_7 \mp  \mathcal{S}_7^2 + 2 c_1 \mathcal{S}_9 - \mathcal{S}_9^2)\,,\\
\mathbb{Z}_2:\quad& n_1 =  4(3 c_1 - \mathcal{S}_7) (c_1 + \mathcal{S}_7) \, , \\
\mathbb{Z}_3:\quad& n_1 =  3( 6 c_1^2 +\mathcal{S}_7 \mathcal{S}_9  - \mathcal{S}_9^2- \mathcal{S}_7^2 + c_1 (\mathcal{S}_7+\mathcal{S}_9))\,. \\
	\end{split}
\end{align}
Indeed, the sum of all singlets solves the integrality conditions.

\section{Conclusions}
In this work we have presented a novel technique to determine base independent expressions for the matter loci of families of complete intersection fibers in toric ambient spaces.
To this end we exploit the duality between F-theory and M-theory and the Gopakumar-Vafa invariants that are encoded in the toplogical string partition function which we in turn calculate using mirror symmetry.
The matter loci have previously been calculated only for the sixteen families of hypersurfaces~\cite{Klevers:2014bqa} and incomplete data was obtained for two of the 3134 codimension two complete intersections~\cite{Cvetic:2013qsa,Braun:2014qka,Oehlmann:2016wsb}.
One reason for this is that the classical approach often depends on ideal decompositions with a computational complexity that quickly grows with the number of complex structure coefficients of the family of fibers.
However, the number of coefficients is large in particular for families of fibers that generically lead to multiple sections or multi-sections and for fibers that are complete intersections in higher codimension.

Our technique, that we refer to as GV-spectroscopy, is applicable to complete intersections in arbitrary codimensions.
It is particularly efficient when the rank of the gauge group, and therefore the number of K\"ahler moduli, is small~\footnote{Without much optimization we were able to study families with up to five K\"ahler moduli that are induced from the fiber.
This is the amount of moduli that arise in fibers that engineer the standard model gauge group.}.
This opens up a vast new class of geometries for geometric engineering of F-theory vacua.
Let us also note that the procedure is algorithmic to the point that it can be implemented on a computer and could be used to perform systematic scans.

We have provided several examples of families of fibers and determined their matter loci, the most complicated engineering a gauge group $G=SU(3)\times SU(2)\times \mathbb{Z}_2$.
In every case we found that all of the six-dimensional supergravity anomalies cancel for any choice of base and fibration.
This provides a strong consistency check of our technique.

As an application we studied two Higgs chains that are engineered by extremal transitions among complete intersection fibers and end on a vacuum with $G=\mathbb{Z}_4$.
The field theoretic analysis of the Higgs transitions provides another non-trivial check of the spectra that we determined with GV-spectroscopy.
Moreover, following~\cite{Cvetic:2015moa}, we commented on the implications for the Tate-Shafarevich group associated to fibrations of bi-quadrics.
We also discussed the cancellation of the discrete anomalies that have been studied in~\cite{Monnier:2018nfs}.

In the future we hope to apply GV-spectroscopy in various contexts.
As we already noted in Section~\ref{sec:su3su2z2model}, the technique can also be used to understand the structure of non-flat fibers.
The latter engineer superconformal matter that is coupled to the supergravity theory and are a topic of intense research~\cite{Apruzzi:2018nre,Apruzzi:2019vpe,Apruzzi:2019opn,Apruzzi:2019enx}.
Another exotic structure that we want to investigate are multiple fibers that appear in certain genus one fibrations.
The Tate-Shafarevich group is then replaced by the Weil-Ch\^atelet group and it would be very interesting to study the relation to Higgs transitions with fluxes.

Finally, as we already pointed out in the introduction, the fiber GV invariants that encode the F-theory spectrum are analogous to the fiber GV invariants of K3-fibrations
and the latter encode the reduced Gromov-Witten theory of the K3 fiber.
Similarly, our results can be interpreted as an attempt to construct an enumerative theory for families of Calabi-Yau onefolds.
It would be very interesting to investigate this further.

\appendix
\section{Base independent intersection calculus}
\label{app:baseindependentintersection}
As we have explained in Section~\ref{sec:gvcalculation_baseindependent}, base independent expressions for the triple intersection numbers as well as for the Euler characteristic can be obtained
by imposing the periodicity of the fiber Gopakumar-Vafa invariants.
However, it is often not practical to calculate the derivatives of the free energy up to sufficiently high orders.
It is then necessary to be able to calculate the triple intersection numbers and, if one wants to check cancellation of the pure gravitational anomalies, also the Euler characteristic with an independent method.

In this appendix we will present such a base independent intersection calculus for fibers that are complete intersections in toric ambient spaces.
Our algorithm is applicable to arbitrary complete intersection fibers and all steps can be automated.
We will first develop the formal ingredients.
This leads to a step-by-step procedure to evaluate intersections which we then illustrate at the hand of an example.
Note that related ideas have been used by~\cite{Cvetic:2013uta,Klevers:2014bqa} for hypersurfaces in toric ambient spaces, see also~\cite{Esole:2019ynq}.

Let us assume that the fiber is a codimension $r$ complete intersection in a $d+r$-dimensional toric variety $\mathbb{P}_\Delta$ with fan $\Sigma_\Delta$ and denote the homogeneous coordinates on $\mathbb{P}_\Delta$ by
\begin{align}
	[x_1:\dots:x_{d'}]\in\mathbb{P}_\Delta\,,
\end{align}
where we introduced $d'=d+r+k$.
The corresponding ray generators will be denoted by $\rho_i\in\Sigma_\Delta(1)$ and $k$ is the number of linear relations among those generators.
The nef partition determines a decomposition of the indices $i=1,...,d'$ into $r$ sets $\mathcal{I}_1,...,\mathcal{I}_r$ with associated divisors $D_{\Delta_1},\dots,D_{\Delta_r}$.

Recall that the Stanley-Reisner ideal $\mathcal{SRI}\subset\mathbb{Z}[x_1,\dots,x_{d'}]$ is defined as
\begin{align}
	\mathcal{SRI}=\langle x_{i_1}\cdot\dots\cdot x_{i_s}\, |\, \rho_{i_1},\dots,\rho_{i_s}\text{ are not part of a common cone in }\Sigma\rangle\,,
\end{align}
and that every monomial
\begin{align}
	x_{i_1}\cdot\dots\cdot x_{i_s}\in\mathcal{SRI}\,,\quad i_1<\dots<i_s\,,
\end{align}
encodes a non-intersection $[x_{i_1}]\cdot\dots\cdot[x_{i_s}]=0$.
We will denote the set of monomials of degree $s$ in $\mathcal{SRI}$ by $\mathcal{M}_s\subset\mathcal{SRI}$ .
Moreover, every point $m\in \mathbb{Z}^{d+r}$ corresponds to a linear equivalence relation
\begin{align}
	\sum\limits_{i=1}^{d'}\langle m,\rho_i\rangle\cdot[x_i]\sim 0\,,
	\label{eqn:linrel}
\end{align}
where $[x_i]$ denotes the divisor associated to $\{x_i=0\}$.
The corresponding linear polynomials in the homogeneous coordinates generate another ideal $\mathcal{J}\subset\mathbb{Z}[x_1,\dots,x_{d'}]$.
The Jurkiewicz-Danilov theorem states~\cite{cox2011toric} that the ring
\begin{align}
	R(\Sigma)=\mathbb{Z}[x_1,\dots,x_{d'}]/(\mathcal{SRI}+\mathcal{J})\,,
\end{align}
is isomorphic to the cohomology ring $H^*(\mathbb{P}_\Delta,\mathbb{Z})$ under the identification $x_i\mapsto [x_i]$.

Note that $H^{n,n}(\mathbb{P}_\Delta)=\empty$ for $n>d+r$ and therefore all polynomials of degree greater than $d+r$ have
to be equivalent to zero in $R(\Sigma)$.
Furthermore, we can use the relations $\mathcal{J}$ to express all elements of $\mathbb{Q}[x_1,\dots,x_{d'}]$ in terms of $k$ generators.
This determines an isomorphism
\begin{align}
	\phi:\,\mathbb{Q}[x_1,\dots,x_{d'}]/\mathcal{J}\rightarrow \mathbb{Q}[y_1,\dots,y_{k}]\,.
	\label{eqn:xyiso}
\end{align}
and we can identify $\mathbb{Q}\otimes R(\Sigma)$ with
\begin{align}
	R'=\mathbb{Q}[y_1,...,y_{k}]/\phi(\mathcal{SRI})\,.
\end{align}
We will denote the class $[y_i]$ by $J_i$ for $i=1,\dots,k$.

To parametrize fibrations, we introduce the classes $\mathcal{S}_i,\,i=1,\dots,d+r$ and $\tilde{\mathcal{S}}_j,\,j=1,...,r$ of line bundles on some base $B$ such
that the relation~\eqref{eqn:linrel} is modified to
\begin{align}
	\sum\limits_{i=1}^{d'}\langle e_j,\rho_i\rangle\cdot[x_i]\sim \mathcal{S}_j\,,
	\label{eqn:linrelmod}
\end{align}
where $e_j,\,j=1,...,r$ corresond to the standard basis of $\mathbb{Z}^{d+r}$,
and the divisors associated to the nef partition are shifted as
\begin{align}
	D_{\Delta_i}=\sum\limits_{j=1}^{d+r}c_{i,j}J_j\rightarrow \sum\limits_{j=1}^{d+r}c_{i,j}J_j+\tilde{\mathcal{S}}_i\,.
\end{align}
The modified relations~\eqref{eqn:linrelmod} determine an ideal
\begin{align}
	\tilde{\mathcal{J}}=\left\langle \sum_{i=1}^{d'}\langle e_j,\rho_i\rangle x_i-s_j,\, j=1,\dots,d+r\right\rangle\subset\mathbb{Q}[x_1,\dots,x_{d'},s_1,\dots,s_{d+r}]\,,
\end{align}
and the isomorphism~\eqref{eqn:xyiso} can be extended to
\begin{align}
	\tilde{\phi}:\,\mathbb{Q}[x_1,\dots,x_{d'},s_1,\dots,s_{d+r}]/\tilde{\mathcal{J}}\rightarrow \mathbb{Q}[y_1,\dots,y_{k},s_1,\dots,s_{d+r}]\,.
	\label{eqn:xyiso2}
\end{align}
We can also define the ring
\begin{align}
	\tilde{R}=\mathbb{Q}[y_1,\dots,y_{k},s_1,\dots,s_{d+r}]/\tilde{\phi}(\mathcal{SRI})\,,
\end{align}
and note that
\begin{align}
	R'=\tilde{R}/\langle s_1,\dots,s_{d+r}\rangle\,.
\end{align}
The crucial point is the following.
Every polynomial in $R'$ of degree greater than $d+r$ is equivalent to zero and this implies that every polynomial in $\tilde{R}$ of degree greater than $d+r$ with respect to $y_1,\dots,y_{k}$ is equivalent to a polynomial
that contains only $d+r$ powers of $y_1,\dots,y_{k}$ or less.
The map $y_i\mapsto J_i,\,s_i\mapsto \mathcal{S}_i$ translates this into relations in the cohomology of the fibration.

Let us denote the dimension of the base by $d_B=\text{dim}(B)$.
Then a second source of relations comes from the fibration structure which implies that any monomial in $\mathcal{S}_i$ and $\tilde{\mathcal{S}_j}$ of degree greater than $d_B$ is equivalent to zero.
Together this leads to the following algorithm.

\paragraph{The algorithm} Let us assume that we want to calculate the intersection of $d$ divisors
\begin{align}
	D_1,\dots,D_{d}\in\{J_1,...,J_{k},\mathcal{S}_{1},\dots,\mathcal{S}_{d+r},\tilde{\mathcal{S}}_1,\dots,\tilde{\mathcal{S}}_r,c_1,\dots c_{d_B}\}\,,
\end{align}
on the genus one fibered Calabi-Yau complete intersection, where $c_i=\pi^{-1}c_i(B)$.
This can be done as follows.
\begin{enumerate}
	\item Multiply the polynomial with $D_{\Delta_1}\cdot\dots\cdot D_{\Delta_r}$.
	\item Use the relations from $\tilde{\phi}(\mathcal{SRI})$ to replace all monomials that contain more than $d+r$ factors of $J_1,\dots,J_{k}$.
	\item Set all monomials that contain more than $d_B$ factors $\mathcal{S}_i,\tilde{\mathcal{S}}_j,c_k$, where $c_k$ counts $k$-times, to zero.
	\item Replace all products of $J_1,\dots, J_{k}$ by the corresponding values in the intersection ring \textit{of the fiber}.
\end{enumerate}
The result is a homogeneous polynomial of degree $d_B$ in $\mathcal{S}_{i},\tilde{\mathcal{S}}_j,c_k$ that provides a base independent expression for the result.

\paragraph{Example} We will now illustrate this procedure at the hand of an example.
Consider again the bi-quadric in $\mathbb{P}^3$ with homogeneous coordinates $[x_1:\dots:x_4]$.
The Stanley-Reisner ideal is given by
\begin{align}
	\mathcal{SRI}=\langle x_1x_2x_3x_4\rangle\,,
\end{align}
and the ideal of linear relations is
\begin{align}
	\mathcal{J}=\langle x_1-x_4,x_2-x_4,x_3-x_4\rangle\,.
\end{align}
We can therefore identify $\mathbb{Q}\otimes R(\Sigma)$ with
\begin{align}
	R'=\mathbb{Q}[y_1]/\langle y_1^4\rangle\,,
\end{align}
and this is isomorphic to the rational cohomology ring of $\mathbb{P}^3$.
With $[y_1]=J_1$ the nef partition corresponds to
\begin{align}
	D_{\Delta_1}=2J_1\,,\quad D_{\Delta_2}=2J_1\,.
\end{align}
We now lift this to a fibration by introducing the classes $\mathcal{S}_1,...,\mathcal{S}_4,\tilde{\mathcal{S}}_1,\tilde{\mathcal{S}}_2$ of line bundles on a two-dimensional base $B$ and shift $J_1$ as well as $D_{\Delta_1},D_{\Delta_2}$.
The modified relations are
\begin{align}
	\tilde{\mathcal{J}}=\langle x_1-x_4-s_1,x_2-x_4-s_2,x_3-x_4-s_3\rangle\,.
\end{align}
By adjunction, the Euler characteristic of a threefold fibration $M$ can be written as
\begin{align}
	\chi(M)=\int\limits_M\frac{(1+c_1+c_2)(1+J_1)\prod\limits_{i=1}^3(1+J_1+\mathcal{S}_i)}{(1+2J_1+\tilde{\mathcal{S}}_1)(1+2J_1+\tilde{\mathcal{S}}_2)}\,,
\end{align}
where in the integrand we implicitly restrict the divisors to the complete intersection.
Expanding the quotient and already dropping terms that are of degree larger than two with respect to the base classes one arrives at the integrand
\begin{align}
	\begin{split}
		P=&-4J_1^3+J_1^2(2 c_1 + 3\mathcal{S}_1 + 3\mathcal{S}_2 + 3\mathcal{S}_3 - 6 \tilde{\mathcal{S}}_1 - 6 \tilde{\mathcal{S}}_2)\\
		&+J_1(-c_1\mathcal{S}_1 - c_1\mathcal{S}_2 - 2\mathcal{S}_1\mathcal{S}_2 - c_1\mathcal{S}_3 - 2\mathcal{S}_1\mathcal{S}_3 - 2\mathcal{S}_2\mathcal{S}_3 + 2 c_1 \tilde{\mathcal{S}}_1 + 3\mathcal{S}_1 \tilde{\mathcal{S}}_1\\
		&+ 3\mathcal{S}_2 \tilde{\mathcal{S}}_1 + 3\mathcal{S}_3 \tilde{\mathcal{S}}_1 - 4 \tilde{\mathcal{S}}_1^2 + 2 c_1 \tilde{\mathcal{S}}_2 + 3\mathcal{S}_1 \tilde{\mathcal{S}}_2 + 
		3\mathcal{S}_2 \tilde{\mathcal{S}}_2 + 3\mathcal{S}_3 \tilde{\mathcal{S}}_2 - 4 \tilde{\mathcal{S}}_1 \tilde{\mathcal{S}}_2 - 4 \tilde{\mathcal{S}}_2^2)\,.
	\end{split}
\end{align}
We can evaluate the integral over $P$ according to the algorithm.

In the first step we multiply with the shifted divisors $D_{\Delta_1}\cdot D_{\Delta_2}$ to obtain
\begin{align}
	P'=(2J_1+\tilde{\mathcal{S}}_1)(2J_1+\tilde{\mathcal{S}}_2)P\,.
\end{align}
Using the relation from the Stanley-Reisner ideal we can replace
\begin{align}
	J_1^4\rightarrow -\left(J_1^3 \mathcal{S}_1 + J_1^3 \mathcal{S}_2 + J_1^2 \mathcal{S}_1 \mathcal{S}_2 + J_1^3 \mathcal{S}_3 + J_1^2 \mathcal{S}_1 \mathcal{S}_3 + J_1^2 \mathcal{S}_2 \mathcal{S}_3 + J_1 \mathcal{S}_1 \mathcal{S}_2 \mathcal{S}_3\right)\,,
\end{align}
and again set all monomials of degree larger then two with respect to the base classes to zero.
Finally, setting $J_1^3=1$ we obtain
\begin{align}
	\begin{split}
		\chi(M)=&-2 \left(6 c_1\mathcal{S}_1 + 14\mathcal{S}_1^2 + 6 c_1\mathcal{S}_2 + 24\mathcal{S}_1\mathcal{S}_2 + 14\mathcal{S}_2^2 + 6 c_1\mathcal{S}_3 + 24\mathcal{S}_1\mathcal{S}_3 + 24\mathcal{S}_2\mathcal{S}_3\right. \\
		&+ 14\mathcal{S}_3^2 - 6 c_1 \tilde{\mathcal{S}}_1 - 25\mathcal{S}_1 \tilde{\mathcal{S}}_1 - 25\mathcal{S}_2 \tilde{\mathcal{S}}_1 - 25\mathcal{S}_3 \tilde{\mathcal{S}}_1 + 14 \tilde{\mathcal{S}}_1^2 - 6 c_1 \tilde{\mathcal{S}}_2 - 25\mathcal{S}_1 \tilde{\mathcal{S}}_2 \\
		&\left.- 25\mathcal{S}_2 \tilde{\mathcal{S}}_2 - 25\mathcal{S}_3 \tilde{\mathcal{S}}_2 + 22 \tilde{\mathcal{S}}_1 \tilde{\mathcal{S}}_2 + 14 \tilde{\mathcal{S}}_2^2\right)\,.
	\end{split}
\end{align}
After identifying the bundles according to~\eqref{eqn:ex2sident} this is equal to $-\tilde{n}^0_4$ in~\eqref{eqn:ex2gv}.

\section{Example: Multiple fibers}
\label{sec:multi}
Here we discuss the family of complete intersection fibers (5,3)~\footnote{\url{http://wwwth.mpp.mpg.de/members/jkeitel/weierstrass/data/5_3.txt}} that generically leads to fibrations with multiple fibers.
The generic gauge group appears to be $G=SU(2)\times\mathbb{Z}_4$ and the theory also exhibits a Higgs transition to a vacuum with $G=\mathbb{Z}_4$ that is engineered by the fibrations of bi-quadrics.
However, the F-theory lift of the associated M-theory vacuum is subtle due to the presence of non-split fibers in codimension one which is closely related to the presence of multiple fibers in codimension two.
The associated issues will be discussed in~\cite{AndersonA} (see also~\cite{Baume:2017hxm}).

We directly provide the toric data associated to the auxilliary polytope
\begin{align}
	\begin{blockarray}{rrrrrrrrrl}
		& & & & & & & \\
		\begin{block}{r(rrrrr|rrr)l}
			e_1&   1&   0&   0& 0& 0& 1&-1&  0&\rho_1\\
			  x&  -1&   0&   0& 0& 0& 1& 0&s_1&\rho_2\\
			  y&   0&   1&   0& 0& 0& 0& 1&s_2&\rho_3\\
			  z&   0&   0&   1& 0& 0& 0& 1&s_3&\rho_4\\
			  w&   1&  -1&  -1& 0& 0& 0& 1&  0&\rho_5\\
			   &   0&   0&   0& 1& 0& 0& 0&  1&\rho_6\\
			   &   0&   0&   0& 0& 1& 0& 0&  1&\rho_7\\
			   &   *&   *&   *&-1&-1& 0& 0&  1&\rho_8\\
		\end{block}
	\end{blockarray} \,.
	\label{eqn:mfadata}
\end{align}
The fourth nef partition can be parametrized as
\begin{align}
	D_{\nabla_1}=D_{\rho_1}+D_{\rho_2}+D_{\rho_3}+\tilde{s}_1\cdot \pi^{-1}(H)\,,\quad D_{\nabla_2}=D_{\rho_4}+D_{\rho_5}+\tilde{s}_2\cdot \pi^{-1}(H)\,,
\end{align}
and corresponds to the polynomials
\begin{align}
	\begin{split}
		p_1=&s_{1,1}\cdot w^2 + s_{1,3}\cdot w y+ s_{1,4}\cdot w z + s_{1,8}\cdot y^2  + s_{1,9}\cdot y z + s_{1,10}\cdot z^2\,,\\
		p_2=&s_{2,1}\cdot e_1^2 w^2 + s_{2,2}\cdot e_1 w x + s_{2,3}\cdot e_1^2 w y+ s_{2,4}\cdot e_1^2 w z + s_{2,5}\cdot x^2 \\
		&+ s_{2,6}\cdot e_1 x y+ s_{2,7}\cdot e_1 x z + s_{2,8}\cdot e_1^2 y^2   + s_{2,9}\cdot e_1^2 y z + s_{2,10}\cdot e_1^2 z^2\,.
	\end{split}
\end{align}
Note that a different nef partition of the same polytope is discussed in Section~\ref{sec:nef51}.
The bundles for the homogeneous coordinates are summarized in Figure~\ref{fig:Nef51} and for the coefficients they are given in Table~\ref{tab:s1p3_coefficients}.

We choose a triangulation that corresponds to the Stanley-Reisner ideal
\begin{align}
	\mathcal{SRI}=\langle xe_1,\,yzw\rangle\,.
\end{align}
It is then easy to see that $[e_1]$ does not intersect the generic fiber.
In fact, it restricts to a fibral divisor that resolves $I_2$ fibers over $\{s_{1,5}=0\}$ which is in the class
\begin{align}
	[s_{1,5}]=-c_1+\mathcal{S}_2+\mathcal{S}_6+\mathcal{S}_7\,.
\end{align}
The other divisors $[x],[y],[z]$ and $[w]$ all restrict to four-sections.
A transition to nef (0,0) can be performed by blowing down $[e_1]$ and deforming the polynomial $p_2$.

We can calculate the GV invariants from the auxilliary data and perform the replacement~\eqref{eqn:ex2sident} to match the conventions for the bi-quadric from Section~\ref{fig:fiberchains}.
The spectrum is particularly easy to interpret when we introduce the degrees
\begin{align}
	d_{e_1}=[e_1]\cdot \beta\,,\quad d_x=[x]\cdot \beta\,,
\end{align}
and denote the corresponding GV invariants by $\tilde{n}^0_{d_{e_1},d_x}$.
\begin{table}
\begin{align*}
	\begin{array}{|rr|c|}
	\hline
	\multirow{2}{*}{$d_{e_1}$}&\multirow{2}{*}{$d_x$}&\multirow{2}{*}{$\tilde{n}^0_{d_{e_1},d_x}$}\\
	&&\\\hline
	\multicolumn{2}{|c}{\three_0}&\\\hline
	-2&0&\multirow{1}{*}{$(-c_1 + \mathcal{S}_2 + \mathcal{S}_6 + \mathcal{S}_7) (-2 c_1 + \mathcal{S}_2 + \mathcal{S}_6 + \mathcal{S}_7)$}\\\hline
	\multicolumn{2}{|c}{\two_0}&\\\hline
	-1& 0&\multirow{1}{*}{$2 (-c_1 + \mathcal{S}_2 + \mathcal{S}_6 + \mathcal{S}_7) (2 c_1 - \mathcal{S}_6 - \mathcal{S}_7 + \mathcal{S}_9)$}\\\hline
	\multicolumn{2}{|c}{\two_1}&\\\hline
	-1& 1&\multirow{2}{*}{$2 (-c_1 + \mathcal{S}_2 + \mathcal{S}_6 + \mathcal{S}_7) (3 c_1 - \mathcal{S}_2 - \mathcal{S}_9)$}\\
	 1& 1&\\\hline
	\multicolumn{2}{|c}{\two_2}&\\\hline
	-1& 2&\multirow{2}{*}{$2 (-c_1 + \mathcal{S}_2 + \mathcal{S}_6 + \mathcal{S}_7) (2 c_1 - \mathcal{S}_6 - \mathcal{S}_7 + \mathcal{S}_9)$}\\
	 1& 2&\\\hline
	\multicolumn{2}{|c}{\two_3}&\\\hline
	-1& 3&\multirow{2}{*}{$2 (-c_1 + \mathcal{S}_2 + \mathcal{S}_6 + \mathcal{S}_7) (3 c_1 - \mathcal{S}_2 - \mathcal{S}_9)$}\\
	 1& 3&\\\hline
	\multicolumn{2}{|c}{\one_1}&\\\hline
	 0& 1&\multirow{1}{*}{$4 (3 c_1 - \mathcal{S}_2 - \mathcal{S}_9) (2 c_1 - \mathcal{S}_6 - \mathcal{S}_7 + \mathcal{S}_9)$}\\\hline
	\multicolumn{2}{|c}{\one_2}&\\\hline
		\multirow{2}{*}{0}& \multirow{2}{*}{2}&\multirow{2}{*}{$\begin{array}{c}2 (10 c_1^2 - 5 c_1 \mathcal{S}_2 + \mathcal{S}_2^2 - 2 c_1 \mathcal{S}_6 - 2 c_1 \mathcal{S}_7 + 4 \mathcal{S}_6 \mathcal{S}_7 \\+ c_1 \mathcal{S}_9 + 2 \mathcal{S}_2 \mathcal{S}_9 - 4 \mathcal{S}_6 \mathcal{S}_9 - 4 \mathcal{S}_7 \mathcal{S}_9 + \mathcal{S}_9^2)\end{array}$}\\&&\\\hline
	\multicolumn{2}{|c}{\one_3}&\\\hline
	 0& 3&\multirow{1}{*}{$4 (3 c_1 - \mathcal{S}_2 - \mathcal{S}_9) (2 c_1 - \mathcal{S}_6 - \mathcal{S}_7 + \mathcal{S}_9)$}\\\hline
	\end{array}
\end{align*}
\caption{Base independent fiber Gopakumar-Vafa invariants for fibrations with generic fiber realized as nef partition $(5,3)$.}
\label{tab:53gv}
\end{table}
However, we have to carefully take into consideration that sometimes geometrically distinct curves are numerically equivalent.
This happens for the curves counted by $\tilde{n}^0_{-1,0}=\tilde{n}^0_{-1,1}=\tilde{n}^0_{1,1}$.
Over the corresponding locus the fiber is of $I_4$ type and there are two fibral curves for each of the degrees $(-1,0)$, $(-1,1)$ and $(1,1)$.
The multiplicity of the corresponding representations is therefore
\begin{align}
	n_{\two_0}=\frac12\tilde{n}^0_{-1,0}\,,\quad n_{\two_2}=\frac12\tilde{n}^0_{-1,0}\,.
\end{align}
The same happens for $\tilde{n}^0_{0,2}$ which leads to the multiplicity $n_{\one_2}=\frac12\tilde{n}^0_{0,2}$.
The Euler characteristic is identical to $\tilde{n}^0_{0,4}$ and reads
\begin{align}
	\begin{split}
		\chi_{(5,3)}=&-2 (14 c_1^2 - 8 c_1 \mathcal{S}_2 + 2 \mathcal{S}_2^2 - 7 c_1 \mathcal{S}_6 + 2 \mathcal{S}_2 \mathcal{S}_6 + 3 \mathcal{S}_6^2 - 7 c_1 \mathcal{S}_7 \\
		&+ 2 \mathcal{S}_2 \mathcal{S}_7 + 2 \mathcal{S}_6 \mathcal{S}_7 + 3 \mathcal{S}_7^2 - 5 c_1 \mathcal{S}_9 + 2 \mathcal{S}_2 \mathcal{S}_9 + 3 \mathcal{S}_9^2)\,.
	\end{split}
\end{align}
All in all we find the multiplicities of hypermultiplets
\begin{align}
	\begin{split}
	n_{\two_0}=n_{\two_2}=& (-c_1 + \mathcal{S}_2 + \mathcal{S}_6 + \mathcal{S}_7) (2 c_1 - \mathcal{S}_6 - \mathcal{S}_7 + \mathcal{S}_9)\,,\\
	n_{\two_1}=&2 (-c_1 + \mathcal{S}_2 + \mathcal{S}_6 + \mathcal{S}_7) (3 c_1 - \mathcal{S}_2 - \mathcal{S}_9)\,,\\
	n_{\one_1}=&4 (3 c_1 - \mathcal{S}_2 - \mathcal{S}_9) (2 c_1 - \mathcal{S}_6 - \mathcal{S}_7 + \mathcal{S}_9)\,,\\
		n_{\one_2}=&10 c_1^2 - 5 c_1 \mathcal{S}_2 + \mathcal{S}_2^2 - 2 c_1 \mathcal{S}_6 - 2 c_1 \mathcal{S}_7 + 4 \mathcal{S}_6 \mathcal{S}_7 \\&+ c_1 \mathcal{S}_9 + 2 \mathcal{S}_2 \mathcal{S}_9 - 4 \mathcal{S}_6 \mathcal{S}_9 - 4 \mathcal{S}_7 \mathcal{S}_9 + \mathcal{S}_9^2\,,\\
		n_{\one_0}=&13-c_1^2-\frac12\chi_{(5,3)}\,.
	\end{split}
	\label{eqn:multispec}
\end{align}
It is easy to check that all of the gauge-, gravitational- and mixed-anomalies as well as the discrete anomalies cancel.

The difference in the Euler numbers of $(5,3)$ and $(0,0)$ is given by
\begin{align}
	\frac12 (\chi_{(5,3)}-\chi_{(0,0)})=2n_{\two_0}+2(g_{SU(2)}-1)\,,
\end{align}
where $g_{SU(2)}$ is the genus of the curve $\{s_{1,5}=0\}\subset B$,
\begin{align}
	g_{SU(2)}=1+\frac12 [s_{1,5}]([s_{1,5}]-c_1)\,,
\end{align}
and corresponds to the number of hypermultiplets in the adjoint representation of $SU(2)$.
The difference reflects the fact that the $SU(2)$ is broken by a vev of the scalars in the $\two_0$ representation.
Moreover, we can match the multiplicites
\begin{align}
	n_{(1)}=2n_{\two_1}+n_{\one_1}\,,\quad n_{(2)}=2n_{\two_2}+n_{\one_2}\,,
\end{align}
where $n_{(1)}$ and $n_{(2)}$ are given in Table~\ref{tab:nef00Spectrum}.

As has been discussed in~\cite{Oehlmann:2016wsb}, there is a second family of $I_2$ fibers over the locus
\begin{align}
	V=\{s_{1,10} s_{1,3}^2 - 4 s_{1,1} s_{1,10} s_{1,8} + s_{1,4}^2 s_{1,8} - s_{1,3} s_{1,4} s_{1,9} + s_{1,1} s_{1,9}^2=0\}\subset B\,,
\end{align}
that is in the class
\begin{align}
	[V]=2c_1-\mathcal{S}_6-\mathcal{S}_7+\mathcal{S}_9\,.
\end{align}
Over this locus the polynomial $p_1$ factorizes as
\begin{align}
	p_1\sim (d s_{1,8}y+b_+z+c_+w)(d s_{1,8}y+b_-z+c_-w)\,,
\end{align}
with
\begin{align}
	b_\pm=\left(b_0\pm b_1\sqrt{d}\right)\,,\quad c_\pm=\frac{d}{2}\left(s_{1,3}\pm\sqrt{d}\right)\,,
\end{align}
and
\begin{align}
	\begin{split}
		b_0=&s_{1,3}^2s_{1,9}-4s_{1,1}s_{1,8}s_{1,9}\,,\quad b_1=s_{1,3}s_{1,9}-2s_{1,4}s_{1,8}\,,\quad d=s_{1,3}^2-4s_{1,1}s_{1,8}\,.
	\end{split}
\end{align}
The components of the fiber experience monodromy when moving around $\{d=0\}$.
Moreover, at $V\cap\{d=0\}$ the polynomial $p_1$ becomes a perfect square and one finds \textit{multiple fibers} with multiplicity two.
The class of $\{d=0\}$ is
\begin{align}
	[d]=2c_1-2\mathcal{S}_7\,.
\end{align}
Perhaps surprisingly, this does not seem to affect the spectrum of the theory.
However, let us stress that the Gopakumar-Vafa invariants correspond to traces over multiplicities of representations.
It could be that contributions to the BPS spectrum from the multiple fibers, perhaps due to the torsional normal sheaf~\cite{Bhardwaj:2015oru}, cancel.
On the other hand, the corresponding degrees of freedom should be relevant for the gravitational anomaly which cancels given the spectrum~\eqref{eqn:multispec}.
Since we do not have a clear understanding of this phenomenon, and due to the subtleties associated with non-split fibers in codimension one, we relegated the discussion of this family to the appendix.

\section{List of Euler numbers}
\label{app:eulers}
In this appendix we provide the generic Euler numbers for the fibers studied in Section~\ref{sec:CICYFI}.
This determines the number of neutral singlets via
\begin{align}
\label{eq:neutral}
	n_{(0)}=h^{2,1}(X)+1 = 12+\text{rank}(G)+\sum_i \text{dim}(C_i) n_i -c_1^2  -\frac12 \chi \, ,
\end{align}
where $G$ is the generic gauge group.
We also included the contribution for $n_i$ superconformal subsectors \cite{Buchmuller:2017wpe,Dierigl:2018nlv} that arise in the smooth geometry from non-flat fibers (e.g. see~\cite{Apruzzi:2018nre}).
Each non-flat fiber contributes an additional set of divisors and the corresponding K\"ahler moduli parametrize the $\text{dim}(C_i)$ dimensional Coulomb branch of the SCFT in five dimensions~\cite{Apruzzi:2019vpe,Apruzzi:2019vpe, Apruzzi:2019enx}. 
The Euler numbers were computed via two independent methods.
One is the base independent intersection calculus that we described in the previous Appendix~\ref{app:baseindependentintersection} and the other is GV-spectroscopy.
Both lead to the results
\begin{align} 
	\begin{split}
		\chi_{(67,0)}=&   -2   \left(12c_1^2 - 6c_1\mathcal{S}_2 + 2\mathcal{S}_2^2 - 6c_1\mathcal{S}_6 + 2\mathcal{S}_2\mathcal{S}_6 + 2\mathcal{S}_6^2 - 6c_1\mathcal{S}_7 + 2\mathcal{S}_2\mathcal{S}_7 \right. \\ 
		&\quad \left.+ \mathcal{S}_6\mathcal{S}_7 + 2\mathcal{S}_7^2 - 9 c_1\mathcal{S}_9 + 3\mathcal{S}_2\mathcal{S}_9 + 3\mathcal{S}_6\mathcal{S}_9 + 3\mathcal{S}_7\mathcal{S}_9 + 3\mathcal{S}_9^2\right)\,,    \\ 
\chi_{(21,0)}=& -2  (12 c_1^2 - 6 c_1 \mathcal{S}_2 + 2 \mathcal{S}_2^2 - 4 c_1 \mathcal{S}_6 + \mathcal{S}_2 \mathcal{S}_6 + 2 \mathcal{S}_6^2 - 6 c_1 \mathcal{S}_7  \\ & \quad+ 
   2 \mathcal{S}_2 \mathcal{S}_7 + 2 \mathcal{S}_7^2 - 8 c_1 \mathcal{S}_9 + 3 \mathcal{S}_2 \mathcal{S}_9 + 3 \mathcal{S}_6 \mathcal{S}_9 + 2 \mathcal{S}_7 \mathcal{S}_9 + 3 \mathcal{S}_9^2)\,,	\\	
	\chi_{(6,1)} =&-2  \left(8 c_1^2 - 5 c_1 \mathcal{S}_2 + 3 \mathcal{S}_2^2 - 4 c_1 \mathcal{S}_6 + 3 \mathcal{S}_2 \mathcal{S}_6 + 2 \mathcal{S}_6^2 - 2 c_1 \mathcal{S}_7 \right.\\
	&\quad \left.+ \mathcal{S}_2 \mathcal{S}_7 + 2 \mathcal{S}_7^2 - 3 c_1 \mathcal{S}_9 + 3 \mathcal{S}_2 \mathcal{S}_9 + \mathcal{S}_6 \mathcal{S}_9 + \mathcal{S}_7 \mathcal{S}_9 + 2 \mathcal{S}_9^2\right)\,,\\ 
		\chi_{(5,1)}=&-2 ( 12 c_1^2 - 6 c_1 \mathcal{S}_2 + 2 \mathcal{S}_2^2 - 4 c_1 \mathcal{S}_6 + \mathcal{S}_2 \mathcal{S}_6 + 2 \mathcal{S}_6^2 - 4 c_1 \mathcal{S}_7 + \mathcal{S}_2 \mathcal{S}_7 \\ 
	& \quad- \mathcal{S}_6 \mathcal{S}_7 + 2 \mathcal{S}_7^2 - 7 c_1 \mathcal{S}_9 + 3 \mathcal{S}_2 \mathcal{S}_9 + 2 \mathcal{S}_6 \mathcal{S}_9 + 2 \mathcal{S}_7 \mathcal{S}_9 + 3 \mathcal{S}_9^2)\,, \\ 
			\chi_{(0,0)}=&-2 ( 12 c_1^2 - 7 c_1 \mathcal{S}_2 + 3 \mathcal{S}_2^2 - 4 c_1 \mathcal{S}_6 + 2 \mathcal{S}_2 \mathcal{S}_6 + 2 \mathcal{S}_6^2 - 4 c_1 \mathcal{S}_7  \\&\quad + 2 \mathcal{S}_2 \mathcal{S}_7+ 2 \mathcal{S}_7^2 - 7 c_1 \mathcal{S}_9 + 4 \mathcal{S}_2 \mathcal{S}_9 + 2 \mathcal{S}_6 \mathcal{S}_9 + 2 \mathcal{S}_7 \mathcal{S}_9 + 3 \mathcal{S}_9^2)\,. \\ 
	\end{split}
\end{align}

\addcontentsline{toc}{section}{References}
\bibliographystyle{utphys}
\bibliography{names}
\end{document}